\documentclass[twocolumn,floatfix,aps,rmp]{revtex4-1}
\usepackage{graphicx}
\usepackage{amsmath}
\usepackage{amssymb}
\usepackage{bm}

\begin{document}
\title{Random-matrix theory of Majorana fermions and topological superconductors}
\author{C. W. J. Beenakker}
\affiliation{Instituut-Lorentz, Universiteit Leiden, P.O. Box 9506, 2300 RA Leiden, The Netherlands}
\date{February 2015}
\begin{abstract}
The theory of random matrices originated half a century ago as a universal description of the spectral statistics of atoms and nuclei, dependent only on the presence or absence of fundamental symmetries. Applications to quantum dots (artificial atoms) followed, stimulated by developments in the field of quantum chaos, as well as applications to Andreev billiards --- quantum dots with induced superconductivity. Superconductors with topologically protected subgap states, Majorana zero-modes and Majorana edge modes, provide a new arena for applications of random-matrix theory. We review these recent developments, with an emphasis on electrical and thermal transport properties that can probe the Majorana fermions.
\end{abstract}
\maketitle
\tableofcontents

\section{Introduction}
\label{intro}

\subsection{What is new in RMT}
\label{whatsnew}

Random matrices made their first appearance in physics in the 1950's \cite{Wig56}, to explain the statistical properties of scattering resonances observed in nuclear reactions \cite{Por65}. Random-matrix theory (RMT) has since found applications in many branches of physics \cite{Ake11,Mehta,Forrester}. In condensed matter physics, RMT can describe the universal properties of disordered metals and superconductors \cite{Bee97,Guh98}, dependent only on the presence or absence of fundamental symmetries in 10 symmetry classes --- the socalled ``ten-fold way'' \cite{Alt97}. 

It was recently discovered that condensed matter with an excitation gap can be in different phases that are not distinguished by a broken symmetry, but by the value of a topological invariant \cite{Has10,Qi11}. Some of these topological superconductors and insulators have been realized in the laboratory, many others are being searched for. In this review we will discuss how RMT can be extended to account for topological properties.

Topological invariants count the number of protected subgap states, either bound to a defect or propagating along a boundary. In a superconductor these are Majorana fermions, described by a real rather than a complex wave function \cite{Maj37,Wil09}. The absence of complex phase factors fundamentally modifies the random-matrix description, notably, the scattering matrix at the Fermi level is real orthogonal rather than complex unitary. The circular ensemble of random orthogonal matrices, and the corresponding Gaussian ensemble of real antisymmetric matrices,  was treated in text books \cite{Mehta,Forrester} for its mathematical elegance --- without applications in quantum physics. Topological superconductors now provide these applications.

Experimentally, the search for Majorana zero-modes (bound to a vortex core or to the end of a superconducting wire) and Majorana edge modes (propating along the boundary of a two-dimensional superconductor) is still at an initial stage \cite{Ali13}. But much is understood from the theoretical point of view \cite{Ali12,Lei12,Sta13,Bee13}, so the time seems right for a review of the RMT of Majorana fermions in the context of topological superconductivity.

The outline of the review is as follows. We continue this introductory section with some background information on superconducting quasiparticles and how they appear in experimental systems that will figure later on in the review. In Section \ref{topsuperc} the key features of topological superconductivity are introduced for the simplest --- and paradigmatic --- example, the Kitaev chain. In Section \ref{Bqp} we then discuss in some more generality the basic symmetries that govern the random-matrix ensembles of topological superconductors, distinguished from the original Wigner-Dyson ensembles by the role played by particle-hole symmetry in addition to time-reversal symmetry. 

The random matrix can be the Hamiltonian (Section \ref{Hens}) or the scattering matrix (Section \ref{Smatrixens}) --- we discuss both, but in the applications to transport properties we focus on the scattering matrix ensembles. Electrical and thermal transport properties are considered separately in Sections \ref{elG} and \ref{thG}. We pay particular attention to experimental signatures of the topological quantum numbers, from a broader perspective than just the search for Majoranas. Topological phase transitions play a central role in the Josephson effect, as we discuss in Section \ref{Jeffect}. We conclude in Section \ref{conclude}.

\subsection{Superconducting quasiparticles}
\label{supercondqp}

The fermionic excitations $\Psi$ of a superconductor are called Bogoliubov quasiparticles \cite{Bog58}. Unlike the electron and hole excitations $\psi_{e}$ and $\psi_{h}$ of a normal metal, the state $\Psi$ has no definite charge --- it is a coherent superposition of $\psi_{e}$ (negatively charged, filled state at energy $E$ above the Fermi level $E_{\rm F}$) and $\psi_{h}$ (positively charged, empty state at $E$ below $E_{\rm F}$). The $\pm 2e$ fluctuations in the quasiparticle charge are absorbed by Cooper pairs of the superconducting condensate. 

The fact that the charge of Bogoliubov quasiparticles is only conserved modulo $2e$ is at the origin of the symmetry of charge conjugation, also called particle-hole symmetry. It expresses the ambiguity that a quasiparticle excitation can be thought of either as a Cooper pair missing a particle or as a Cooper pair having an extra particle. 

A consequence of particle-hole symmetry is the existence of a correspondence between Bogoliubov quasiparticles and Majorana fermions \cite{Cha10,Wil14,Ell14}, a concept from particle physics referring to a particle that is its own antiparticle \cite{Maj37}. The correspondence breaks down if Coulomb interactions become important, because these remove the equivalence modulo $2e$ of charge $+e$ and charge $-e$ excitations. Since Coulomb interactions are strongly screened in a superconductor, the Majorana representation remains a useful starting point to describe physical phenomena such as the mutual annihilation of two colliding Bogoliubov quasiparticles \cite{Bee14}.

Bogoliubov quasiparticles can be bound by a magnetic vortex or an electrostatic defect \cite{Car64}. Particle-hole symmetry requires that the bound states come in pairs at $\pm E$, with the possibility of an unpaired state at $E=0$. The creation and annihilation operators are related by $a_E=a^{\dagger}_{-E},\label{aadagger}$ so they are identical at $E=0$ (at the Fermi level). This self-conjugate bound state, $a_0=a_0^{\dagger}$, is called a Majorana zero-mode or Majorana bound state (or sometimes just Majorana fermion, when no confusion with unbound Bogoliubov quasiparticles can arise). 

A Majorana zero-mode has a certain stability, it cannot be displaced away from the Fermi level without breaking the $\pm E$ symmetry of the spectrum \cite{Vol99}. If a vortex contains a nondegenerate state at $E=0$, then it will remain pinned to the Fermi level if we perturb the system. This robustness is called ``topological protection'' and a superconductor that supports Majorana zero-modes is called a topological superconductor (or a ``topologially nontrivial'' superconductor). For an overview of the ongoing search for Majorana zero-modes in superconductors, see \textcite{Ali12,Lei12,Sta13,Bee13}.

The ground state of $2n$ vortices containing Majorana zero-modes is $2^n$-fold degenerate and the exchange of pairs of vortices is a unitary operation on the ground-state manifold \cite{Rea00,Iva01}. Such non-commuting exchange operations (``non-Abelian statistics'') are at the basis of proposals to store and manipulate quantum information in Majorana zero-modes \cite{Kit01,Kit03,Nay08}. We will not address this application here, for a recent review see \textcite{Das15}.

\subsection{Experimental platforms}
\label{platforms}

Random-matrix theory is designed for an ensemble of chaotic scatterers,\footnote{A scatterer is called chaotic if it uniformly mixes the incoming and outgoing degrees of freedom. Mathematically, this uniformity is expressed by the Haar measure on the unitary group of scattering matrices, see Sec.\ \ref{chaotic_scat}.} without making specific assumptions on how this ensemble is realized. In a typical application the scattering is due to disorder and the different members of the ensemble have different disorder configurations. Alternatively, the chaotic dynamics may result from some irregularly shaped boundary and the ensemble is produced by varying the boundary shape or the energy. 

In the electronic context the chaotic scatterer is referred to as a quantum dot or quantum billiard, pointing to the quantization of the energy spectrum by the confinement. When superconductivity enters the chaotic dynamics is governed by the interplay of normal scattering from the electrostatic potential and Andreev scattering from the pair potential. One then speaks of Andreev billiards, with a spectrum of Andreev levels, see \textcite{Bee05} for a review.

Since random-matrix theory addresses universal properties, there is a great variety of experimental systems to which it might be applied. We give a brief overview of platforms that seem most promising for applications of RMT to Majorana fermions.

\begin{figure}[tb]
\centerline{\includegraphics[width=0.9\linewidth]{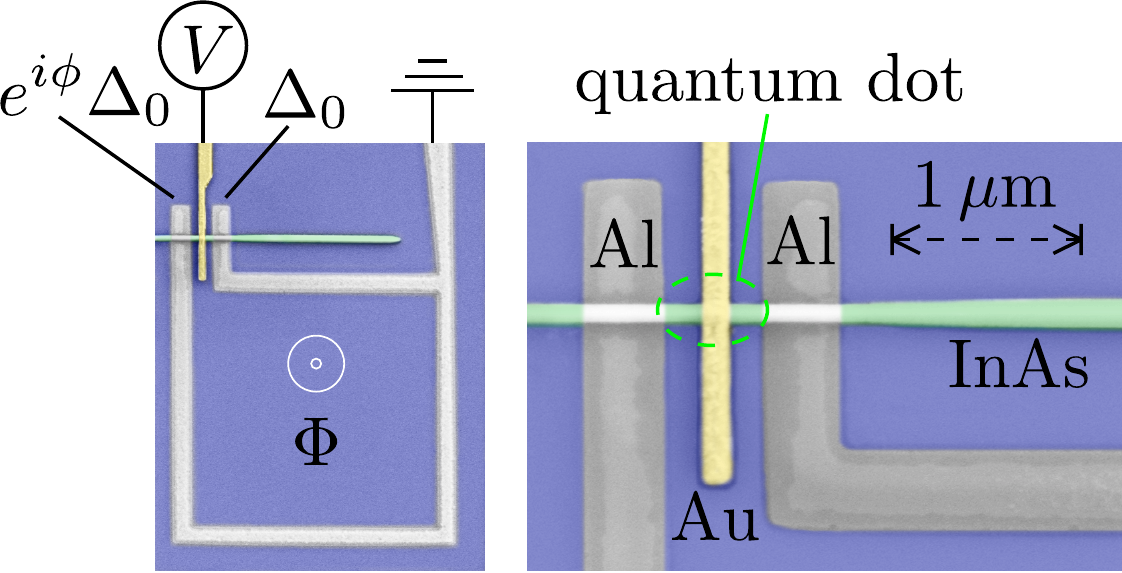}}
\caption{
Quantum-dot Josephson junction formed by a segment of a semiconducting wire (InAs) in a superconducting ring (Al), enclosing a magnetic flux $\Phi$. A weakly coupled tunnel probe (Au) at bias voltage $V$ measures the excitation spectrum of electron and hole quasiparticles confined to the quantum dot, as a function of the phase difference $\phi=\Phi\times 2e/\hbar$ across the junction. Electron micrographs (entire device and enlarged region) from \textcite{Cha12}.
}
\label{fig_quantumdot}
\end{figure}

\begin{figure}[tb]
\centerline{\includegraphics[width=0.6\linewidth]{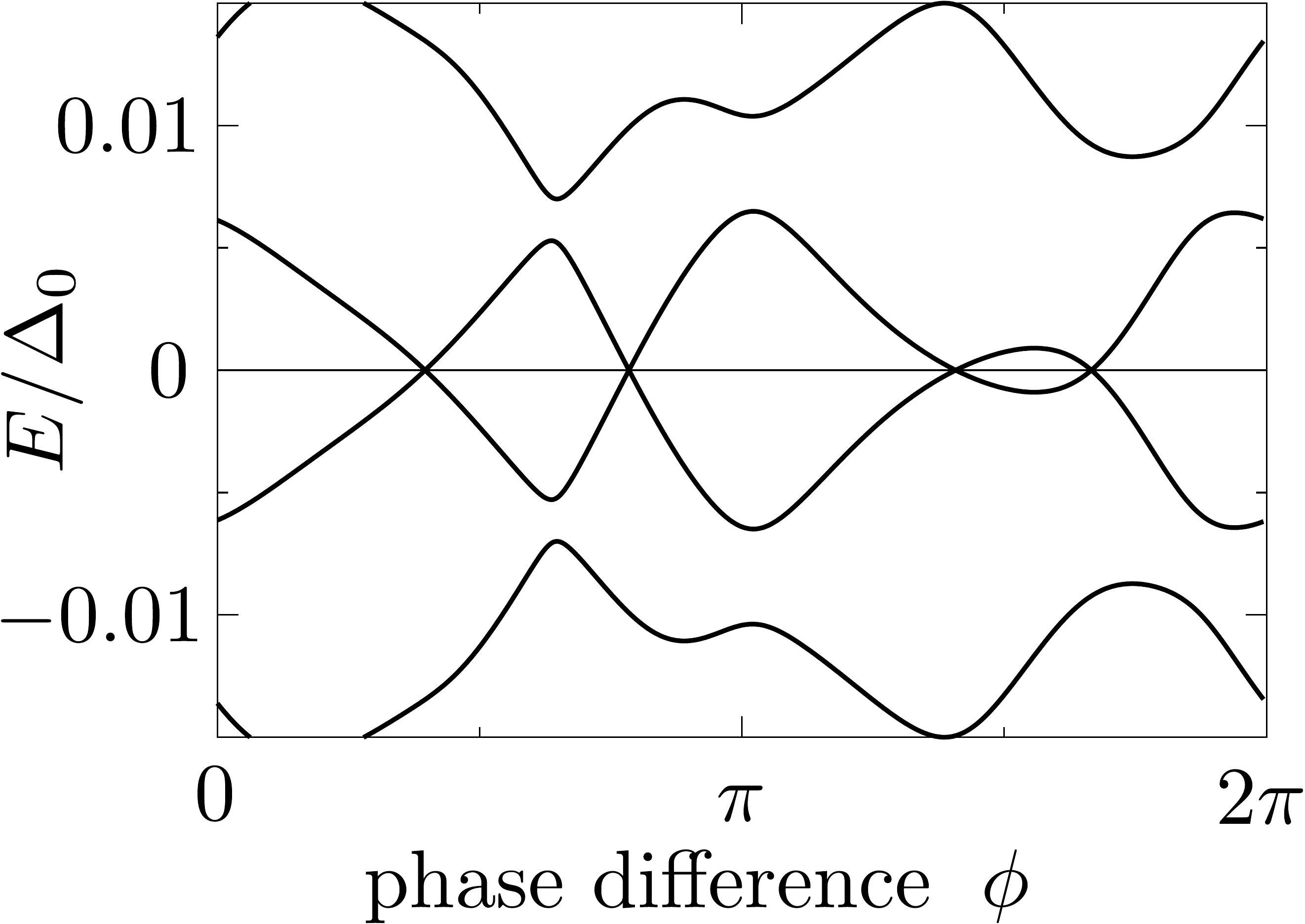}}
\caption{Model calculation of the excitation spectrum of a Josephson junction similar to Fig.\ \ref{fig_quantumdot}, but of larger dimensions in order to increase the flux inside the junction (about $3\,h/e$ for a disordered normal region of size $1\,\mu{\rm m}\times 2\,\mu{\rm m}$). The spin-orbit coupling length is $l_{\rm so}=250\,{\rm nm}$, the superconducting gap $\Delta_0=0.4\,{\rm meV}$, and the Fermi energy $E_{\rm F}=2.5\,{\rm meV}$ (corresponding to $N=20$ electronic modes in the nanowire). Level repulsion at nonzero $E$ coexists with level crossings at $E=0$. Here the number of crossings in a $2\pi$ phase increment is \textit{even}, but it can be \textit{odd} in a topological superconductor. Data provided by M. Wimmer.}
\label{fig_repulsion}
\end{figure}

\textit{Nanowire SNS geometry ---}
The nanowire device of Fig.\ \ref{fig_quantumdot} is a good starting point to introduce a characteristic feature of topological superconductivity. A Josephson junction, or SNS junction, is formed by two superconducting electrodes (S) connected via a normal region (N), in this case a segment of a semiconducting wire representing a quantum dot \cite{Cha12}. Electron and hole quasiparticles (filled states above the Fermi level or empty states below it) are confined to the quantum dot by the superconducting gap $\Delta_0$ that the superconductor induces locally in the wire. Andreev reflection at the normal-superconductor (NS) interface couples the electron and hole quasiparticles.

A model calculation of the spectrum in a similar device is shown in Fig.\ \ref{fig_repulsion}, as a function of the superconducting phase difference $\phi$ across the SNS junction. Notice the special role played by the Fermi level $E=0$. As $\phi$ is varied, pairs of levels that approach each other repel without crossing if they stay away from the Fermi level, however, level crossings appear at $E=0$. In Fig.\ \ref{fig_repulsion} there is an even number $N_X=4$ of level crossings between $\phi=0$ and $\phi=2\pi$. It $N_X$ is odd, the superconductor is called topologically nontrivial. 

Level crossings appear generically if a magnetic field is applied to close the excitation gap in the quantum dot, and if spin-orbit coupling removes the spin degeneracy of the Andreev levels \cite{Alt97}. (In the InAs nanowire of Fig.\ \ref{fig_quantumdot} the spin-orbit coupling is produced by the Rashba effect.) The transition from $N_X$ even to $N_X$ odd requires a Zeeman energy $E_{\rm Z}$ larger than $\Delta_0$ \cite{Lut10,Ore10}. This topological phase transition has observable consequences \cite{Kit01}: The supercurrent $I_n(\phi)\propto dE_n/d\phi$ carried by a level $E_n$ with an odd number of crossings has $4\pi$-periodicity in $\phi$, meaning a doubling of the flux-periodicity from $h/2e$ to $h/e$. 

Sec.\ \ref{Jeffect} explains the connection between level crossings and switches in the ground-state fermion parity and describes their statistics in the framework of RMT. 

\begin{figure}[tb]
\centerline{\includegraphics[width=0.9\linewidth]{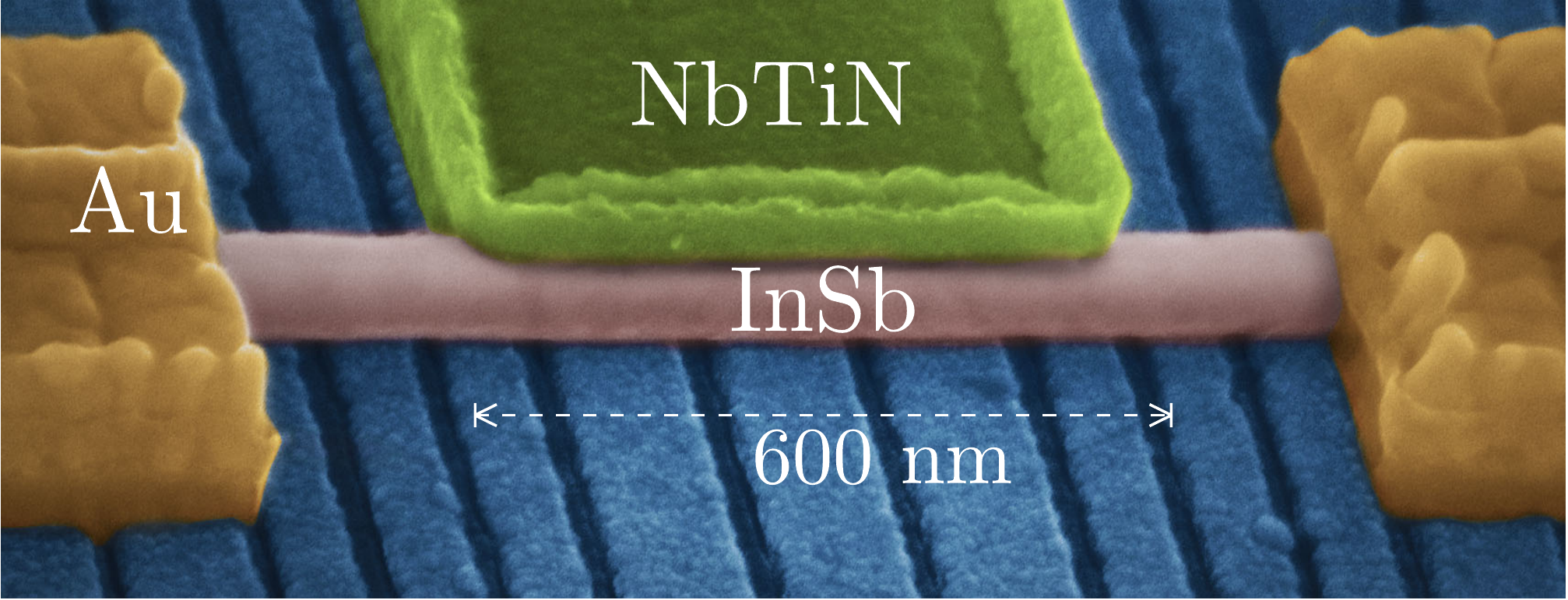}}
\caption{Scanning electron microscope image of a device similar to the one studied in  \textcite{Mou12}, where superconductivity is induced in an InSb nanowire by proximity to a NbTiN superconductor. The Au electrodes are normal-metal Ohmic contacts, used to measure the electrical conductance of the wire. Image provided by V. Mourik.}
\label{fig_nanowire}
\end{figure}

\textit{Nanowire NSN geometry ---}
Each level crossing in the SNS junction of Fig.\ \ref{fig_quantumdot} produces a pair of zero-modes between the superconducting electrodes. The NSN junction of Fig.\ \ref{fig_nanowire}, with one rather than two superconducting electrodes, allows for the study of a single zero-mode in isolation. One can imagine gradually decoupling the superconductors by inserting a tunnel barrier in the Josephson junction. As one raises the barrier height the Andreev levels are pushed away from $E=0$ and the $\phi$-dependence becomes flat. The level crossings can annihilate pairwise without affecting the ground-state fermion parity. However, if $N_X$ is odd one level crossing must remain no matter how high the tunnel barrier has become. This remaining level crossing for a topologically nontrivial superconductor corresponds to a pair of isolated Majorana zero-modes, one for each NS interface.

\begin{figure}[tb]
\centerline{\includegraphics[width=0.9\linewidth]{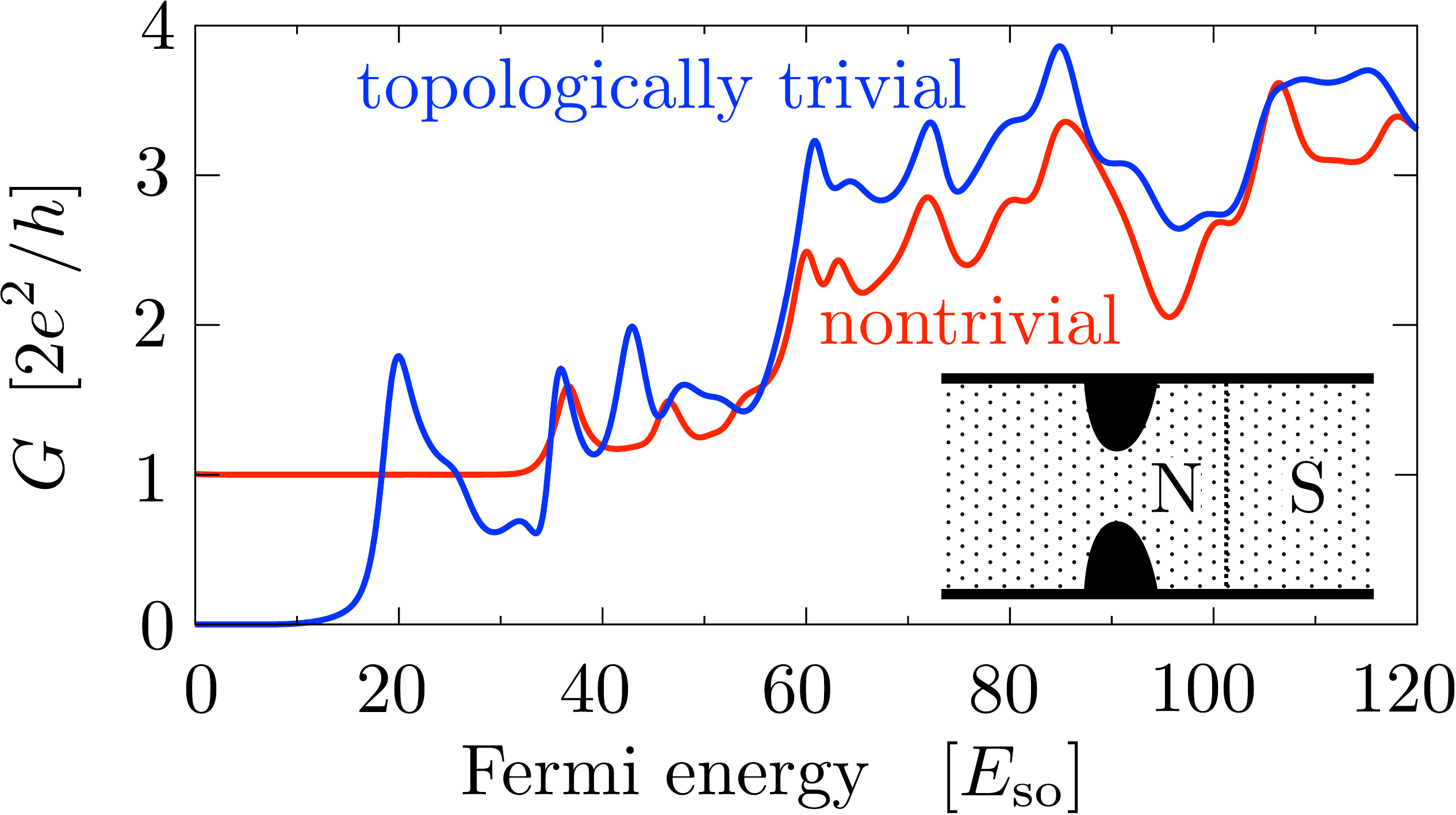}}
\caption{Model calculation of a device similar to Fig.\ \ref{fig_nanowire}, with an additional point contact (as shown in the inset) to allow for variation of the number of transmitted modes. The ratio of the spin-orbit coupling energy $E_{\rm so}$ and superconducting gap $\Delta_0$ is adjusted (at fixed Zeeman energy $E_{\rm Z}=6\,E_{\rm so}$), so that the superconductor is either in the topologically trivial phase ($\Delta_0/E_{\rm so}=8$) or non-trivial phase ($\Delta_0/E_{\rm so}=4$). By varying the Fermi energy inside the point contact the number $N$ of transmitted electronic modes is varied between 0 and 8. The first conductance plateau in the topologically nontrivial phase remains precisely quantized, notwithstanding the presence of a large amount of disorder in the simulation. Figure adapted from \textcite{Wim11}.}
\label{fig_QPC}
\end{figure}

The conductance $G$ of the NS interface reveals the presence of the zero-mode, as shown in the model calculation of Fig.\ \ref{fig_QPC}. Notice that the Majorana fermion in such a device is only weakly bound: The wave function leaks out into the normal electrode and the zero-mode is a broad resonance centered at the Fermi energy rather than a discrete level at $E=0$. Still, the distinction between a topologically trivial and nontrivial superconductor remains clearly visible in the conductance, in particular in the single-mode regime, when $G=2e^2/h$ or 0 depending on the presence or absence of a quasi-bound Majorana fermion.

In the context of RMT, the distinction appears because of a difference in the sign of the determinant of the reflection matrix $r$ of the NS interface: ${\rm Det}\,r=+1$ or $-1$ for a topologically trivial or nontrivial superconductor, respectively. The consequence of this sign change for the conductance statistics is discussed in Sec.\ \ref{elG}.

\begin{figure}[tb]
\centerline{\includegraphics[width=0.8\linewidth]{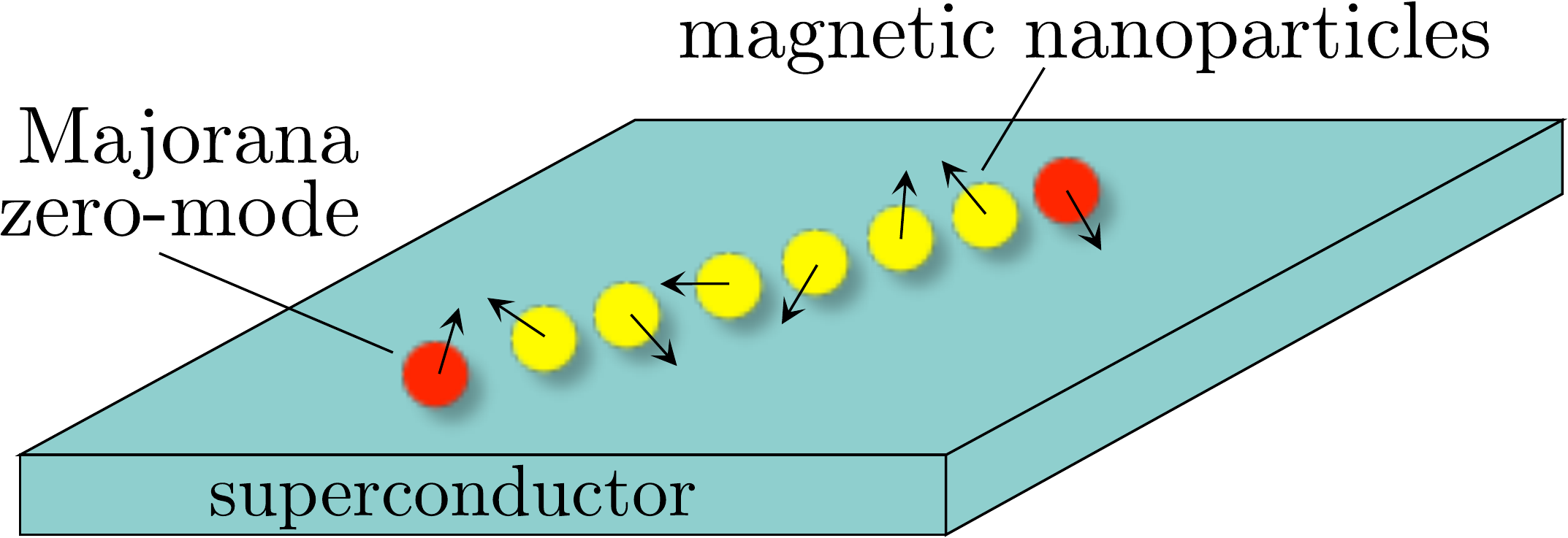}}
\caption{
Chain of magnetic nanoparticles on a superconducting substrate. If the direction of the magnetization (arrows) varies along the chain, there can be Majorana zero-modes at the ends without any spin-orbit coupling in the superconductor.
}
\label{fig_chain}
\end{figure}

\textit{Chain of magnetic nanoparticles ---}
An atomic, single-channel, variation on the semiconductor nanowire is formed by a chain of magnetic atoms on a superconducting substrate (Fig.\ \ref{fig_chain}). In such a system no Rashba spin-orbit coupling is required to produce Majorana zero-modes at the end points of the chain, it is sufficient if the magnetization varies in direction from atom to atom \cite{Cho11,Nad13,Pie13,Kli13}. In the experimental realization of Fe atoms on Pb by \textcite{Nad14}, the magnetic moments are aligned ferromagnetically, so there spin-orbit coupling in the superconductor is still needed.

\begin{figure}[tb]
\centerline{\includegraphics[width=1\linewidth]{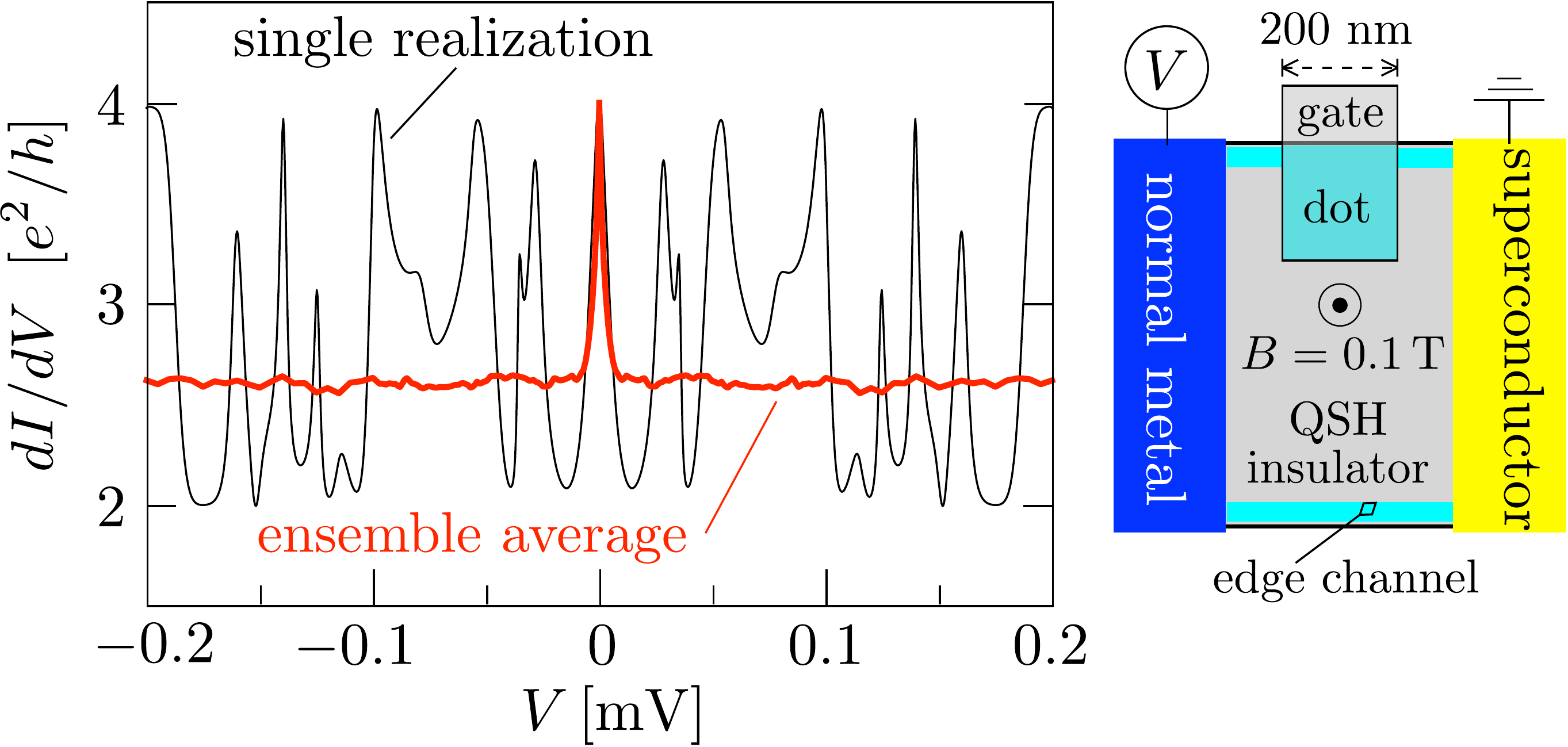}}
\caption{
Right panel: Schematic of a quantum dot, created by a gate electrode at the edge of a quantum spin-Hall (QSH) insulator in a perpendicular magnetic field $B$. A current $I$ is passed between metallic and superconducting contacts, and the differential conductance $dI/dV$ is determined as a function of the bias voltage $V$. Results of a model calculation for an InAs/GaSb quantum well are shown in the left panel. The Majorana zero-mode produces a resonant peak at $V=0$, which survives the average over disorder realizations. Figure adapted from \textcite{Bee13b}.
}
\label{fig_QSHdot}
\end{figure}

\textit{Quantum spin-Hall edge ---}
An alternative single-channel conductor is formed by the edge of a quantum spin-Hall insulator \cite{Fu09}, see Fig.\ \ref{fig_QSHdot}. This is a two-dimensional quantum well in a semiconductor heterostructure (HgTe or InAs/GaSb), with an inverted band gap in the bulk that closes at the edge \cite{Kon08}. Electrons and holes propagate along the edge in a helical mode, meaning that the direction of motion is tied to the spin direction. The reflection matrix at the interface with a superconducting electrode has determinant $-1$, indicating the presence of a Majorana zero-mode. In the geometry of Fig.\ \ref{fig_QSHdot} the zero-mode is weakly confined by a quantum dot and produces a resonant conductance peak at $V=0$.

\textit{Topological insulator surface ---}
The two-dimensional counterpart of the quantum spin-Hall edge is the conducting surface of a three-dimensional topological insulator, such as ${\rm Bi}_2{\rm Se}_3$ \cite{Has10,Qi11}. The proximity effect from a superconductor opens an inverted band gap in the conducting surface. This heterostructure then behaves as a topologically nontrivial superconductor, with Majorana zero-modes bound to the core of a magnetic vortex \cite{Fu08}. 

\begin{figure}[tb]
\centerline{\includegraphics[width=0.7\linewidth]{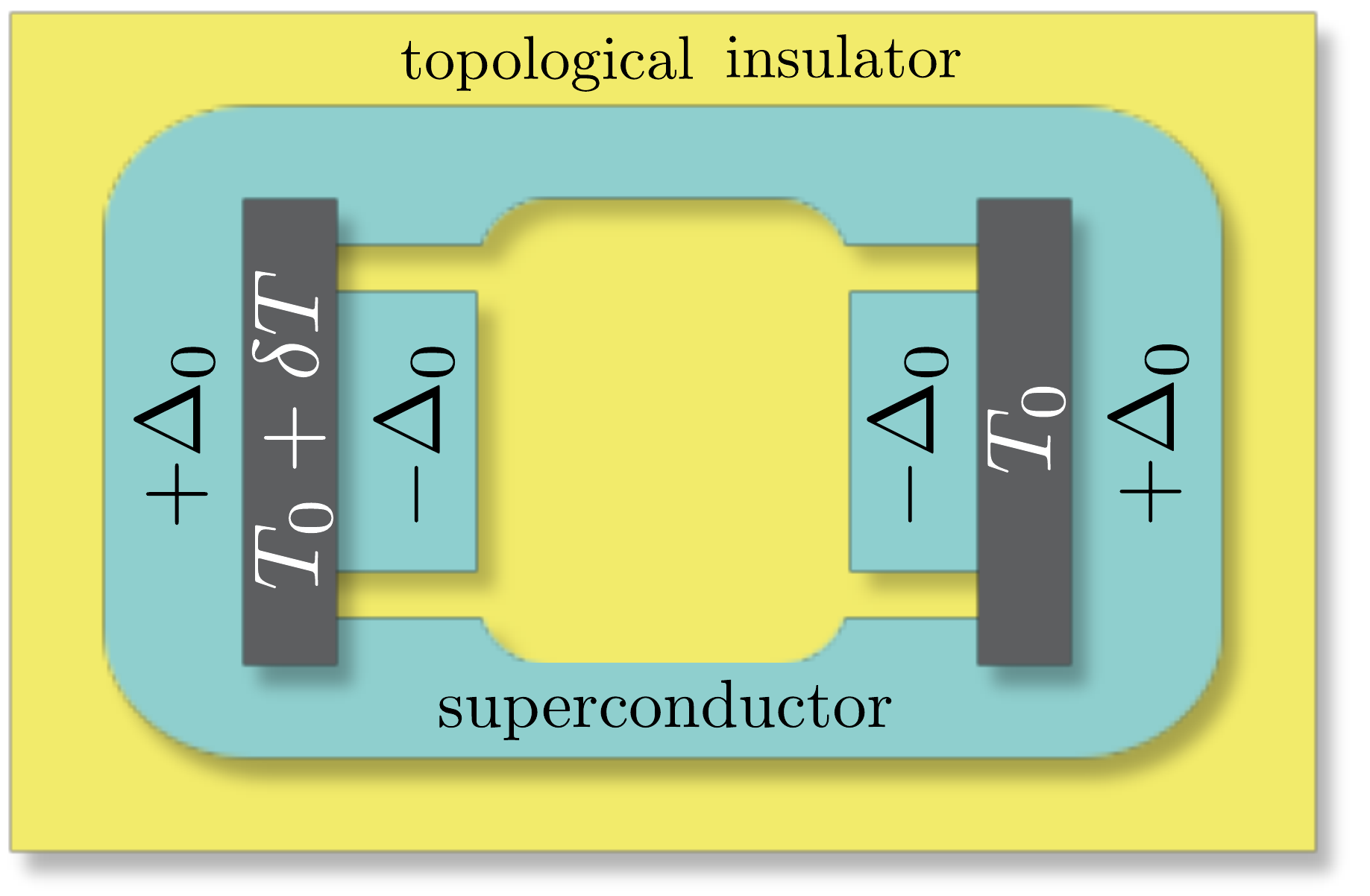}}
\caption{
Andreev billiard on the surface of a topological insulator. A sign change of the superconducting pair potential $\pm\Delta_0$ closes the excitation gap inside the billiard without breaking time-reversal symmetry. The statistics of the thermal conductance between two metal electrodes at temperature difference $\delta T$ is governed by a random-matrix ensemble in symmetry class DIII (in zero magnetic field), in class BDI (nonzero field at the Dirac point), or class D (nonzero field away from the Dirac point). Figure adapted from \textcite{Dah10}.}
\label{fig_DIIIbilliard}
\end{figure}

The electrons and holes on the surface of a topological insulator are massless Dirac fermions, similar to graphene --- but without the spin and valley degeneracy of graphene. (This is why the superconducting proximity effect does not produce Majorana zero-modes in graphene \cite{Gha12}.) Dirac fermions cannot be confined by an electrostatic potential, because of Klein tunneling, but they can be confined by a superconducting pair potential \cite{Bee08}. A quantum dot can therefore be constructed on the surface of a topological insulator by suitably patterning the superconductor, as indicated in Fig.\ \ref{fig_DIIIbilliard}. 

Chaotic scattering in this Andreev billiard cannot be probed electrically, because the superconductor acts as a short, but since the superconductor is a thermal insulator one can rely on heat transport as a probe. The statistics of the thermal conductance, dependent on the presence or absence of the fundamental symmetries of time-reversal, particle-hole conjugation, and chirality, is discussed in Sec.\ \ref{thG}.

\textit{Chiral \textit{p}-wave superconductor ---}
In all these systems the superconductor itself has the conventional spin-singlet \textit{s}-wave pairing, and it is the proximity effect that provides the topologically nontrivial phase. A two-dimensional superconducting layer with spin-triplet pairing and $p_x\pm ip_y$ orbital symmetry can be topologically nontrivial on its own. Strontium ruthenate is a candidate material for such chiral \textit{p}-wave pairing \cite{Mac03,Kal09}. This material is predicted to host Majorana zero-modes, localized by magnetic vortices, as well as Majorana edge modes, propagating along the boundaries.

\begin{figure}[tb]
\centerline{\includegraphics[width=0.6\linewidth]{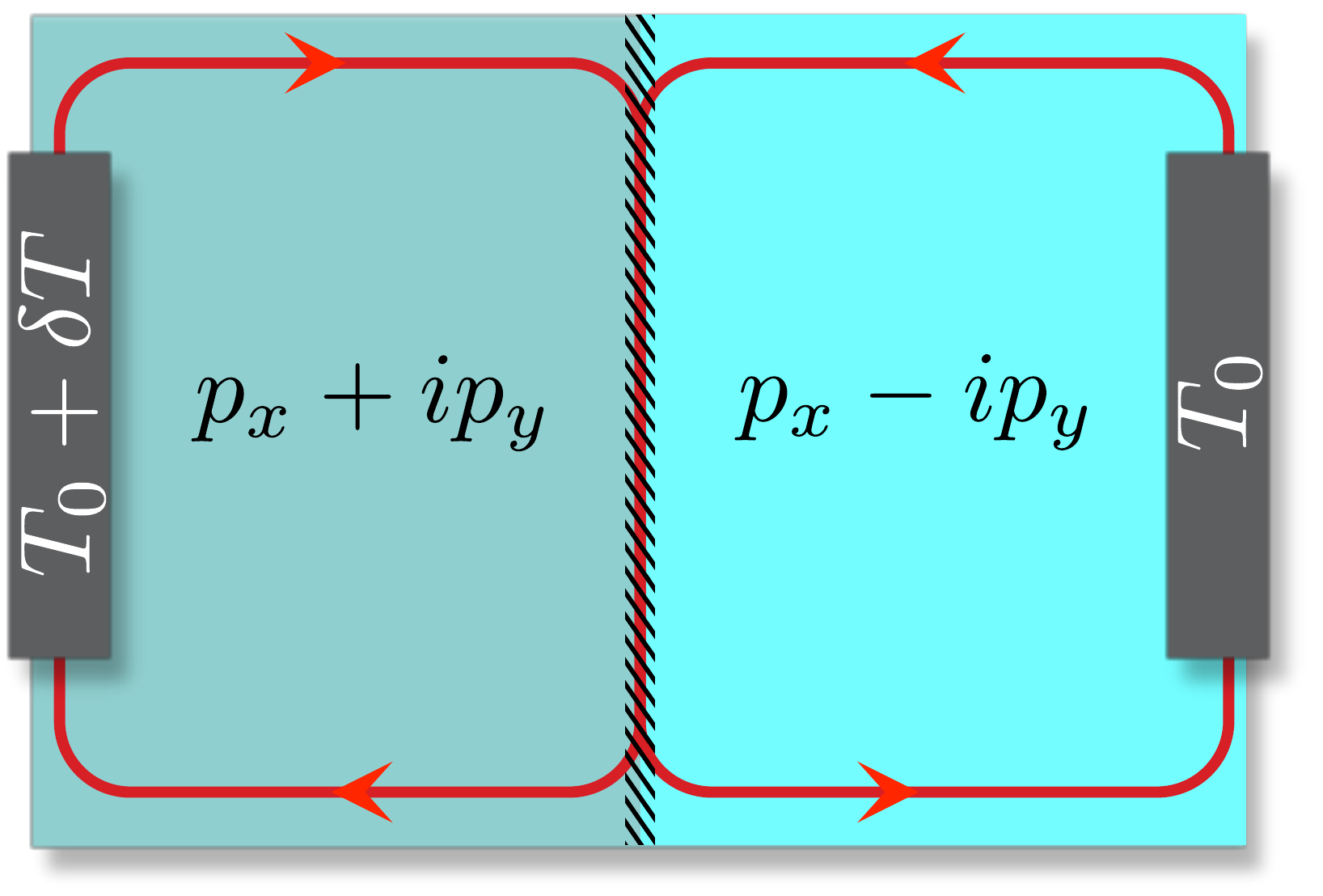}}
\caption{
Chiral \textit{p}-wave superconductor with two domains of opposite chirality. Arrows indicate the direction of propagation of the Majorana edge modes. If the two modes along the domain wall are uniformly mixed, the statistics of the thermal conductance is determined by the probability distribution of a $2\times 2$ orthogonal random matrix.
}
\label{fig_Dbilliard}
\end{figure}

The direction of propagation is opposite in domains of opposite chirality, so that a domain wall forms a conducting pathway for pairs of Majorana modes (see Fig.\ \ref{fig_Dbilliard}). The Majorana edge modes are analogous to the chiral edge modes of the quantum Hall effect, and the thermal conductance of the domain wall is then analogous to the electrical conductance of a bipolar junction between electron-doped and hole-doped regions. The random-matrix statistics is different, because the scattering matrix of the bipolar junction is complex unitary, rather than real orthogonal. This difference stems from the Majorana nature of the quasiparticle excitations of a superconductor.

\section{Topological superconductivity}
\label{topsuperc}

\subsection{Kitaev chain}
\label{Kitaevintro}

The application of RMT to superconductivity is based on a connection between the quasiparticle excitation spectrum and the eigenvalues of a real antisymmetric matrix. By way of introduction we present this formulation in the context of one of the physical systems from the previous section, the chain of magnetic nanoparticles on a superconducting substrate (Fig.\ \ref{fig_chain}). This provides possibly the simplest realization of the Kitaev chain, a paradigm for Majorana zero-modes and topological superconductivity \cite{Kit01}.

Each magnetic nanoparticle (labeled $n=1,2,\ldots M$) binds a fermionic state near the Fermi level in the superconducting gap \cite{Yu65,Shi68, Rus69}, through a competition of the magnetic exchange energy $\bm{m}_n\cdot\bm{\sigma}$ (favoring a spin-polarized state aligned with the magnetization $\bm{m}_n$) and the pairing energy $\Delta_0$ (favoring a spin-singlet state). Adjacent nanoparticles are coupled by a hopping energy $t_0$ and feel a chemical potential $\mu_0$. The mean-field Hamiltonian is
\begin{align}
&H=-\sum_{n,\alpha} \left(t_{0} a^\dagger_{n\alpha} a^{\vphantom{\dagger}}_{n+1,\alpha}+\text{H.c}\right) - \sum_{n,\alpha} \mu_0 a^\dagger_{n\alpha} a^{\vphantom{\dagger}}_{n\alpha}\nonumber\\
&\quad+ \sum_{n,\alpha,\beta} (\bm{m}_n \cdot\bm{\sigma})_{\alpha\beta} a^\dagger_{n\alpha}  a^{\vphantom{\dagger}}_{n\beta}+ \sum_{n} \left(\Delta_{0}a_{n\uparrow} a_{n\downarrow} + \text{H.c.} \right),
\label{h0}
\end{align}
where the abbreviation H.c.\ stands for Hermitian conjugate. The operator $a_{n\alpha}$ is the fermion operator for a spin-$\alpha$ electron on the $n$-th nanoparticle and $\bm{\sigma}=(\sigma_{x},\sigma_{y},\sigma_{z})$ denotes the vector of Pauli matrices.

For large magnetization, the electron spin on the $n$-th nanoparticle is polarized along $\bm{m}_{n}$. The Hamiltonian \eqref{h0} can then be projected onto the lowest spin band, resulting in an effective spinless Hamiltonian \cite{Cho11}
\begin{align}
H ={}& \sum_n \biggl[ -\bigl( {t}_{n} a^\dagger_n a_{n+1}^{\vphantom{\dagger}}+{t}'_{n} a^\dagger_n a_{n+2} ^{\vphantom{\dagger}} + \text{H.c.} \bigr) - {\mu}_{n} a^\dagger_n a_n \nonumber\\
&+ \bigl( {\Delta}_n  a_{n} a_{n+1} + \text{H.c.} \bigr) \biggr].\label{proj}
\end{align}
The coefficients $t_n,\mu_n,\Delta_n$ have become site-dependent and an additional next-nearest-neighbor hopping energy $t'_n$ has appeared. More importantly, the pairing energy now couples adjacent sites in the chain, with effective pair potential ${\Delta}_{n}$ of order $\Delta_{0}t_{0}/m_{n}$, dependent on the relative angle between $\bm{m}_{n}$ and $\bm{m}_{n+1}$. For parallel magnetic moments ${\Delta}_{n}$ vanishes.

A Hamiltonian of the form \eqref{proj} was introduced by Kitaev as a toy model for a \textit{p}-wave superconductor. In the magnetic chain the \textit{p}-wave pairing is obtained from \textit{s}-wave pairing due to the coupling of the electron spin to local magnetic moments. (Rashba spin-orbit coupling has the same effect in the semiconductor nanowires of Figs.\ \ref{fig_quantumdot} and \ref{fig_nanowire}.)  

\subsection{Majorana operators}
\label{Mfermion}

Majorana operators are defined by
\begin{equation}
\left.
\begin{array}{l}
\gamma_{n1}=a_n^{\vphantom{\dagger}}+a_n^\dagger\\
\gamma_{n2}=ia_n^{\vphantom{\dagger}}-ia_n^\dagger
\end{array}
\right\}
\Leftrightarrow 
\left\{
\begin{array}{l}
a_{n}=(\gamma_{n1}-i\gamma_{n2})/2\\
a_{n}^\dagger=(\gamma_{n1}+i\gamma_{n2})/2
\end{array}
\right.
\label{gammanpmdef}
\end{equation}
Each site (and each spin band) of the Kitaev chain is associated with a pair of Majorana operators. By construction these are Hermitian operators, $\gamma_{ns}=\gamma_{ns}^\dagger$, with anticommutation relation
\[
\gamma_{ns}\gamma_{n's'}+\gamma_{n's'}\gamma_{ns}=2\delta_{nn'}\delta_{ss'}.\label{gammacomm}
\]
We collect the Majorana operators in one large vector $\Gamma=(\gamma_{11},\gamma_{12},\gamma_{21},\gamma_{22},\ldots)$. 

In the Majorana representation the Hamiltonian becomes the bilinear form
\begin{equation}
H={\rm const.}+\sum_{n\neq m}i\Gamma_{n}\Gamma_{m}A_{nm},\;\;A=A^\ast=-A^{\rm T}.\label{HKM}
\end{equation}
The constant first term (arising from the product $\Gamma_{n}^2=1$) is an irrelevant energy offset and may be ignored. The Hamiltonian is thus represented by a  matrix $A$ which is antisymmetric (because $\Gamma_n$ and $\Gamma_m$ anticommute) and real (because ${H}$ is Hermitian). This matrix is the object to which the methods of RMT are applied.

Any real antisymmetric matrix of even dimension can be factored as \cite{You61}
\begin{equation}
A=O
\begin{pmatrix}
           0 &  E_1    &       &   &       \\
          -E_1 &   0    &      & \emptyset &            \\
       &  & \ddots &  &    \\
               & \emptyset &  &   0    &  E_{M}     \\
               &        &  &  -E_{M}    &   0    
  \end{pmatrix}O^{\rm T},\label{Youla}
\end{equation}
with $O\in{\rm O}(2M)$ a real orthogonal matrix and $E_{n}\geq 0$. The eigenvalues of $A$ come in inverse pairs $\pm E_n$. This charge-conjugation or particle-hole symmetry is intrinsic of the mean-field theory of superconductivity. For thermodynamic properties it suffices to retain only the positive $E_n$'s, but for dynamical properties one needs a complete basis for $A$ and so both positive and negative $E_n$'s need to be retained.

\begin{figure}[tb]
\centerline{\includegraphics[width=0.7\linewidth]{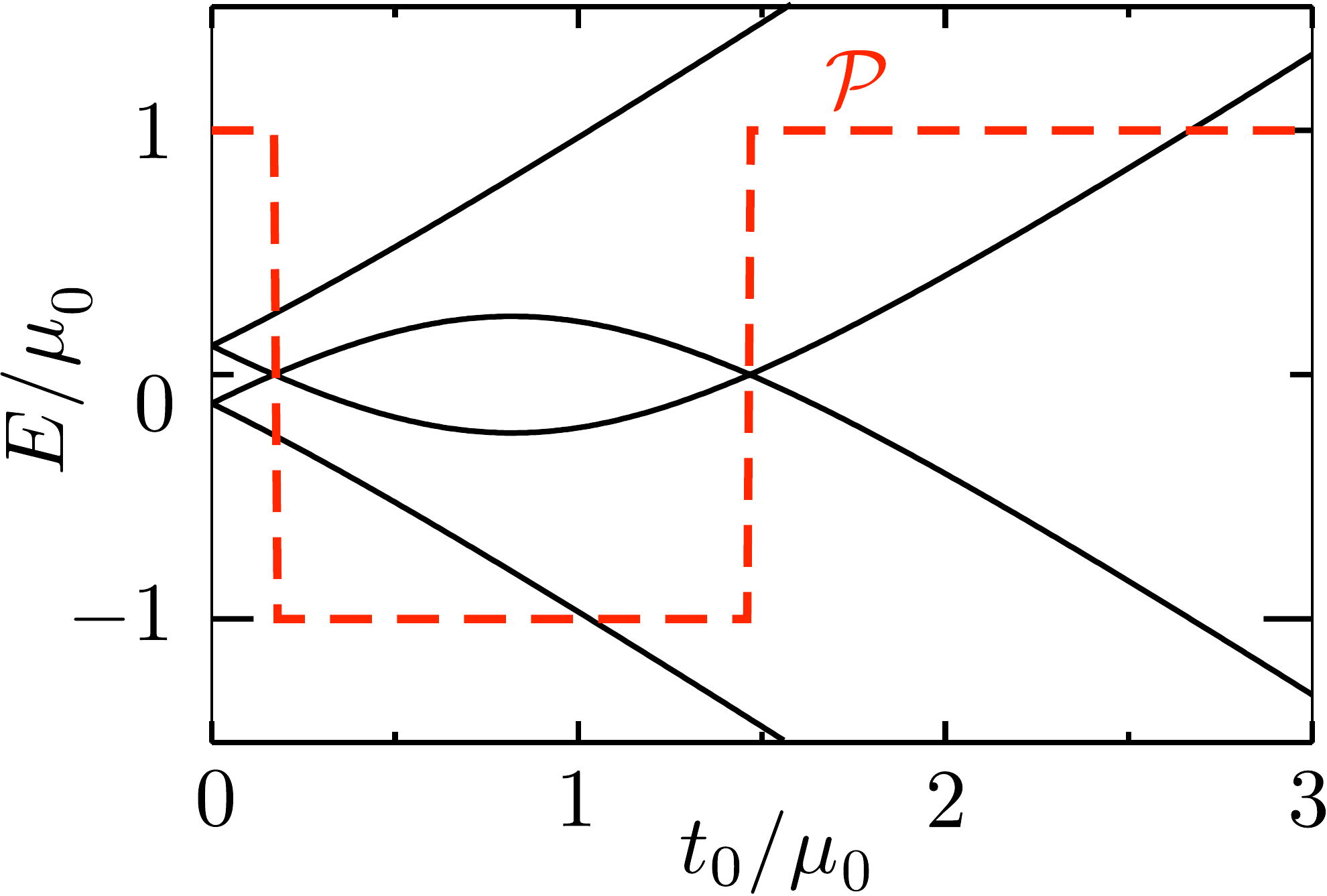}}
\caption{
Excitation spectrum (solid curves) of two magnetic particles with an angle $\theta=70^{\circ}$ between their magnetic moments, calculated from the Hamiltonian \eqref{h0} at fixed $\mu_0=|m_{n}|=2\Delta_{0}$ as a function of the hopping energy $t_{0}$. The level crossings at $E=0$ do not split because the ground states at the two sides of the crossing differ in fermion parity ${\cal P}$ [dashed curve, calculated from Eq.\ \eqref{PdetO}]. Figure adapted from \textcite{Cho11}.
}
\label{fig_2sites}
\end{figure}

The determinant of the matrix $O$ in the Youla decomposition \eqref{Youla} equals $\pm 1$ and can only change sign when one of the eigenvalues crosses zero, see Fig.\ \ref{fig_2sites}. Because of the identity ${\rm Pf}\,A={\rm Det}\,O\,\prod_{n}E_n$ one can also compute this sign directly from the Pfaffian of $A$,
\begin{equation}
{\cal P}\equiv {\rm Det}\,O={\rm sign}\,{\rm Pf}\,A=\pm 1.\label{PdetO}
\end{equation}
The physical interpretation of the quantum number ${\cal P}$ is that it gives the fermion parity of the superconducting ground state \cite{Kit01}: All electrons are paired in the ground state for ${\cal P}=+1$, while there is one unpaired electron for ${\cal P}=-1$.

\subsection{Majorana zero-modes}
\label{Mzm}

As mentioned in connection with Fig.\ \ref{fig_repulsion}, one can count level crossings to determine whether the superconductor is topologically trivial or not: One would then close the chain into a ring, pass a flux $\Phi$ through it and compare the fermion parity ${\cal P}(\Phi)$ at $\Phi=0$ and $\Phi=h/2e$. If the two fermion parities differ, the superconductor is topologically nontrivial \cite{Kit01}:
\begin{equation}
Q={\cal P}(0){\cal P}(h/2e)=\begin{cases}
+1&{\rm trivial},\\
-1&{\rm nontrivial}.
\end{cases}\label{QP0Ph2e}
\end{equation}
This is a thermodynamic signature of topological superconductivity. 

\begin{figure}[tb]
\centerline{\includegraphics[width=0.7\linewidth]{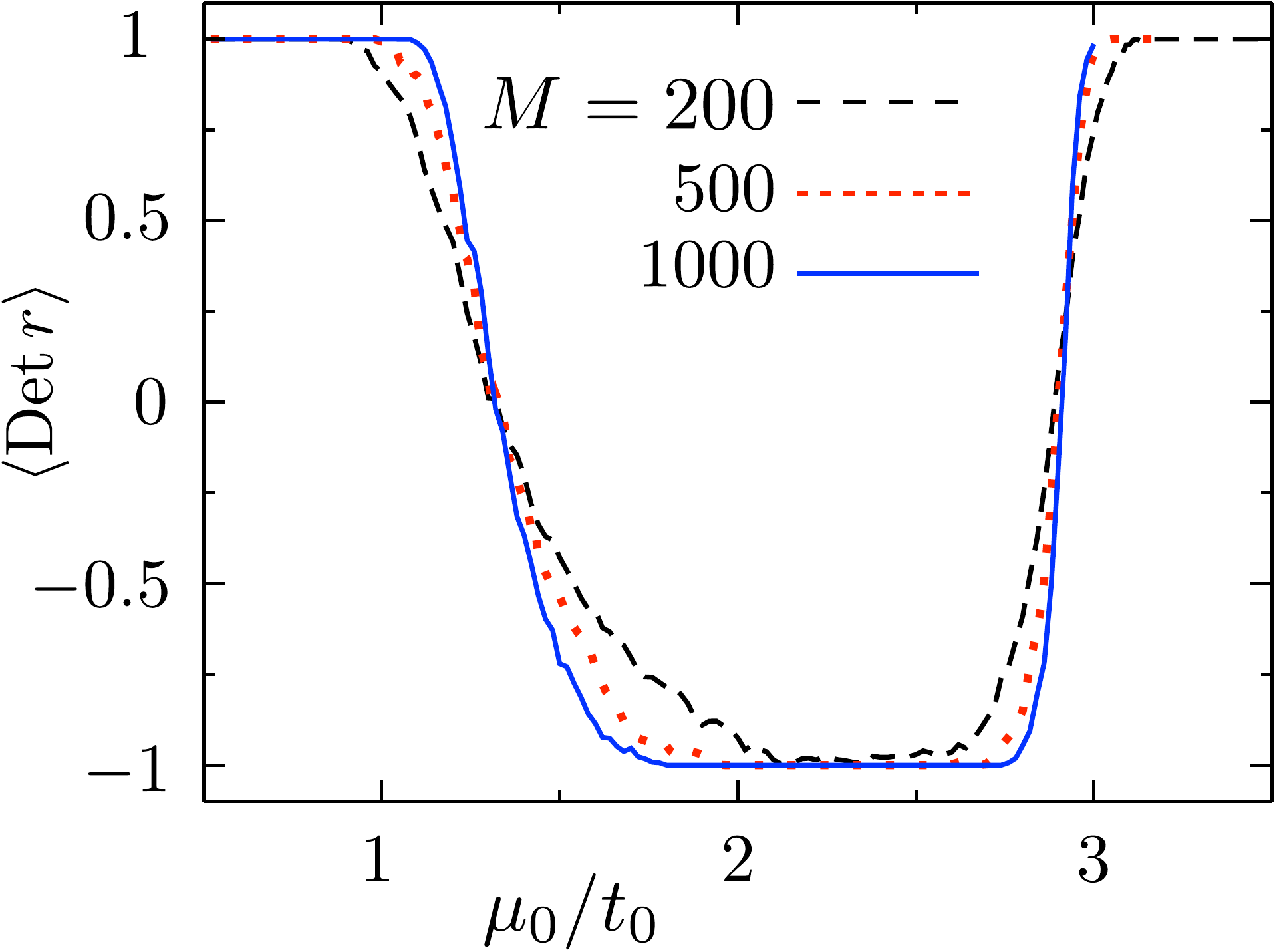}}
\caption{
Determinant of the reflection matrix at one end of a chain of magnetic nanoparticles [Hamiltonian from Eq.\ \eqref{h0}], ensemble averaged over random and uncorrelated orientations of the magnetic moments (with fixed magnitude $|m_n|=2\,t_0$ and $\Delta_0=0.9\,t_{0}$). The transition into the topologically nontrivial phase becomes sharper with increasing length $M$ of the chain. Figure adapted from \textcite{Cho11}.
}
\label{fig_ensemble}
\end{figure}

For a transport signature one would keep the chain open and determine the reflection matrix $r$ from the left or right end.\footnote{Which end does not matter, because the full scattering matrix $S$ of the open chain has determinant $+1$ and ${\rm Det}\,S=({\rm Det}\,r_{\rm left})({\rm Det}\,r_{\rm right})$ in the absence of any transmission through the chain.} At the Fermi level ($E=0$), the matrix $r\in{\rm O}(2N)$, assuming the chain is sufficiently long that transmission through it can be neglected. (The factor of two in the dimension $2N$ of $r$ refers to the electron-hole degree of freedom.) The transport equivalent of the condition \eqref{QP0Ph2e} is \cite{Akh11}:
\begin{equation}
Q={\rm sign}\,{\rm Det}\,r=\begin{cases}
+1&{\rm trivial},\\
-1&{\rm nontrivial}.
\end{cases}\label{QDetr}
\end{equation}
The transition ${\rm Det}\,r=+1\mapsto -1$ happens via a closing of the excitation gap, at which an eigenvalue of $r$ passes through zero. Such a topological phase transition is illustrated in Fig.\ \ref{fig_ensemble} for the model Hamiltonian \eqref{h0}.

To make the connection between ${\rm Det}\,r=-1$ and the appearance of a Majorana zero-mode at the end of the chain, one can argue as follows \cite{Ful11}. Upon termination of one end with a barrier, the condition for a bound state at $E=0$ is
\begin{equation}
{\rm Det}\,(1-r_{\rm B}r)=0.\label{DetrBr}
\end{equation} 
The reflection matrix $r_{\rm B}$ of the barrier has ${\rm Det}\,r_{\rm B}=1$, irrespective of ${\rm Det}\,r=\pm 1$. The number $N_0$ of bound states is the number of eigenvalues $+1$ of the orthogonal matrix $r_{\rm B}r$, while the other $2N-N_0$ eigenvalues are either equal to $-1$ or come in conjugate pairs $e^{\pm i\phi}$. Hence ${\rm Det}\,(r_{\rm B} r)={\rm Det}\, r = (-1)^{N_0}$, so if ${\rm Det}\,r=-1$ there is an unpaired ($\equiv$ Majorana) zero-mode at the end of the chain.

\subsection{Phase transition beyond mean-field}
\label{beyondMFT}

Although random-matrix theory is not inherently limited to single-particle Hamiltonians (it was originally developed for strongly interacting nuclei), the application to superconductors relies on an effective single-particle description in which the pairing interaction is treated at the mean-field level. While the full Hamiltonian conserves the total number ${\cal N}$ of electrons in the system, the mean-field Hamiltonian only conserves the fermion parity: the pairing terms $\Delta aa$ and $\Delta^* a^\dagger a^\dagger$ change ${\cal N}$ by $\pm 2$. The resulting correspondence between charge $+e$ and charge $-e$ quasiparticles is at the origin of particle-hole symmetry, which can be thought of as an emergent symmetry of the mean-field description.

It is reassuring that for special choices of the pairing interaction the Kitaev model of topological superconductivity can be solved exactly, without recourse to the mean-field approximation \cite{Ort14}. Considering a chain of $L$ sites with nearest-neighbor hopping energy $t_0$ and pairing interaction $g_0\eta(n-m)$ between sites $n$ and $m$, the Hamiltonian takes the form
\begin{subequations}
\label{RGK1}
\begin{align}
&H = -t_0\sum_{n=1}^L  \bigl( a^\dagger_n a_{n+1}^{\vphantom{\dagger}}+ \text{H.c.} \bigr) - \frac{4g_0}{L} I^\dagger I,\label{RGKa}\\
&I=\sum_{1=m<n}^L  \eta(n-m) \, a_{n}a_{m},\;\;a_{n+L}=e^{i\phi/2}a_n.\label{Iaadef}
\end{align}
\end{subequations}
The chain is closed in a ring containing a flux $\Phi=\phi\times\hbar/2e$, corresponding to periodic or antiperiodic boundary conditions for $\Phi=0$ or $\Phi=h/2e$. 

To arrive at the mean-field Kitaev Hamiltonian \eqref{proj}, one would substitute $2 I^\dagger I^{\;}\rightarrow \langle I^\dagger\rangle I +I\langle I^\dagger\rangle $, and add a chemical potential term $-\mu_0\sum_{n} a_n^\dagger a_n^{\vphantom{\dagger}}$ to control the electron density $\rho={\cal N}/L$. Because the Hamiltonian \eqref{RGK1} conserves the particle number, the appropriate ensemble is canonical rather than grand-canonical and no chemical potential term is needed.

An exact solution is possible if the pairing interaction has the \textit{p}-wave form $\sin(k/2)$ in momentum space, corresponding in real space to the long-range coupling
\begin{equation}
\eta(s)=\frac{1}{\pi}\int_{-\pi}^\pi e^{iks}\sin(k/2)\,dk= \frac{8i}{\pi}\frac{(-1)^ss}{1-4s^2}.\label{etas}
\end{equation}
(All lengths are measured in units of the lattice constant.) For this choice of interaction the Kitaev chain belongs to a class of exactly solvable pairing Hamiltonians first studied by \textcite{Ric63,Gau76}.

The mean-field Hamiltonian has a transition into a topologically nontrivial ``weak pairing'' phase \cite{Rea00} at a critical pairing energy
\begin{equation}
g_{\rm c}=\frac{t_0}{1-2\rho}.\label{gcdef}
\end{equation}
Majorana zero-modes appear at the end points of the open chain for $g_0<g_{\rm c}$. Because of the long range of the interaction potential, they only decay algebraically into the bulk \cite{Pie13,DeG13,Vod14}.

\begin{figure}[tb]
\centerline{\includegraphics[width=0.9\linewidth]{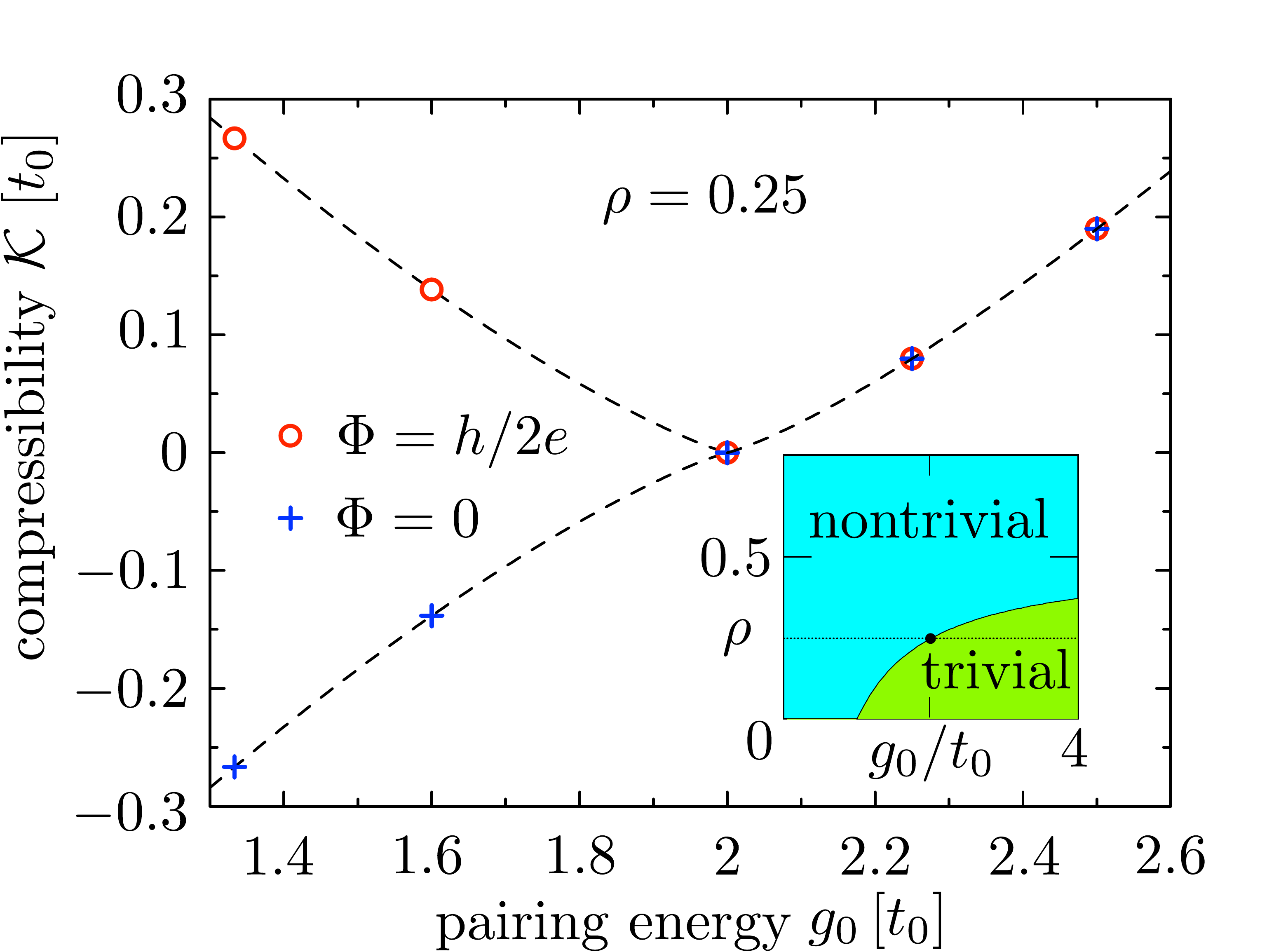}}
\caption{Dependence of the (inverse) compressibility \eqref{calKdef} on the pairing strength $g_0$ of the Kitaev chain, for periodic boundary conditions ($+$ data points) and for antiperiodic boundary conditions ($\circ$ data points), calculated from the particle-number conserving Hamiltonian \eqref{RGK1} for ${\cal N}=2N=512$, $L=2048$. The dashed curves result in the thermodynamic limit ${\cal N},L\rightarrow\infty$ at fixed $\rho= {\cal N/L}=0.25$. The inset shows the mean-field phase diagram. Figure adapted from \textcite{Ort14}.
}
\label{fig_RGK}
\end{figure}

To establish the transition from a topologically trivial to nontrivial state in a particle-conserving Hamiltonian, we determine the ground-state fermion parity from the (inverse) compressibility
\begin{equation}
{\cal K}(\Phi)=\tfrac{1}{2}E_0(2N+1,\Phi)+\tfrac{1}{2}E_0(2N-1,\Phi)-E_0(2N,\Phi),\label{calKdef}
\end{equation}
where $E_0({\cal N},\Phi)$ is the lowest eigenvalue of the Kitaev chain containing ${\cal N}$ electrons and enclosing a flux $\Phi$. The topological invariant \eqref{QP0Ph2e} then follows from
\begin{equation}
Q={\rm sign}\left[{\cal K}(0){\cal K}(h/2e)\right].\label{calQcalK}
\end{equation}
The exact results shown in Fig.\ \ref{fig_RGK} demonstrate a topological phase transition at $g_0/t_0=2$ for $\rho=0.25$, as predicted by the mean-field Hamiltonian.

\section{Fundamental symmetries}
\label{Bqp}

In the disordered systems to which RMT is applied, translational and rotational symmetries are broken. If a unitary symmetry remains, so if $H=UHU^\dagger$ for some unitary operator $U$, then the Hamiltonian $H$ can be decomposed into blocks acting on subspaces within which $U$ is the identity operator (times a phase factor). The unitary symmetry can thus be ignored if we restrict ourselves to one block. Constraints on the Hamiltonian that cannot be removed in this way arise from particle-hole and time-reversal symmetry. These are anti-unitary symmetries $H=\pm UH^\ast U^\dagger$, involving a complex conjugation.  

\subsection{Particle-hole symmetry}
\label{phsym}

As mentioned in Sec.\ \ref{Mfermion} when we introduced the Majorana operators, particle-hole (or charge conjugation) symmetry is a property of the mean-field theory of superconductivity. For spin-singlet \textit{s}-wave pairing the mean-field Hamiltonian has the general form \cite{Bog58,DeG66}
\begin{align}
&{\cal H}=\hat{\Psi}^\dagger H\hat{\Psi},\;\;\hat{\Psi}=(\hat{\psi},\hat{\psi}^\dagger)=(\hat{\psi}_{\uparrow}^{\vphantom{\dagger}},\hat{\psi}_{\downarrow}^{\vphantom{\dagger}},\hat{\psi}_{\uparrow}^\dagger,\hat{\psi}_{\downarrow}^\dagger),\label{hatPsidef}\\
&H=\begin{pmatrix}
H_{0}-E_{\rm F}&-i\sigma_{y}\Delta\\
i\sigma_{y}\Delta^{\ast}&E_{\rm F}-H_{0}^{\ast}
\end{pmatrix}.\label{HBdG}
\end{align}
The Hermitian operator $H$ acts on the four-component Nambu spinor $\hat{\Psi}$, which is a field operator in second quantization.\footnote{
The Hamiltonian \eqref{HBdG} is sometimes given in the alternative basis $(\hat{\psi}_{\uparrow},\hat{\psi}_{\downarrow},-\hat{\psi}^\dagger_{\downarrow},\hat{\psi}^\dagger_{\uparrow})$, when it has the form 
\begin{equation}
\tilde{H}=\begin{pmatrix}
H_{0}-E_{\rm F}&\Delta\\
\Delta^{\ast}&E_{\rm F}-\sigma_y H_{0}^{\ast}\sigma_y
\end{pmatrix},\nonumber
\end{equation}
with a scalar off-diagonal block. The charge conjugation operator then equals $\tilde{\cal C}=(\sigma_y\otimes\tau_y){\cal K}$. We prefer the equivalent representation \eqref{HBdG} because of the simpler ${\cal C}=\tau_x{\cal K}$. Note that both ${\cal C}$ and $\tilde{C}$ square to $+1$.
}

In first quantization, one can interpret $H$ as the Hamiltonian that governs the dynamics of Bogoliubov quasiparticles,
\begin{align}
&H\Psi(\bm{r},t)=-i\hbar\frac{\partial}{\partial t}\Psi(\bm{r},t),\label{HBdG3}\\
&\Psi=(\psi_e,\psi_h)=(\psi_{e\uparrow},\psi_{e\downarrow},\psi_{h\uparrow},\psi_{h\downarrow}).\label{psidef}
\end{align} 
This matrix wave equation is called the Bogoliubov-De Gennes (BdG) equation. The upper-left block $H_0-E_{\rm F}$ of $H$ acts on the electron component $\psi_{e}$, while the lower-right block $E_{\rm F}-H_0^\ast$ acts on the hole component $\psi_{h}$. The off-diagonal blocks couple electrons and holes in opposite spin bands $\uparrow,\downarrow$ (switched by the Pauli matrix\footnote{
The matrices $\sigma_\alpha$ and $\tau_\alpha$ act on, respectively, the spin and particle-hole degree of freedom, according to
\[
\sigma_0=
{\scriptstyle\begin{pmatrix}
1&0\\
0&1
\end{pmatrix}},\;\;
\sigma_x={\scriptstyle\begin{pmatrix}
0&1\\
1&0
\end{pmatrix}},\;\; 
\sigma_y={\scriptstyle\begin{pmatrix}
0&-i\\
i&0
\end{pmatrix}},\;\; 
\sigma_z={\scriptstyle\begin{pmatrix}
1&0\\
0&-1
\end{pmatrix}}.
\]
}
$\sigma_{y}$), through the (complex) pair potential $\Delta$. 

Each eigenfunction $\Psi$ of $H$ at energy $E>0$ has a copy $\tau_{x}\Psi$ at $-E$. (The Pauli matrix $\tau_{x}$ switches electrons and holes.) The corresponding symmetry of $H$,
\begin{equation}
H=-{\cal C}H{\cal C}^{-1}=-\tau_{x}H^{\ast}\tau_{x},\label{HBdGsymmetry}
\end{equation} 
is called particle-hole symmetry.\footnote{
The term ``particle-hole symmetry'' is also used in semiconductor physics, with a different meaning: The symmetry \eqref{HBdGsymmetry} expresses the fact that creation and annihilation operators are each others Hermitian conjugate, see Eq.\ \eqref{hatPsidef}. It holds at all energies, irrespective of the band structure. Particle-hole symmetry in semiconductors holds if we linearize the band structure near the Fermi level, so that filled states above the Fermi level and empty states below it have the same dispersion. In the context of superconductors this is called the Andreev approximation, it is unrelated to Eq.\ \eqref{HBdGsymmetry}.}
The charge conjugation operator ${\cal C}=\tau_{x}{\cal K}$, with ${\cal K}$ the operator of complex conjugation, is anti-unitary and squares to $+1$.

If $H_{0}$ is spin independent, then the Hamiltonian \eqref{HBdG} decouples into the two blocks
\begin{equation}
H_{\pm}=\begin{pmatrix}
H_{0}-E_{\rm F}&\pm\Delta\\
\pm\Delta^{\ast}&E_{\rm F}-H_{0}^{\ast}
\end{pmatrix},\label{HBdG2}
\end{equation}
acting separately on $(\psi_{e\downarrow},\psi_{h\uparrow})$ and $(\psi_{e\uparrow},\psi_{h\downarrow})$. The charge conjugation operator ${\cal C}=i\tau_{y}{\cal K}$ for each block now squares to $-1$.

\subsection{Majorana representation}
\label{Mfermionrepr}

The Majorana nature of Bogoliubov quasiparticles is hidden in the electron-hole basis \eqref{psidef}, but becomes apparent upon a unitary transformation,
\begin{subequations}
\label{HPsiOmega}
\begin{align}
&H\mapsto \Omega H\Omega^{\dagger},\;\;\Omega=\sqrt{\tfrac{1}{2}}\begin{pmatrix}
1&1\\
i&-i
\end{pmatrix},\label{HPsiOmegaa}\\
&\Psi\mapsto\Omega\Psi=\sqrt{\frac{1}{2}}\begin{pmatrix}
\psi_{e}+\psi_{h}\\
i\psi_{e}-i\psi_{h}
\end{pmatrix}.\label{HPsiOmegab}
\end{align}
\end{subequations}
This is the Majorana representation introduced for the Kitaev chain in Sec.\ \ref{Mfermion}. The particle-hole symmetry relation \eqref{HBdGsymmetry} now reads simply
\begin{equation}
{\cal C}={\cal K},\;\;H=-H^{\ast},\label{HBdGsymmetry2}
\end{equation}
so $H=iA$ is given by a real antisymmetric matrix $A$ [as in Eq.\ \eqref{HKM}]. 

The BdG equation \eqref{HBdG3} becomes a real wave equation,
\begin{equation}
A\Psi(\bm{r},t)=-\hbar\frac{\partial}{\partial t}\Psi(\bm{r},t).\label{realwaveeq}
\end{equation}
The corresponding field operator is self-conjugate, $\hat{\Psi}(\bm{r},t)=\hat{\Psi}^{\dagger}(\bm{r},t)$, so creation and annihilation operators are one and the same. In this sense a Bogoliubov quasiparticle is a Majorana fermion.

All of this refers to the four-component BdG Hamiltonian \eqref{HBdG}, with ${\cal C}^{2}=+1$, so ${\cal C}\mapsto {\cal K}$ can be achieved by a unitary transformation. The wave equation for the reduced two-component BdG Hamiltonian \eqref{HBdG2}, with ${\cal C}^{2}=-1$, cannot be brought to a real form by any unitary transformation.

Superconductor quasiparticles are typically probed in the energy domain, rather than in the time domain. The Fourier transform
\begin{equation}
{\Psi}_{E}(\bm{r})=\int dt\,e^{iEt/\hbar}{\Psi}(\bm{r},t)={\Psi}_{-E}^{\dagger}(\bm{r})\label{PsiEdef}
\end{equation}
is real at $E=0$, so for quasiparticles at the Fermi level. Transport experiments at small voltage and low temperature can therefore probe the Majorana nature of Bogoliubov quasiparticles.

\subsection{Time-reversal and chiral symmetry}
\label{Tchsym}

Anti-unitary symmetries come in two types, the Hamiltonian $H$ may commute or anti-commute with an anti-unitary operator. The particle-hole symmetry discussed in Section \ref{phsym} is the anti-commutation, $H{\cal C}=-{\cal C}H$, while the commutation $H{\cal T}={\cal T}H$ is called time-reversal symmetry. The physical operation of time reversal should reverse the spin, ${\cal T}\sigma_{k}{\cal T}^{-1}=-\sigma_{k}$, as well as the momentum, ${\cal T}\bm{p}{\cal T}^{-1}=-\bm{p}$. The corresponding operator ${\cal T}=i\sigma_{y}{\cal K}$ squares to $-1$. The Hamiltonian \eqref{HBdG} commutes with ${\cal T}$ if $\Delta$ is real and
\begin{equation}
H_{0}={\cal T}H_{0}{\cal T}^{-1}=\sigma_{y}H_{0}^{\ast}\sigma_{y}.\label{H0calT}
\end{equation}

For a real Hamiltonian we can take ${\cal T}={\cal K}$ squaring to $+1$. The combination of this fake time-reversal symmetry with the particle-hole symmetry \eqref{HBdGsymmetry} implies that
\begin{equation}
H\tau_{x}=-\tau_{x}H.\label{Htaux}
\end{equation}
Such anti-commutation of the Hamiltonian with a unitary operator, $HU=-UH$, is called a chiral symmetry. 

If we change basis such that $\tau_x\mapsto\tau_z$, the chiral symmetry \eqref{Htaux} implies that the Hamiltonian in the new basis has the block structure
\begin{equation}
H\mapsto\begin{pmatrix}
0&h\\
h^\dagger&0
\end{pmatrix}.\label{Hnewbasis}
\end{equation}
Because the nonzero blocks are not on the diagonal, we cannot restrict ourselves to a single block, as we could have done if $H$ would commute rather than anticommute with $U$.

\section{Hamiltonian ensembles}
\label{Hens}

\subsection{The ten-fold way}
\label{ten-foldway}

When Wigner conceived of random-matrix theory in the context of nuclear physics, there was only a single ensemble of real Gaussian Hamiltonians \cite{Wig56,Wig67}. How this number grew to ten is a remarkable development in mathematical physics, starting with Dyson's threefold way \cite{Dys62}: The Gaussian orthogonal, unitary, and symplectic ensembles (GOE, GUE, GSE) of Hermitian matrices $H$ with Gaussian elements that are real, complex, or quaternion, respectively. (The names refer to the type of transformation that diagonalizes the Hamiltonian.) Which of the three Wigner-Dyson ensembles applies, is determined by the symmetry under time reversal. 

Dyson understood the algebraic reason for the trinity: There are only three types of numbers, real -- complex -- quaternion, that can be used to construct a vector space for a quantum theory.\footnote{The technical statement is that there are only three associative normed division algebras \cite{Bae12}.} Dyson also appreciated that the three circular ensembles (COE, CUE, CSE) obtained by exponentiating $S=e^{iH}$ correspond to 3 of the 10 compact symmetric spaces of differential geometry \cite{Dys70}.\footnote{The ten compact symmetric spaces consist of the orthogonal, unitary, and symplectic groups ${\rm O}(N)$, ${\rm U}(N)$, and ${\rm Sp}(2N)$, the three cosets ${\rm X}(p+q)/{\rm X}(p)\times {\rm X}(q)$ with X\,=\,O,U,Sp, and four more cosets ${\rm U}(N)/{\rm O}(N)$, ${\rm U}(2N)/{\rm Sp}(2N)$, ${\rm O}(2N)/{\rm U}(N)$, ${\rm Sp}(2N)/{\rm U}(N)$ \cite{Zir96,Cas04}. The CUE, COE, and CSE correspond, respectively, to ${\rm U}(N)$, ${\rm U}(N)/{\rm O}(N)$, and ${\rm U}(2N)/{\rm Sp}(2N)$.}

It would take several decades before all 10 ensembles were identified in a physical context. The number of ensembles was first expanded to 6 by adding chiral symmetry in the context of QCD \cite{Ver94,Ver00}, and then completed to 10 by adding particle-hole symmetry in the context of superconductivity \cite{Alt97,Hei05}. 

To understand why considerations of symmetry produce a ten-fold classification one searches for unitary or anti-unitary operators that commute or anti-commute with the Hamiltonian:
\begin{itemize}
\item
A unitary operator $U$ that commutes with $H$ can be removed from consideration by restricting $H$ to an eigenspace of $U$. (This is possible because $H$ and $U$ can be diagonalized simultaneously if $HU=UH$.)
\item
An anti-unitary operator ${\cal T}$ that commutes with $H$ produces 2 symmetry classes, depending on whether ${\cal T}^2=+1$ or $-1$. This takes care of time-reversal symmetry without particle-hole symmetry.
\item
An anti-unitary operator ${\cal C}$ that anti-commutes with $H$ and squares to $\pm 1$ also produces 2 symmetry classes. This takes care of particle-hole symmetry without time-reversal symmetry.
\item
The combination of ${\cal C}^2=\pm 1$ and ${\cal T}^2=\pm 1$ produces 4 symmetry classes. This takes care of time-reversal symmetry with particle-hole symmetry.
\item 
The product ${\cal CT}$ is a unitary operator that anti-commutes with $H$. The presence or absence of this chiral symmetry produces an additional 2 symmetry classes if neither time-reversal nor particle-hole symmetry apply.
\end{itemize}
In Table \ref{table_AZ} we summarize the ten symmetry classes that follow from this inventory. The (seemingly unsystematic) labeling of each class, shown in the top row of the table, is the Cartan name of the compact symmetric space. 

\begin{table*}
\centering
\begin{tabular}{ | c || c | c | c | c | c | c | c | c | c | c |}
\hline
& \textbf{D} & \textbf{BDI} & \textbf{DIII} & \textbf{C} & \textbf{CI} & \textbf{CII} & \textbf{A} & \textbf{AI} & \textbf{AII} & \textbf{AIII} \\ \hline \hline
$H{\cal C}=-{\cal C}H$, ${\cal C}^{2}=$ &  $+1$ &  $+1$ & $+1$ & $-1$ & $-1$ & $-1$ & $\times$ & $\times$ & $\times$ & $\times$ \\ \hline
$H{\cal T}={\cal T}H$, ${\cal T}^{2}=$ &  $\times$ & $+1$ & $-1$ & $\times$ & $+1$ & $-1$ & $\times$ & $+1$ & $-1$ & $\times$ \\  \hline
$H{\cal CT}=-{\cal CT}H$ &  $\times$ & \checkmark & \checkmark & $\times$ & \checkmark & \checkmark & $\times$ & $\times$ & $\times$ & \checkmark \\  \hline
$\nu$ & 0,1 & 0,1,2,3,\ldots & 0,1 & 0 & 0 & 0,1,2,3,\ldots & 0 & 0 & 0 & 0,1,2,3,\ldots \\ \hline
$d_{E}$ & 1 & 1 & 2 & 1 & 1 &  2 & 1 & 1 & 2 & 1\\ \hline 
$\alpha_{E}$ & 0 & 0 & 1 & 2 & 1 & 3 & 0 & 0 & 0 & 1 \\  \hline
$\beta_{E}$ & 2 & 1 & 4 & 2 & 1 & 4 & 2 & 1 & 4 & 2 \\ \hline \hline
$S^{-1}=$ & \multicolumn{3}{|c|}{$S^{\rm T}$ (orthogonal)} & \multicolumn{3}{|c|}{$\tau_{y}S^{\rm T}\tau_{y}$ (symplectic)} & \multicolumn{4}{|c|}{$S^{\dagger}$ (unitary)}  \\ \hline
$S^{\rm T}=$ &  $\times$ & $+S$ & $-S$ & $\times$ & $+S$ & $-S$ & $\times$ & $+S$ & $-S$ & $S^{\ast}$ \\  \hline
$d_{T}$ & 1 & 1 & 2 & 2 & 2 &  2 & 1 & 1 & 2 & 1 \\ \hline 
$\alpha_{T}$ & $-1$ & $-1$ & $-1$ & $2$ & $1$ & $-1$ & 0 & 0 & 0 & $-1$ \\  \hline
$\beta_{T}$ & 1 & 1 & 2 & 4 & 2 & 4 & 2 & 1 & 4 & 2 \\ \hline
\end{tabular}
\caption{The ``ten-fold way'' classification of Hamiltonians $H$ and scattering matrices $S$ at the Fermi level. The symmetry classes are distinguished by the anti-unitary symmetries ${\cal C}$ (particle-hole) and ${\cal T}$ (time reversal), squaring to $+1$ or $-1$. (A cross indicates that the symmetry is not present.) The third row lists whether or not the product ${\cal CT}$ is a chiral symmetry of the Hamiltonian. The degeneracies $d$ and repulsion exponents $\alpha,\beta$ of the energy and transmission eigenvalues are distinguished by subscripts $E,T$. (Uncoupled spin bands are not included in the degeneracy count.) The integer $\nu$ counts the number of $d_E$-fold degenerate, topologically protected zero-modes (Majorana in class D, BDI, and DIII). 
}
\label{table_AZ}
\end{table*}

For a random-matrix approach the Hamiltonian operator is represented by an ${\cal N}\times {\cal N}$ Hermitian matrix $H=H^{\dagger}$. In the Gaussian ensemble the Hamiltonian has the probability distribution\footnote{The factor-of-two difference in the coefficient $c$ is there on account of the $\pm E$ symmetry of the spectrum in the classes with particle-hole or chiral symmetry, see  \textcite{Mi14}. The mean level spacing $\delta_0$ refers to distinct levels, not counting degeneracies.}
\begin{subequations}
\label{PHGaussian}
\begin{align}
&P(H)\propto\exp\left(-\frac{c}{\cal N}\,{\rm Tr}\,H^2\right),\label{PHGaussiana}\\
&c=\frac{\pi^{2}\beta_{E}}{8\delta_0^{2}}\times\begin{cases}
2& \text{in class A, AI, AII},\\
1& \text{in the other classes},
\end{cases}\label{PHGaussianb}
\end{align}
\end{subequations}
where $\delta_0$ is the mean level spacing of $H$ in the bulk of the spectrum and $\beta_{E}\in\{1,2,4\}$ describes the strength of the level repulsion (see Section \ref{levelrep}). The Gaussian form is chosen for mathematical convenience, in the large-${\cal N}$ limit the spectral correlations depend only on the symmetries of $H$. 

The ``ten-fold way'' of RMT provides the basis for the classification of topologically distinct states of matter \cite{Sch08,Ryu10,Kit09}. Five of the ten symmetry classes allow for a topological invariant $\nu$, integer valued ($\nu\in\mathbb{Z}$) in class BDI, CII, AIII and binary ($\nu\in\mathbb{Z}_2$) in class D and DIII. The existence of a $\mathbb{Z}$ invariant was first noticed in the context of QCD, where $\nu$ is the topological charge of a gauge field configuration \cite{Shu93,Ver93}. The $\mathbb{Z}_2$ invariant first appeared in studies of the spectral statistics of a vortex core in a \textit{p}-wave superconductor \cite{Boc00,Iva02}, as we discuss next.

\subsection{Midgap spectral peak}
\label{spectralpeak}

The $\pm E$ particle-hole symmetry modifies the spectral correlations near $E=0$. This is the Fermi level, in the middle of the superconducting gap, so to allow for states near $E=0$ one needs to locally close the gap, for example by means of a magnetic vortex \cite{Car64}. If spin-rotation symmetry is broken by spin-orbit coupling, the system is in symmetry class D. The particle-hole symmetry relation \eqref{HBdGsymmetry2} requires that $H=iA$ is purely imaginary in the Majorana basis, and since it is also Hermitian it must be an antisymmetric matrix: $A_{nm}=-A_{mn}=A_{nm}^{\ast}$. There are no other symmetry constraints in class D.

In the Gaussian ensemble the upper-diagonal matrix elements $A_{nm}$ ($n>m$) of the real antisymmetric matrix $A$ all have identical and independent distributions,
\begin{equation}
P(\{A_{nm}\})\propto\prod_{n>m}\exp\left(-\frac{\pi^{2}A_{nm}^{2}}{2{\cal N}\delta_0^{2}}\right),\label{GaussEns}
\end{equation}
cf.\ Eq.\ \eqref{PHGaussian} with $\beta_{E}=2$. As for any antisymmetric matrix, the eigenvalues come in $\pm E$ pairs, so if ${\cal N}$ is odd then $H$ necessarily has one eigenvalue pinned at zero --- a Majorana zero-mode, see Section \ref{Mzm}. For the nonzero eigenvalue pairs $\pm E_n$ the Gaussian ensemble gives the probability distribution \cite{Mehta}
\begin{equation}
P(\{E_n\})\propto\sideset{}{'}\prod_{i<j}(E_i^2-E_j^2)^2\sideset{}{'}\prod_{k}E_k^{2\nu}\exp\left(-\frac{\pi^2 E_k^2}{2{\cal N}\delta_0^2}\right).\label{PEnclassD}
\end{equation}
The primed product $\prod'$ is a reminder that only positive energies are included. The number $\nu$ indicates the presence or absence of a Majorana zero-mode: $\nu=1$ if ${\cal N}$ is odd and $\nu=0$ if ${\cal N}$ is even.

\begin{figure}[tb]
\centerline{\includegraphics[width=1\linewidth]{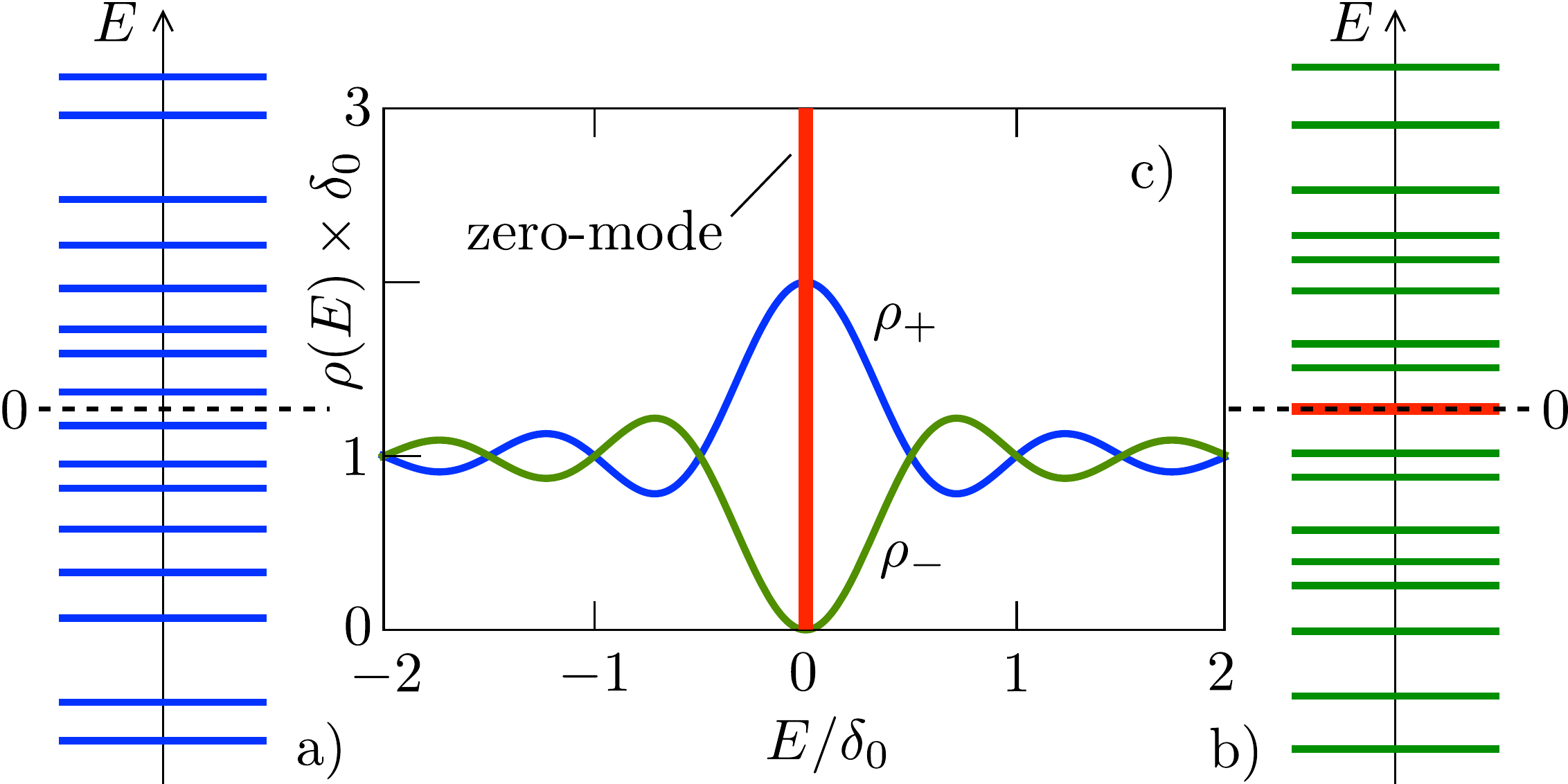}}
\caption{Panels a) and b) show the spectrum of a vortex core in a class-D superconductor. The $\pm E$ symmetric spectrum may or may not have an unpaired Majorana zero-mode, pinned to $E=0$. The ensemble-averaged density of states \eqref{rhopmdef} is plotted in panel c). The delta-function contribution from the zero-mode is accompanied by a dip in the smooth part $\rho_-$ of the density of states. Without the zero-mode there is a midgap spectral peak $\rho_+$. The density of states in symmetry class C is given by $\rho_-$ without the zero-mode contribution.
}
\label{fig_spectralpeak}
\end{figure}

In either case $\nu=0,1$ the ensemble-averaged density of states $\rho(E)$ has a peak at $E=0$ \cite{Mehta,Alt97,Boc00,Iva02},
\begin{align}
&\rho(E)=\begin{cases}
\rho_+(E)&{\rm if}\;\;\nu=0,\\
\rho_-(E)+\delta(E)&{\rm if}\;\;\nu=1,
\end{cases}\label{rhoEresult}\\
&\rho_{\pm}( E)=\delta_{0}^{-1}\pm\frac{\sin(2\pi E/\delta_0)}{2\pi E},\label{rhopmdef}
\end{align}
see Fig.\ \ref{fig_spectralpeak}.

A physical realization of a class D, $\nu=1$ vortex is offered by the surface of a three-dimensional topological insulator (such as ${\rm Bi}_{2}{\rm Te}_{3}$) covered by an \textit{s}-wave superconductor \cite{Fu08}. The small level spacing $\delta_0\simeq \Delta_0^2/E_{\rm F}$ in the vortex core (with superconducting gap $\Delta_0$ much smaller than the Fermi energy $E_{\rm F}$) complicates the detection of the midgap spectral peak at experimentally accessible temperatures \cite{Xu13}. Notice that $\rho=\rho_+$ and $\rho=\delta(E)+\rho_{-}$ have identical spectral weight $\int( \rho-1/\delta_0)\, dE=1/2$, so a thermally smeared density of states is not a prominent signature of a Majorana zero-mode. (The transport signatures discussed in Section \ref{elG} are more reliable for that purpose.)

The midgap spectral peak does serve as an unambiguous distinction between symmetry classes C and D, with and without spin-rotation symmetry. Particle-hole symmetry of the class-C Hamiltonian \eqref{HBdG2} can be expressed as
\begin{equation}
H_{\pm}=-\tau_y H_{\pm}^{\ast} \tau_y=\begin{pmatrix}
0&1\\
-1&0
\end{pmatrix}H_{\pm}^{\ast}\begin{pmatrix}
0&1\\
-1&0
\end{pmatrix},\label{HastC}
\end{equation}
where the subscript $\pm$ labels the spin degree of freedom and the subblocks refer to the electron-hole degree of freedom. Because of the spin degeneracy it is sufficient to consider $H_+\equiv iQ$.

Eq.\ \eqref{HastC} implies that the matrix elements of the ${\cal N}\times {\cal N}$ anti-Hermitian matrix $Q$ are quaternion numbers, of the form
\begin{equation}
Q_{nm}=a_{nm}\tau_0+ib_{nm}\tau_x+ic_{nm}\tau_y+id_{nm}\tau_z,\label{quaterniondef}
\end{equation}
with real coefficients $a,b,c,d$ and indices $n,m=1,2\ldots{\cal N}/2$. (The dimensionality ${\cal N}$ is necessarily even in class C.) The corresponding eigenvalue distribution has the form \eqref{PEnclassD} with $\nu=1$, but without any level pinned at zero. (There are no Majorana zero-modes in class C.) As a consequence, the average density of states is $\rho(E)=\rho_{-}(E)$ without the delta-function contribution, so instead of a midgap spectral peak there is now a midgap spectral dip  \cite{Alt97}.

\subsection{Energy level repulsion}
\label{levelrep}

The random-matrix ensembles of class A,AI,AII introduced by Wigner and Dyson have distinct power laws for the probability to find two levels $E_i$ and $E_j$ close to each other \cite{Wig67,Dys62}. The probability vanishes as $|E_i-E_j|^{\beta_{E}}$, with $\beta_{E}=2$ in the absence of time-reversal symmetry, $\beta_{E}=1$ if ${\cal T}^2=+1$, and $\beta_{E}=4$ if ${\cal T}^2=-1$.

All of this still applies if we add $\pm E$ symmetry, due to particle-hole or chiral symmetry. The probability distribution of the positive eigenvalues then has a factor
\begin{equation}
\sideset{}{'}\prod_{i<j}|E_i^2-E_j^2|^{\beta_{E}}=\sideset{}{'}\prod_{i<j}|E_i-E_j|^{\beta_{E}} |E_i+E_j|^{\beta_{E}},\label{levelrepulsionfactor} 
\end{equation}
with $\beta_{E}$ cycling through 2,1,4, see Table \ref{table_AZ}. This factor describes the repulsion of a level $E_i>0$ with the pair of levels $\pm E_j$.

A new feature of particle-hole symmetry is the repulsion of a level at $+E$ with its counterpart at $-E$. This repulsion introduces a factor $\prod'_k |E_k|^{\alpha_{E}}$ into the probability distribution, with an exponent $\alpha_{E}$ that may be different from $\beta_{E}$. In class C one has $\alpha_{E}=\beta_{E}=2$, but in class D instead $\alpha_{E}=0$, so there is no level repulsion at the Fermi level in a class-D superconductor \cite{Alt97}. Fig.\ \ref{fig_repulsion} illustrates this in a computer simulation: Level crossings are avoided away from the Fermi level, but at the Fermi level pairs of levels may cross. (The physical significance of the level crossings is discussed in Section \ref{parityswitch}.)

\begin{figure}[tb]
\centerline{\includegraphics[width=0.6\linewidth]{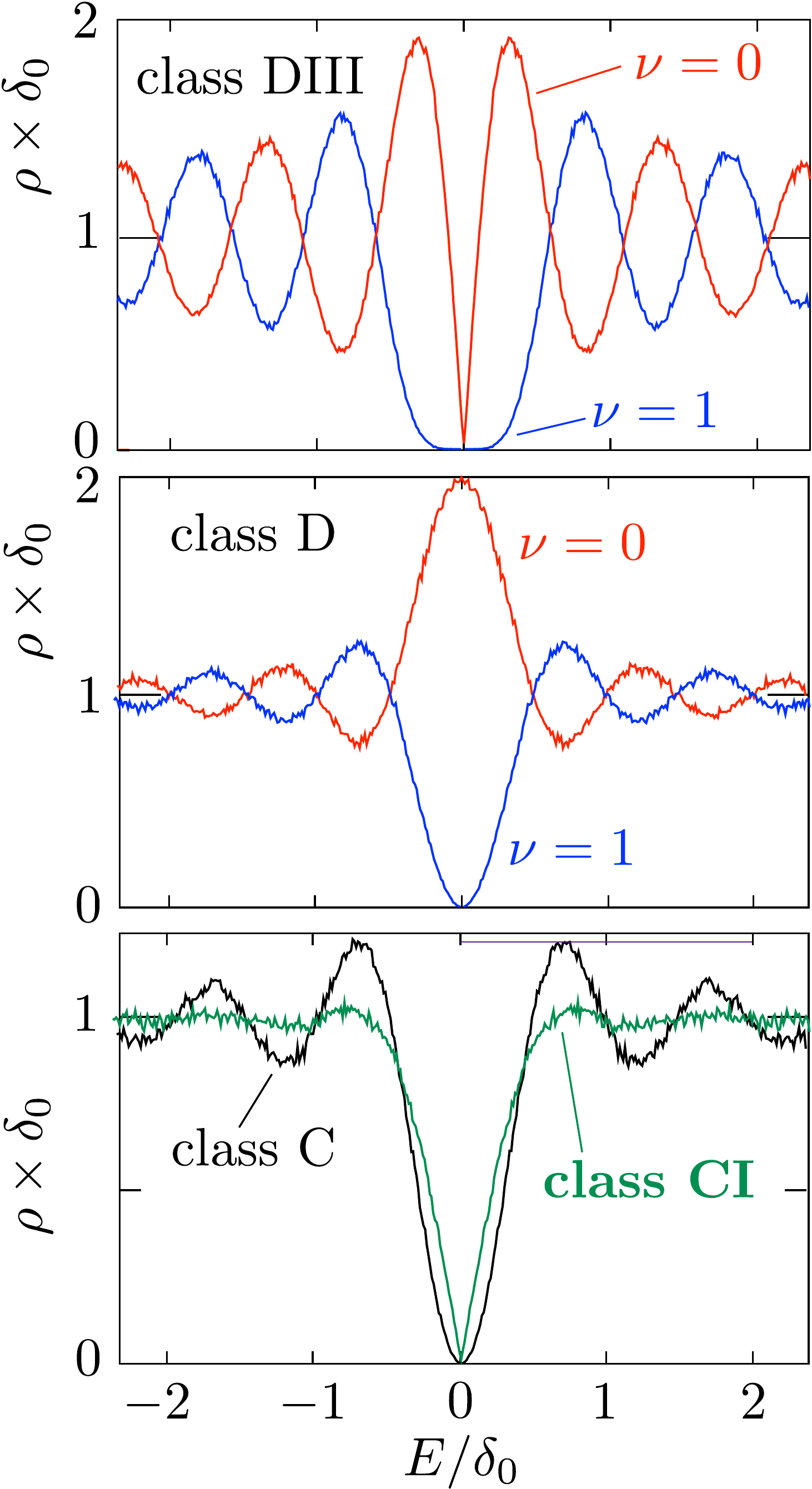}}
\caption{Ensemble-averaged density of states in the four Altland-Zirnbauer ensembles, calculated numerically for Hamiltonians of dimensionality ${\cal N}=60$ in class C and CI, ${\cal N}=60+\nu$ in class D, and ${\cal N}=120+2\nu$ in class DIII. (Analytical expressions in the large-${\cal N}$ limit are collected in  \textcite{Iva02}, where class D, $\nu=1$ is called ``class B''.) The delta-function singularity of the zero-mode for $\nu=1$ is not plotted. Except for class D with $\nu=0$, the density of states vanishes at the Fermi level as $|E|^{\alpha_{E}+\nu\beta_{E}}$. Figure adapted from \textcite{Mi14}.
}
\label{fig_DOS}
\end{figure}

For reference we record the complete expressions \cite{Alt97,Iva02} for the probability distributions of energy levels in the ten symmetry classes from Table \ref{table_AZ}:
\begin{align}
&P(\{E_n\})\propto\prod_{1=i<j}^{{\cal N}/d_{E}}|E_i-E_j|^{\beta_{E}}\prod_{k=1}^{{\cal N}/d_{E}}\exp\left(-\frac{\pi^2\beta_{E}d_{E} E_k^2}{4{\cal N}\delta_0^2}\right)\nonumber\\
&\qquad\text{in class A, AI, AII},\label{PEnWD}\\
&P(\{E_n\})\propto\sideset{}{'}\prod_{k=1}^{({\cal N}-\nu d_{E})/2d_{E}}|E_k|^{\alpha_{E}+\nu\beta_{E}}\exp\left(-\frac{\pi^2\beta_{E}d_{E} E_k^2}{4{\cal N}\delta_0^2}\right)\nonumber\\
&\qquad\times\sideset{}{'}\prod_{1=i<j}^{({\cal N}-\nu d_{E})/2d_{E}}|E_i^2-E_j^2|^{\beta_{E}},\;\;\text{in the other classes}.\label{PEnAZ}
\end{align}
The $d_{E}$-fold degenerate levels are included only once in each product, and the primed product indicates that only positive energies are included (excluding also the $\nu d_{E}$ zero-modes). The effect of the repulsion factor $|E|^{\alpha_{E}+\nu\beta_{E}}$ on the density of states in the Altland-Zirnbauer ensembles is shown in Fig.\ \ref{fig_DOS}.

\section{Scattering matrix ensembles}
\label{Smatrixens}

\subsection{Fundamental symmetries}
\label{Smatrixsym}

We seek to probe Majorana fermions by means of electrical or thermal transport at low voltages and temperatures, near the Fermi level where the Majorana operators \eqref{PsiEdef} are selfconjugate. These transport properties are determined by a quantum mechanical scattering problem, in which a set of incident and outgoing wave amplitudes $\psi_{n}^{\rm in}$, $\psi_{n}^{\rm out}$, $n=1,2,\ldots N$, is linearly related by
\begin{equation}
\psi_{n}^{\rm out}(E)=\sum_{m=1}^{N}S_{nm}(E)\psi_{m}^{\rm in}(E).\label{Spsirelation}
\end{equation}
The scattering is elastic, so incident and outgoing states are at the same energy $E$, and it is conservative, so $\sum_{n}|\psi_{n}^{\rm out}|^{2}=\sum_{n}|\psi_{n}^{\rm in}|^{2}$ and $S\in{\rm U}(N)$ is a unitary $N\times N$ matrix,
\begin{equation}
S^{-1}(E)=S^{\dagger}(E)\Leftrightarrow\sum_{k}S_{nk}(E)S_{mk}^{\ast}(E)=\delta_{nm}.\label{Sunitary}
\end{equation}

The Cayley transform \cite{Mah68,Ver85}
\begin{equation}
S(E)=\frac{1-i\pi {K}(E)}{1+i\pi {K}(E)},\;\;{K}(E)=W^\dagger\frac{1}{E-H}W,\label{Cayley}
\end{equation}
relates the unitary scattering matrix $S(E)$ to the matrix ${K}(E)$, being the Green function $(E-H)^{-1}$ projected onto the scattering states by a coupling matrix $W$. (The $K$-matrix is known as the reaction matrix in the theory of nuclear scattering \cite{Wig47}.) We assume that $W$ commutes both with the charge-conjugation operator ${\cal C}$ and the time-reversal operator ${\cal T}$. (See  \textcite{Ful12} for a more general treatment of the symmetry constraints on the scattering matrix.)

Particle-hole symmetry $H=-{\cal C}H{\cal C}^{-1}$ of the Hamiltonian translates into the scattering-matrix symmetry
\begin{equation}
S(-E)={\cal C}S(E){\cal C}^{-1}=\begin{cases}
S^{\ast}(E)&{\rm if}\;\;{\cal C}={\cal K},\\
\tau_{y}S^{\ast}(E)\tau_{y}&{\rm if}\;\;{\cal C}=i\tau_{y}{\cal K}.
\end{cases}\label{SCsymm}
\end{equation}
At the Fermi level, $E=0$, this implies that $S(0)\in{\rm O}(N)$ is a real orthogonal matrix for ${\cal C}^2=+1$, while $S(0)\in{\rm Sp}(N)$ is a unitary symplectic matrix\footnote{Orthogonal and symplectic matrices are both unitary, but while the matrix elements of an orthogonal matrix are real numbers, the matrix elements of a symplectic matrix are quaternions, of the form \eqref{quaterniondef}.}
for ${\cal C}^{2}=-1$.

Time-reversal symmetry $H={\cal T}H{\cal T}^{-1}$ translates into
\begin{equation}
S(E)={\cal T}S^{\dagger}(E){\cal T}^{-1}=\begin{cases}
S^{\rm T}(E)&{\rm if}\;\;{\cal T}={\cal K},\\
\sigma_{y}S^{\rm T}(E)\sigma_{y}&{\rm if}\;\;{\cal T}=i\sigma_{y}{\cal K},
\end{cases}\label{STsymm}
\end{equation}
where the superscript T denotes the transpose. This may be written more succinctly, upon a change of basis $S\mapsto i\sigma_{y}S$ of the outgoing modes, as a condition of symmetry or antisymmetry,
\begin{equation}
S(E)=\pm S^{\rm T}(E),\label{ST2E}
\end{equation}
depending on whether the anti-unitary operator ${\cal T}$ squares to $+1$ or $-1$. Finally, chiral symmetry $H{\cal CT}=-{\cal CT}H$ translates into
\begin{equation}
S(E)={\cal CT}S^\dagger(-E)({\cal CT})^{-1}.\label{chiralsymmetry}
\end{equation}

The symmetry requirements on the scattering matrix at the Fermi level ($E=0$) are summarized in Table \ref{table_AZ}, for each of the ten symmetry classes.

\subsection{Chaotic scattering}
\label{chaotic_scat}

The approach of random-matrix theory applies if the scattering is ``chaotic''. Chaotic scattering is a concept that originates from classical mechanics, referring to the exponential sensitivity of a trajectory to a slight change in initial condition \cite{Gut90}. This concept was transferred to quantum mechanics \cite{Blu90}, by considering --- at a fixed energy --- the ensemble of scattering matrices produced by slight deformations of the scattering potential. Chaotic scattering then refers to a uniform distribution\footnote{
Uniformity in the unitary group is defined in terms of the Haar measure $dU=d(UU_{0})$ for any fixed $U_{0}\in{\rm U}(N)$. See \textcite{Mez07} for how one can generate random matrices with this uniform distribution, and \textcite{Cre78,Col06} for how one can perform integrals $\int dU$ of polynomials of $U$.}
of this ensemble in the unitary group,
\begin{equation}
P(S)={\rm constant},\;\;S\in{\rm U}(N).\label{PSUNdef}
\end{equation}
This socalled Circular Unitary Ensemble (CUE) was introduced in the early days of RMT \cite{Dys62}, long before the advent of quantum chaos. It has found many applications in the context of microwave cavities \cite{Sto07}, and electronic quantum dots \cite{Bee97}.

The constraint \eqref{ST2E} on the scattering matrix imposed by time-reversal symmetry restricts $S$ to a subset of ${\rm U}(N)$. A \textit{symmetric} scattering matrix $S=UU^{\rm T}=S^{\rm T}$ applies to electrons when their spin is conserved by the scattering potential. The ensemble generated by the uniform distribution of $U\in{\rm U}(N)$ then describes chaotic scattering. Somewhat confusingly, this ensemble is called the Circular Orthogonal Ensemble (COE) although it does not contain orthogonal matrices. The name refers to the fact that unitary symmetric matrices form the coset ${\rm U}(N)/{\rm O}(N)$ of the orthogonal group ${\rm O}(N)$.

In the presence of spin-orbit coupling the constraint of time-reversal symmetry reads $S=U\sigma_{y}U^{\rm T}\sigma_{y}$. The uniform distribution of $U\in{\rm U}(N)$, with $N$ even, then produces the Circular Symplectic Ensemble (CSE), thus called because $S$ is in the coset ${\rm U}(N)/{\rm Sp}(N)$ of the unitary symplectic group ${\rm Sp}(N)$. Equivalently, upon a change of basis $S\mapsto i\sigma_{y}S$, we may describe the CSE by the set of unitary \textit{antisymmetric} matrices \cite{Bar08}, $S=U\sigma_{y}U^{\rm T}=-S^{\rm T}$.

\subsection{Circular ensembles}
\label{CA_ens}

Superconductivity introduces a new type of scattering process, Andreev scattering \cite{And64}, which is the conversion of an electron-like excitation at $E_{\rm F}+E$ into a hole-like excitation at $E_{\rm F}-E$. At the Fermi level, for excitation energy $E\rightarrow 0$, electrons and holes have complex conjugate wave functions. A linear superposition produces quasiparticles with a real wave function, the selfconjugate Majorana fermions discussed in Section \ref{Mfermionrepr}.

Andreev scattering does not conserve charge (the missing charge is accounted for by the superconducting condensate), but it does conserve particle number. The scattering matrix $S(E)$ therefore remains a unitary matrix, of dimension $2N\times 2N$ to accommodate the $N$ electron and $N$ hole degrees of freedom. The constraint of a real scattering amplitude of Majorana fermions restricts $S(0)$ to the orthogonal subgroup ${\rm O}(2N)$. Chaotic scattering then implies a uniform distribution,
\begin{equation}
P(S)={\rm constant},\;\;S\in{\rm O}(2N).\label{PSO2Ndef}
\end{equation}
This extension of Dyson's circular ensembles to include Andreev scattering was introduced by \textcite{Alt97}. 

A few words about nomenclature. The name ``circular orthogonal ensemble'' for the distribution \eqref{PSO2Ndef} would be most logical, but this name is already taken for the ensemble of unitary symmetric matrices (COE). We have become used to calling it the Circular Real Ensemble (CRE) --- another name found in the literature \cite{Poz98} is ``Haar orthogonal ensemble''. An alternative name could be ``class D'' ensemble, referring to the mathematical labeling of symmetric spaces \cite{Cas04}, but as \textcite{Zir10} has argued one should distinguish the symmetry of the matrix space from the uniformity of the matrix ensemble.

The restriction $S=\pm S^{\rm T}$ to symmetric or antisymmetric orthogonal matrices produces two further ensembles, which we will refer to as T$_{+}$CRE (symmetry class BDI) and T$_{-}$CRE (symmetry class DIII). This is analogous to how the CUE produces the COE and CSE, but the physics is different. A Majorana zero-mode is a coherent superposition of electrons and holes from the \textit{same} spin band, while Andreev scattering couples \textit{opposite} spin bands. Spin-orbit coupling is therefore needed to mix the spin bands and realize the CRE --- while the CUE can exist with or without spin-orbit coupling. As a consequence, time-reversal symmetry can realize only the ensemble T$_-$CRE of antisymmetric orthogonal matrices. The symmetry that is responsible for the ensemble T$_+$CRE of symmetric orthogonal matrices is the ``chiral'' symmetry discussed in Section \ref{Tchsym}.

In Table \ref{tab:table2} we summarize the three scattering matrix ensembles that support Majorana zero-modes. The first row lists the name of the ensemble and the second row lists the name of the corresponding symmetric space. The last row lists the topological invariant, discussed next. 

\begin{table*}
\centering
\begin{tabular}{ | l || c | c | c |}
\hline
Ensemble name & \textbf{CRE} & \textbf{T$\bm{_{+}}$CRE} & \textbf{T$\bm{_{-}}$CRE} \\ \hline
Symmetry class &  D &  BDI & DIII \\ \hline
$S$-matrix elements &  real &  real & real \\ \hline
$S$-matrix space & orthogonal & orthogonal symmetric & orthogonal antisymmetric\\ \hline
Topological invariant & ${\rm Det}\,S$ & $\tfrac{1}{2}\,{\rm Tr}\,S$ & ${\rm Pf}\,S$ \\ \hline
\end{tabular}
\caption{The three ensembles that support Majorana zero-modes.}
\label{tab:table2}
\end{table*}

\subsection{Topological quantum numbers}
\label{Top_qn}

Typically, whenever the orthogonal group appears in a physics problem, it is sufficient to consider only matrices with determinant ${\rm Det}\,S=+1$ --- the socalled special orthogonal group (denoted SO or ${\rm O}_{+}$). The remaining orthogonal matrices in ${\rm O}_{-}$ have ${\rm Det}\,S=-1$, they are disconnected from the identity matrix and would seem unphysical.

The advent of topological superconductors \cite{Ryu10,Has10,Qi11} has provided for a physical realization of orthogonal scattering matrices with ${\rm Det}\,S=-1$ \cite{Boc00,Mer02,Akh11}. More generally, the three ensembles from Table \ref{tab:table2} each decompose into disjunct sub-ensembles, distinguished by an integer $Q$ called ``topological quantum number'' or ``topological invariant''. In terms of the scattering matrix, this number is represented by \cite{Ful11}
\begin{align}
&Q={\rm Det}\,S=\pm 1\;\;\mbox{in class D (CRE)},\label{Qformulasa}
\\
&Q=\tfrac{1}{2}\,{\rm Tr}\,S\in\{0,\pm 1,\ldots \pm N\}\;\;\mbox{in class BDI (T$_{+}$CRE)},\label{Qformulasb}
\\
&Q={\rm Pf}\,S=\pm 1\;\;\mbox{in class DIII (T$_{-}$CRE)}.\label{Qformulasc}
\end{align}

In the CRE the two sub-ensembles correspond to a uniform distribution of the orthogonal matrix $S$ in ${\rm O}_{\pm}(2N)$. The orthogonal antisymmetric matrices in the T$_{-}$CRE can be decomposed as
\begin{equation}
S=OJO^{\rm T},\;\;O\in{\rm O}_{\pm}(2N),\;\;J=\begin{pmatrix}
0&1\\
-1&0
\end{pmatrix},\label{SOJ}
\end{equation}
where each block of $J$ has dimension $N\times N$. Again, a uniform distribution of $O\in{\rm O}_{\pm}(2N)$ produces two sub-ensembles, distinguished by the Pfaffian of the scattering matrix,\footnote{The Pfaffian entry in Wikipedia contains a useful collection of formulas. Computer algorithms for the evaluation of Pfaffians can be obtained from  \textcite{Wim12}.}
\begin{equation}
{\rm Pf}\,S=({\rm Pf}\,J)({\rm Det}\,O)=(-1)^{N(N-1)/2}\,{\rm Det}\,O=\pm 1.\label{PfSformula}
\end{equation}

Finally, in the T$_{+}$CRE the scattering matrix is both orthogonal and symmetric, so its eigenvalues are $\pm 1$ and it has the decomposition
\begin{equation}
S=O\Sigma O^{\rm T},\;\;O\in{\rm O}_{+}(2N),\;\;\Sigma={\rm diag}\,(\pm 1,\pm 1,\ldots\pm 1).\label{SOcalS}
\end{equation}
The matrix $\Sigma$ is a socalled signature matrix and the number of $-1$'s on the diagonal represents the signature $\nu(S)$ of $S$. Now we may take $O\in{\rm O}_{+}(2N)$ without loss of generality, while the sub-ensembles are distinguished by the trace (or the signature) of the scattering matrix,
\begin{equation}
\tfrac{1}{2}\,{\rm Tr}\,S=\tfrac{1}{2}\,{\rm Tr}\,\Sigma=N-\nu(S)=0,\pm 1,\ldots \pm N.\label{TrSformula}
\end{equation}

The invariant $Q$ of symmetry class D and DIII is called a $\mathbb{Z}_2$ topological quantum number, because it can take on only two values. Symmetry class BDI has a $\mathbb{Z}$ topological quantum number, because as $N$ is varied $Q$ ranges over all (positive and negative) integer values. These invariants first appeared as winding numbers of Fermi surfaces, hence the adjective ``topological'' \cite{Sch08,Ryu10,Kit09,Has10,Qi11}. In the present context of scattering matrices, where $Q$ results from an operation in linear algebra \cite{Ful11}, the name ``algebraic invariant'' might be more natural.

Class D, DIII, and BDI are the three symmetry classes that support Majorana zero-modes. Two further symmetry classes AIII and CII have a $\mathbb{Z}$ topological quantum number given by the same Eq.\ \eqref{Qformulasb} as class BDI, but the corresponding zero-modes lack the self-conjugate Majorana nature (in class AIII because of the absence of particle-hole symmetry, in class CII because particle-hole symmetry relates different spin bands).

\section{Electrical conduction}
\label{elG}

\subsection{Majorana nanowire}
\label{RZH}

A semiconducting layer on a superconducting substrate is in symmetry class D when both time-reversal symmetry and spin-rotation symmetry are broken. The Bogoliubov-De Gennes Hamiltonian has the form of Eq.\ \eqref{HBdG}, with
\begin{equation}
H_{0}=\frac{\bm{p}^{2}}{2m_{\rm eff}}+U(\bm{r})+\frac{\alpha_{\rm so}}{\hbar}(\sigma_{x}p_{y}-\sigma_{y}p_{x})+\tfrac{1}{2}g_{\rm eff}\mu_{B}B\sigma_{x}.\label{H0RashbaZeeman}
\end{equation}
The first two terms give the kinetic energy and electrostatic potential energy. In the third term the momentum $\bm{p}=-i\hbar\partial/\partial\bm{r}$ in the $x$-$y$ plane of the layer is coupled to spin by the Rashba effect, breaking spin-rotation symmetry with characteristic length $l_{\rm so}=\hbar^{2}(m_{\rm eff}\alpha_{\rm so})^{-1}\simeq 100\,{\rm nm}$ and energy $E_{\rm so}=m_{\rm eff}(\alpha_{\rm so}/\hbar)^{2}\simeq 0.1\,{\rm meV}$. The last term describes the Zeeman effect of a magnetic field $B\hat{x}$, parallel to the layer, breaking time-reversal symmetry with characteristic energy $V_{\rm Z}=\tfrac{1}{2}g_{\rm eff}\mu_{B}B\simeq 1\,{\rm meV}$ at $B=1\,{\rm T}$.

Without the term $\sigma_{x}p_{y}$ the Hamiltonian would be real and hence the chiral symmetry of Section \ref{Tchsym} would promote the system from class D to class BDI \cite{Tew12}. Model calculations in a wire geometry (width $W$ in the $y$-direction), demonstrate that the chiral symmetry is effectively unbroken for $W\lesssim l_{\rm so}/2$ \cite{Die12}. Experimentally realized InSb nanowires \cite{Mou12}, of the type shown in Fig.\ \ref{fig_nanowire}, have $l_{\rm so}$ in the range 100--200~nm, so the crossover from class D to class BDI happens when the wire becomes narrower than about 100~nm.

As discovered by \textcite{Lut10} and by \textcite{Ore10}, the nanowire enters into a topologically nontrivial phase, with Majorana zero-modes at the end points, once the Zeeman energy $V_{\rm Z}$ exceeds the superconducting gap $\Delta_0$ (induced by the proximity effect). This theoretical prediction was a strong motivation for the experiments reviewed by \textcite{Ali12,Lei12,Sta13,Bee13}, as well as for the development of the random-matrix theory reviewed here. 

\subsection{Counting Majorana zero-modes}
\label{countingMzm}

The topological quantum number $Q$ from Section \ref{Top_qn} counts the number $\nu$ of stable Majorana zero-modes at each end of the $N$-mode nanowire, at most one in class D and up to $N$ in class BDI. This number is fully determined by the $2N\times 2N$ matrix $r$ of Fermi-level reflection amplitudes from the end of the nanowire \cite{Ful11}. 

The reflection matrix $r$ is a unitary matrix if the wire is sufficiently long that transmission to the other end can be neglected. It has a block structure of $N\times N$ submatrices,
\begin{equation}
r=\begin{pmatrix}
r_{ee}&r_{eh}\\
r_{he}&r_{hh}
\end{pmatrix}.\label{rsubmatrices}
\end{equation}
Andreev reflection (from electron to hole or from hole to electron) is described by the off-diagonal blocks, while the diagonal blocks describe normal reflection (without change of charge). 

At the Fermi level ($E=0$) the particle-hole symmetry, operative in both class D and BDI, is expressed by $r=\tau_{x}r^{\ast}\tau_{x}$, while the chiral symmetry (or fake time-reversal symmetry) of class BDI is $r=r^{\rm T}$. The corresponding symmetry operators ${\cal C}=\tau_{x}{\cal K}$ and ${\cal T}={\cal K}$ both square to $+1$, in accordance with Table \ref{table_AZ}.  In terms of the submatrices, this corresponds to
\begin{subequations}
\label{rDBDIsymmetries}
\begin{align}
&r^{\vphantom{\ast}}_{ee}=r_{hh}^{\ast},\;\;r^{\vphantom{\ast}}_{he}=r_{eh}^{\ast},\;\;\text{in class D and BDI},\label{rDBDIsymmetriesa}
\\
&r^{\vphantom{\rm T}}_{ee}=r_{ee}^{\rm T},\;\;r^{\vphantom{\dagger}}_{he}=r_{he}^{\dagger},\;\;\text{in class BDI only}.\label{rDBDIsymmetriesb}
\end{align}
\end{subequations}

The determinant of a unitary matrix lies on the unit circle in the complex plane, while $r=\tau_{x}r^{\ast}\tau_{x}$ implies that the determinant is real, hence equal to $\pm 1$. In class BDI the unitary matrix $\tau_{x}r$ squares to the unit matrix, $(\tau_{x}r)^{2}=r^{\ast}r=r^{\dagger}r=\openone$, so its $2N$ eigenvalues are $\pm 1$. The corresponding topological quantum numbers are\footnote{
In Table \ref{tab:table2} the class BDI topological quantum number is defined without the $\tau_x$ matrix, because there the scattering matrix is taken in the Majorana basis, while here we use the electron-hole basis.}
\begin{subequations}
\label{QclassDBDI}
\begin{align}
&Q={\rm Det}\,r=\pm 1,\;\;\nu=\tfrac{1}{2}(1-Q),\;\;\text{in class D},\label{QclassD}\\
&Q=\tfrac{1}{2}{\rm Tr}\,(\tau_{x}r)={\rm Tr}\,r_{he}\in\{0,\pm 1,\pm\ldots N\},\nonumber\\
&\qquad\qquad\qquad\qquad \nu=|Q|,\;\;\text{in class BDI}.\label{QclassBDI}
\end{align}
\end{subequations}

\subsection{Conductance distribution}
\label{GprobeMzm}

\begin{figure}[tb]
\centerline{\includegraphics[width=0.8\linewidth]{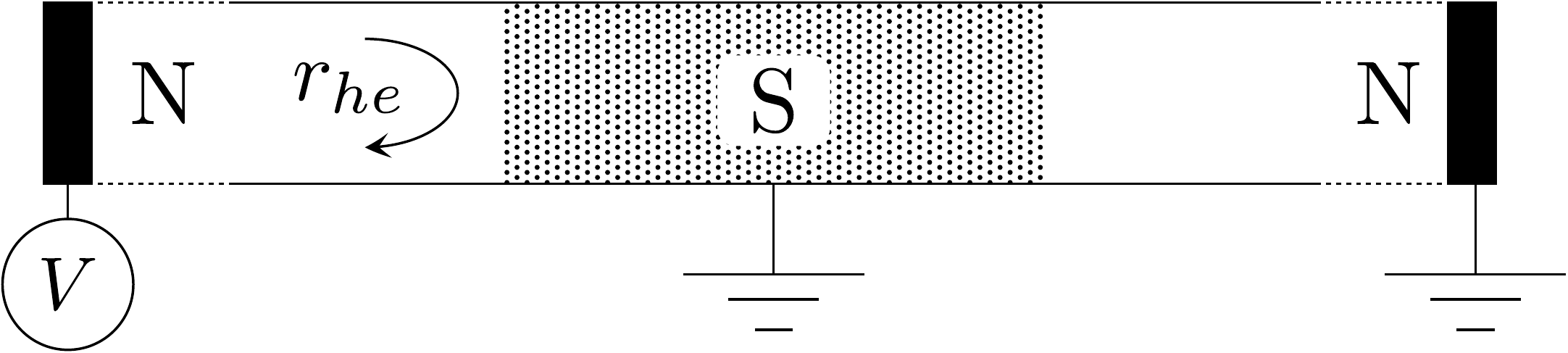}}
\caption{Superconducting wire (S) connected at both ends to a normal metal contact (N), in a geometry similar to Fig.\ \ref{fig_nanowire}. Majorana zero-modes may appear at the two NS interfaces. The current $I$ flowing from the normal metal (at voltage $V$) into the grounded superconductor gives the electrical conductance $G = I/V$, determined by the Andreev reflection matrix $r_{he}$.}
\label{fig_nanowire_setup}
\end{figure}

The electrical conductance $G$ in the nanowire geometry of Fig.\ \ref{fig_nanowire_setup} is determined by the Andreev reflection eigenvalues $A_{n}$ \cite{Tak92},
\begin{equation}
G/G_{0}=2\,{\rm Tr}\,r_{he}^{\vphantom{\dagger}}r_{he}^{\dagger}=2\sum_{n=1}^{N}A_{n},\label{GRnrelation}
\end{equation}
where $G_{0}=e^{2}/h$ is the conductance quantum. The factor of two in front of the sum accounts for the fact that Andreev reflection of an electron doubles the current. The eigenvalues $A_{n}$ of the Hermitian matrix product $r_{he}^{\vphantom{\dagger}}r_{he}^{\dagger}$ lie in the interval $[0,1]$. The $A_{n}$'s different from 0 and 1 are twofold degenerate (B\'{e}ri degeneracy, see App.\ \ref{AppBeri}).

Both the conductance \eqref{GRnrelation} and the number of Majorana zero-modes \eqref{QclassDBDI} are given by the same reflection matrix, so we can try to relate them. The B\'{e}ri degeneracy enforces the upper and lower bounds \cite{Die12}
\begin{equation}
2\nu\leq G/G_{0}\leq 2(N-\zeta),\label{NMGrelation}
\end{equation}
where $\zeta=0$ if $N-\nu$ is even and $\zeta=1$ if $N-\nu$ is odd. For $N=1$ this immediately gives $G/G_{0}=2\nu$, but for $N>1$ there is no one-to-one relation between the two quantities. If we assume that $r$ is distributed according to the circular ensemble, a statistical dependence of $G$ on $\nu$ can be obtained.

\begin{figure}[tb]
\centerline{\includegraphics[width=0.9\linewidth]{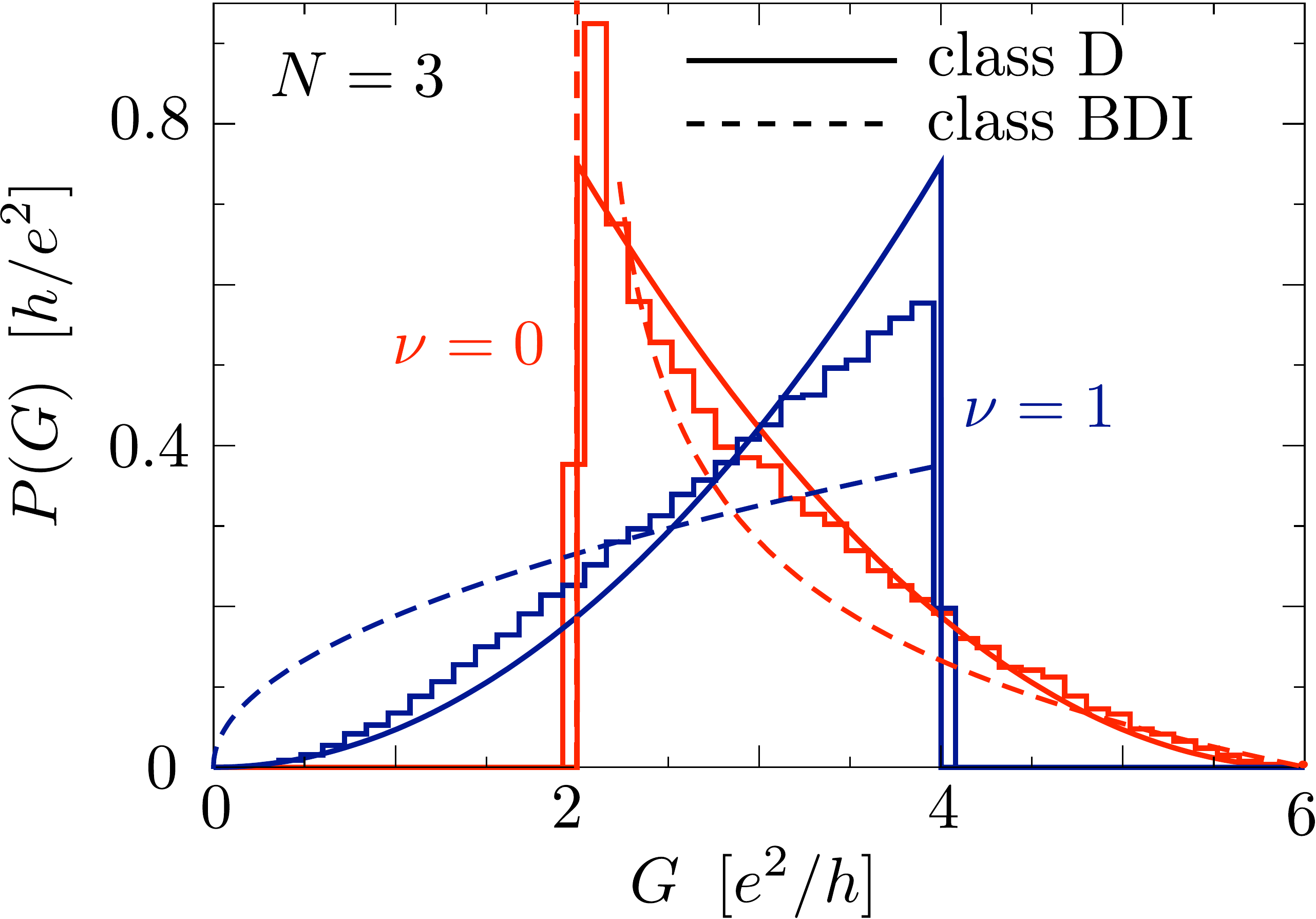}}
\caption{Probability distribution of the electrical conductance for $N=3$ modes, in the circular ensemble of class D (solid curves) and class BDI (dashed curves), either without any Majorana zero-modes ($\nu=0$) or with one zero-mode ($\nu=1$). The curves follow by integration of the probability distribution \eqref{Pcircular} of the Andreev reflection eigenvalues. The histograms are the results of a microscopic calculation for the Rashba-Zeeman Hamiltonian \eqref{H0RashbaZeeman}, in a three-mode nanowire of width $W=l_{\rm so}=100\,{\rm nm}$ (in class D, but close to the crossover into class BDI at $W\lesssim l_{\rm so}/2$). Figure adapted from \textcite{Bee11}.
}
\label{fig_PDBDI}
\end{figure}

For that purpose we need the probability distribution of the $M=\frac{1}{2}(N-\nu-\zeta)$ twofold degenerate Andreev reflection eigenvalues in the class D or BDI circular ensemble. It is given by \cite{Die12,Bee11}
\begin{subequations}
\label{Pcircular}
\begin{align}
&P_{\rm D}\propto\prod_{1=i<j}^{M}(A_{i}-A_{j})^{4}\prod_{k=1}^{M}A_{k}^{2\zeta}(1-A_{k})^{2\nu},\label{PcircularD}\\
&P_{\rm BDI}\propto\prod_{1=i<j}^{M}(A_{i}-A_{j})^{2}\prod_{k=1}^{M}A_{k}^{\zeta-1/2}(1-A_{k})^{\nu}.\label{PcircularBDI}
\end{align}
\end{subequations}
These twofold degenerate $A_{n}$'s are free to vary in the interval $(0,1)$. In addition, there are $\nu$ Andreev reflection eigenvalues pinned at $1$ and $\zeta$ pinned at $0$.  The resulting dependence of the conductance distribution $P(G)$ on $\nu$ is plotted in Fig.\ \ref{fig_PDBDI}, for the case $N=3$.

The sensitivity of $P(G)$ to Majorana zero-modes becomes weaker and weaker with increasing $N$. This happens in a particularly striking (nonperturbative) way in the circular real ensemble of class D, where the $p$-th cumulant of the conductance becomes completely independent of the topological quantum number for $N>p$ \cite{Bee11}. Indeed, the red and blue solid curves in Fig.\ \ref{fig_PDBDI} have different skewness but identical average and variance, as expected for $N=3$.

Fig.\ \ref{fig_PDBDI} also includes histograms of conductance from a microscopic calculation in the Majorana nanowire of Section \ref{RZH}, where the ensemble is generated by varying the disorder potential. The agreement with the predictions from the circular ensemble is quite reasonable, with two reservations. The first is that the nanowire was near the class-D-to-BDI crossover, with a partially broken chiral symmetry. The second is that the diffusive scattering produced by the disorder potential is not the chaotic scattering of the circular ensembles --- the scattering channels are not uniformly mixed by disorder.

\subsection{Weak antilocalization}
\label{MR}

\begin{figure}[tb]
\centerline{\includegraphics[width=0.9\linewidth]{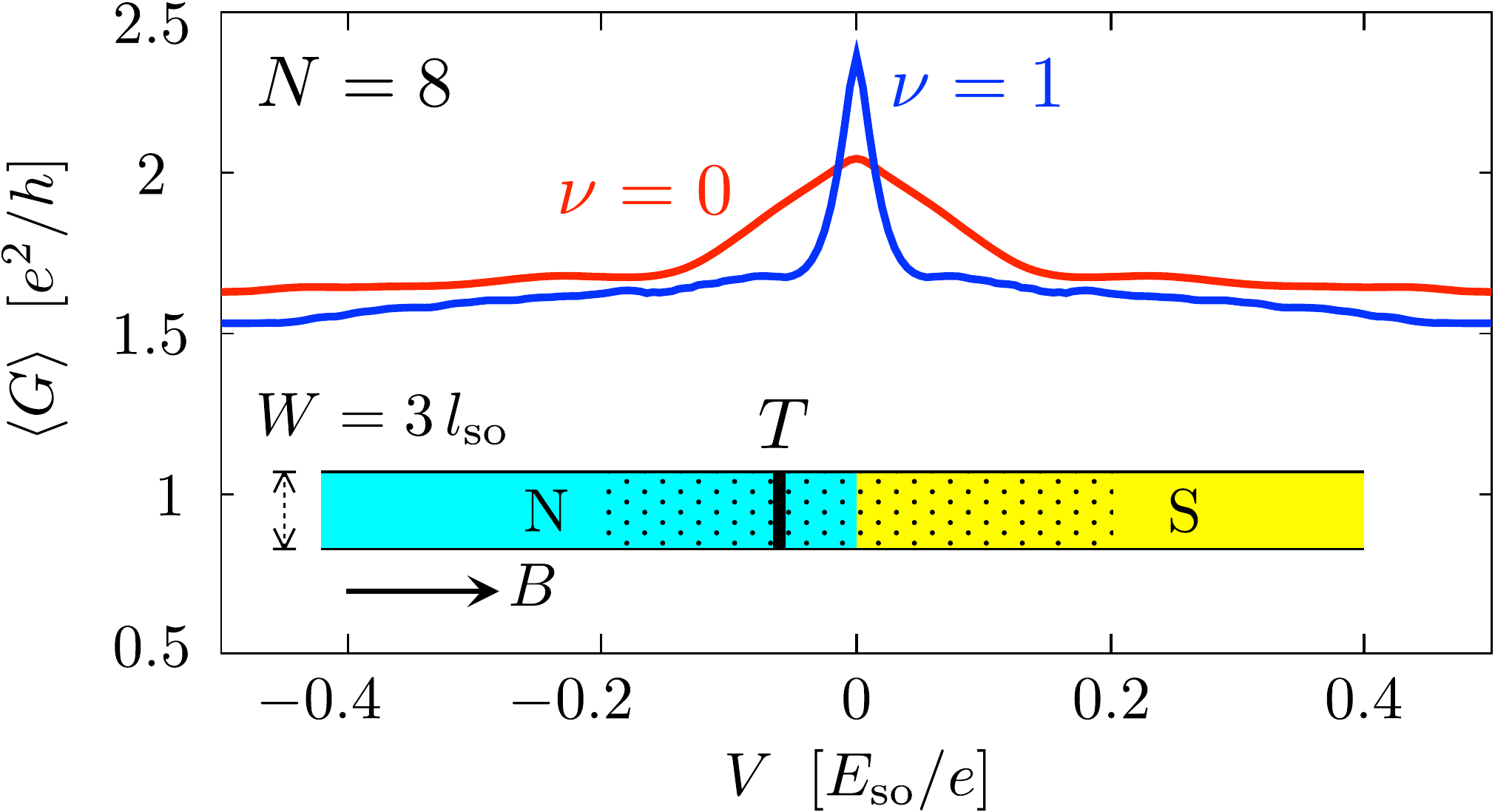}}
\caption{Disorder-averaged differential conductance as a function of bias voltage, calculated for the nanowire shown in the inset. (The solid vertical line indicates the position of the tunnel barrier, transmission probability $T=0.4$ per mode; disordered regions are dotted.) The $\nu=0$ curve is for a weak magnetic field ($E_{\rm Z}=2.5\,E_{\rm so}$), when the system is topologically trivial and the zero-bias peak is due to weak antilocalization. (The corresponding peak for a single disorder realization is shown in Fig.\ \ref{fig_Dsingle}.) The $\nu=1$ curve, in a stronger magnetic field ($E_{\rm Z}=10.5\,E_{\rm so}$), shows the Majorana resonance in the topologically  nontrivial regime. Figure adapted from \textcite{Pik12}.
}
\label{fig_zbp}
\end{figure}

The presence or absence of a Majorana zero-mode is a topological property of the nanowire, irrespective of how the wire is terminated. In particular, the lower bound $G\geq 2\nu e^{2}/h$ holds whether or not there is any tunnel barrier to confine the Majorana at the end of the wire. The barrier serves a purpose in providing a resonant peak in the differential conductance $G(V)=dI/dV$ around zero voltage \cite{Sen01,Bol07,Law09,Sau10,Fle10}. This resonant peak, reported in several experiments \cite{Mou12,Das12,Den12,Fin13,Chu13,Nad14}, is shown in the computer simulation of Fig.\ \ref{fig_zbp}.

\begin{figure}[tb]
\centerline{\includegraphics[width=0.9\linewidth]{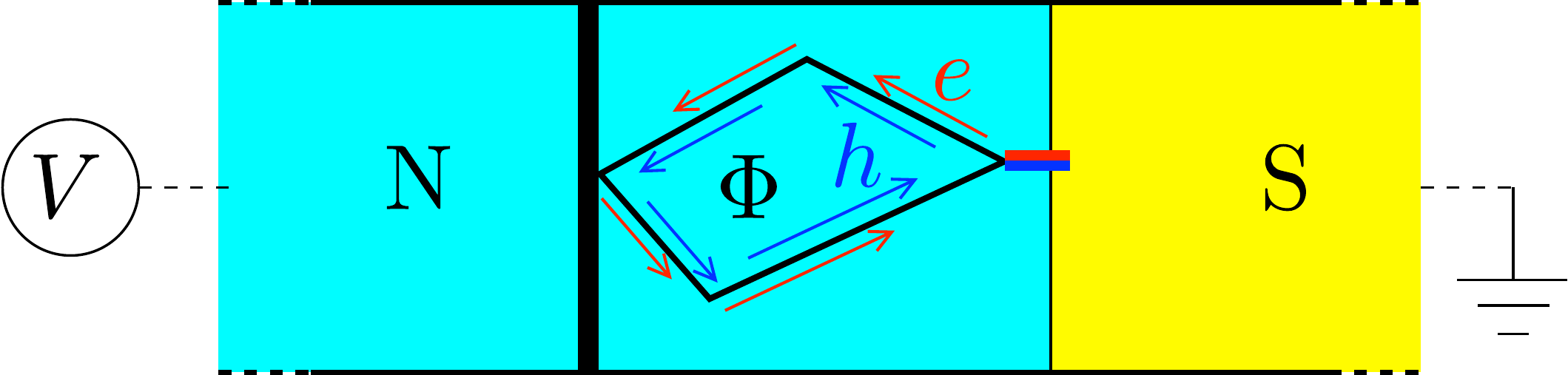}}
\caption{Example of a phase-conjugate series of scattering events responsible for the weak antilocalization effect in an NS junction. The same loop is traversed once as an electron ($e$) and once as a hole ($h$), with an intermediate Andreev reflection. Since electron and hole encircle the same flux $\Phi$ with opposite charge, the total accumulated phase shift vanishes even though time-reversal symmetry is broken by the magnetic field. The systematic constructive interference shows up as a zero-bias conductance peak that survives a disorder average. Figure adapted from \textcite{Pik12}.
}
\label{fig_NS}
\end{figure}

One sees from that simulation (based on the Rashba-Zeeman Hamiltonian of Section \ref{RZH}), that a broader and smaller zero-bias peak appears also in the disorder-averaged conductance of a topologically {\em trivial} nanowire --- without any Majorana zero-modes (compare blue and red curves). In that case the origin of the peak is the weak-antilocalization effect \cite{Pik12}: the constructive interference of phase-conjugate scattering sequences (see Fig.\ \ref{fig_NS}). In normal metals this interference effect requires time-reversal symmetry, but in the presence of a superconductor particle-hole symmetry suffices \cite{Bro95,Alt96}. The same interference effect is responsible for a midgap peak in the density of states \cite{Bag12,Nev13}, see Sec.\ \ref{spectralpeak}.

The two distinct origins of a zero-bias conductance peak in the average conductance can be compared in a random-matrix model, by including the effect of a tunnel barrier on the circular ensemble. In the zero-voltage limit, so at the Fermi level, we take for the reflection matrix $r_{0}$ without the barrier a uniform distribution in ${\rm O}_{+}(2N)$ for the topologically trivial system (no Majoranas, $\nu=0$) and in ${\rm O}_{-}(2N)$ for the nontrivial system (with a Majorana zero-mode, $\nu=1$). This is the circular real ensemble (CRE) of symmetry class D, in the Majorana basis (see Table \ref{tab:table2}). Away from the Fermi level, at voltages large compared to the Thouless energy, the constraint from particle-hole symmetry is ineffective and $r_{0}$ is distributed uniformly over the entire unitary group ${\rm U}(2N)$. This is the circular unitary ensemble (CUE). 

The tunnel barrier (transmission probability $T$ per mode) transforms $r_{0}$ into
\begin{equation}
r=\sqrt{1-T}+Tr_{0}(1+\sqrt{1-T}\,r_{0})^{-1}.\label{RTR0}
\end{equation}
The resulting nonuniform distribution of $r$ is known as the Poisson kernel of the circular ensemble \cite{Ber09a,Bro95b},
\begin{equation}
P(r)\propto|{\rm Det}\,(1-\sqrt{1-T}\,r)|^{-p}.
\label{PRPoisson}
\end{equation}
The exponent equals $p=4N$ in the CUE and $p=2N-1$ in the CRE.

\begin{figure}[tb]
\centerline{\includegraphics[width=0.9\linewidth]{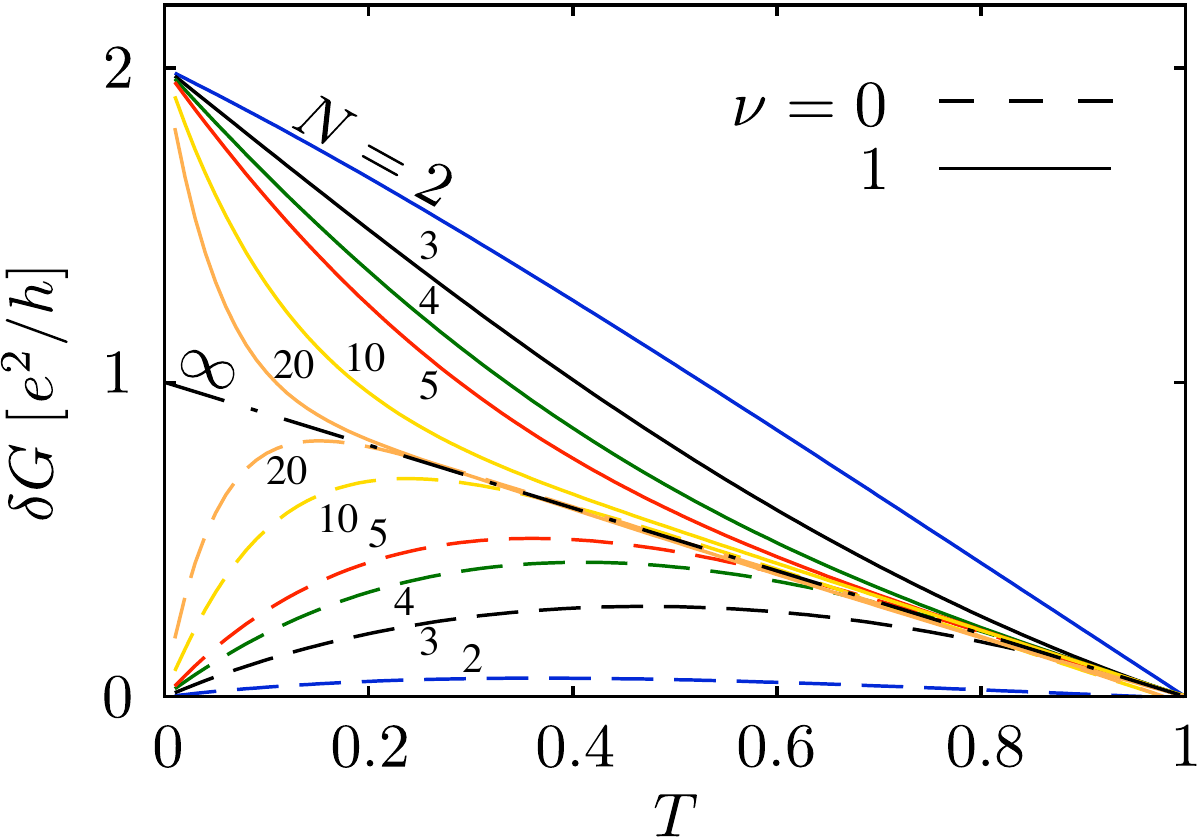}}
\caption{Amplitude $\delta G$ of the average zero-voltage conductance peak as a function of (mode-independent) transmission probability $T$ through the tunnel barrier, calculated numerically for the class D circular ensemble. The dashed and solid curves represent, respectively, the topologically trivial and nontrivial superconductor. The dash-dotted curve is the topology-independent large-$N$ limit \eqref{deltaGlargeNAZ}. Figure adapted from \textcite{Pik12}.
}
\label{fig_deltaGcircular}
\end{figure}

The formula \eqref{GRnrelation} for the conductance in the electron-hole basis can be rewritten in the Majorana basis, by carrying out the unitary transformation
\begin{equation}
r\mapsto \Omega r\Omega^{\dagger},\;\; \Omega=\sqrt{\tfrac{1}{2}}\begin{pmatrix}
1&1\\
i&-i
\end{pmatrix}.\label{RMajoranabasis}
\end{equation}
The result is
\begin{equation}
G/G_{0}=N-\tfrac{1}{2}\,{\rm Tr}\,r\tau_{y}r^{\dagger}\tau_{y},\;\;\tau_{y}=\begin{pmatrix}
0&-i\\
i&0
\end{pmatrix}.\label{Gtauyrelation}
\end{equation}

The zero-bias conductance peak is then given by the difference $\delta G=\langle G\rangle_{\rm CRE}-\langle G\rangle_{\rm CUE}$ of the average of $r_{0}$ over ${\rm O}_{\pm}(2N)$ (for the CRE) and over ${\rm U}(2N)$ (for the CUE). Results are shown in Fig.\ \ref{fig_deltaGcircular}. The large-$N$ limit has the $\nu$-independent value \cite{Alt97}
\begin{equation}
\delta G/G_{0}=1-T+{\cal O}(N^{-1}).\label{deltaGlargeNAZ}
\end{equation}

\subsection{Andreev resonances}
\label{AR}

\begin{figure}[tb]
\centerline{\includegraphics[width=0.9\linewidth]{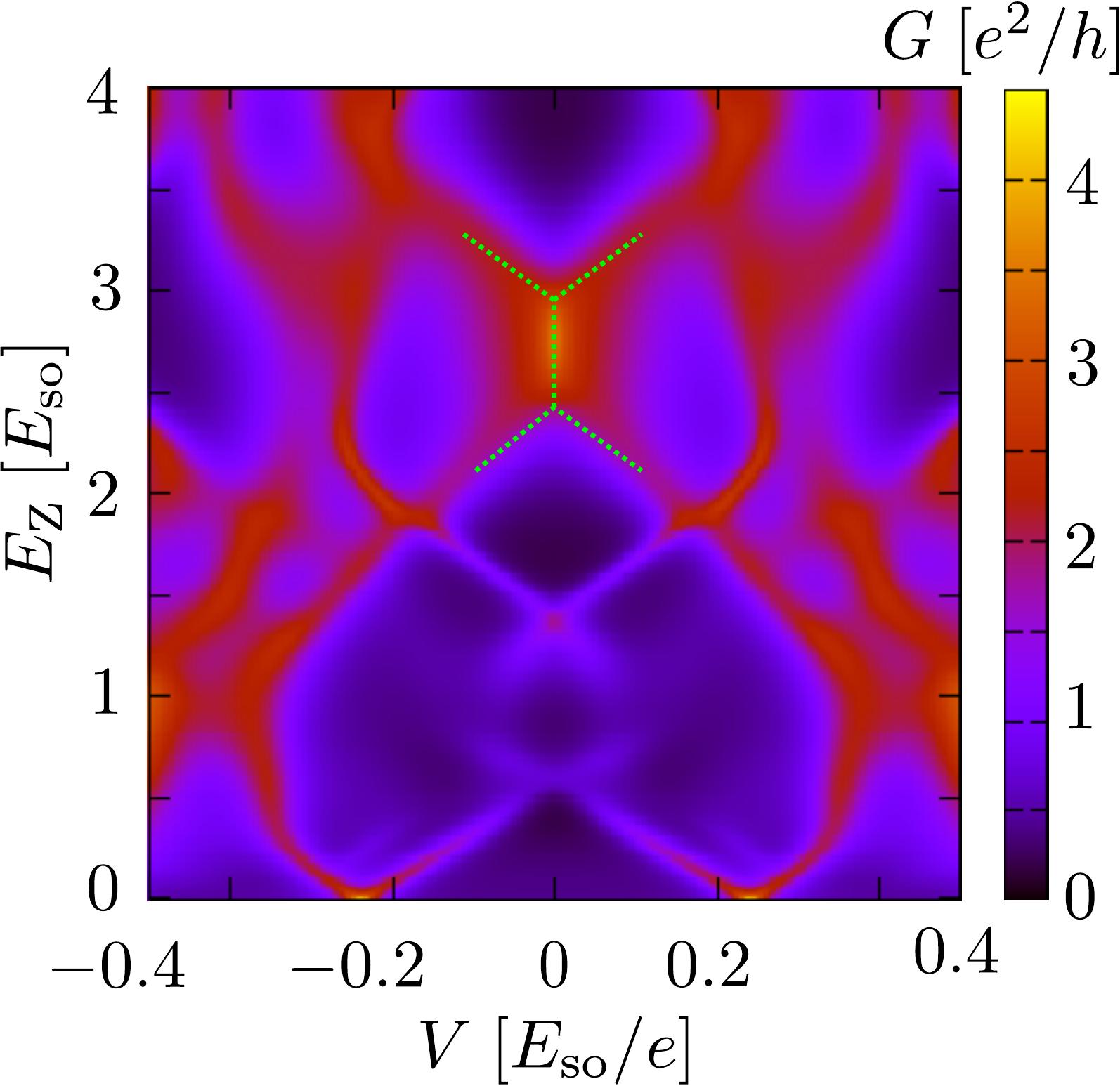}}
\caption{Voltage and magnetic-field dependence of the conductance in the same nanowire as in Fig.\ \ref{fig_zbp}, but now for a single disorder realization. The magnetic field range is in the topologically trivial regime, without Majorana zero-modes. The conductance peak indicated by the dotted line is pinned to zero voltage over a range of magnetic field values because of the accumulation of reflection matrix poles on the imaginary axis in a class-D superconductor, see Fig.\ \ref{fig_poles}. Figure adapted from \textcite{Pik12}.
}
\label{fig_Dsingle}
\end{figure}

The weak antilocalization effect explains the appearance of a zero-bias peak in the {\em disorder-averaged} conductance. {\em Sample-specific} zero-bias peaks in the same nanowire geometry are shown in Fig. \ref{fig_Dsingle}. These are due to resonant Andreev reflection from quasibound states near the Fermi level. Level crossings produce an X-shaped pattern when two resonant peaks meet and split again, but there is also a Y-shaped pattern of peaks that merge and remain pinned to $V=0$ over a range of magnetic field values.

The center $E$ and width $2\gamma$ of the Andreev resonances are encoded by the poles $\varepsilon=E-i\gamma$ of the  reflection matrix in the complex energy plane. Referring to the scattering matrix expression \eqref{Cayley}, which can also be written as
\begin{equation}
r(E)=1-2\pi i W^\dagger(E-H+i\pi WW^\dagger)^{-1}W,\label{Cayley2}
\end{equation}
the reflection matrix poles are eigenvalues of the ${\cal N}\times {\cal N}$ non-Hermitian matrix
\begin{equation}
{\cal H}=H-i\pi WW^{\dagger}.\label{Heffdef}
\end{equation}
Because the coupling matrix product $WW^{\dagger}$ is positive definite, the poles all lie in the lower half of the complex plane ($\gamma>0$), as required by causality. Particle-hole symmetry requires that the poles are symmetrically arranged around the imaginary axis ($\varepsilon$ and $-\varepsilon^{\ast}$ are both poles). 

\begin{figure}[tb]
\centerline{\includegraphics[width=1\linewidth]{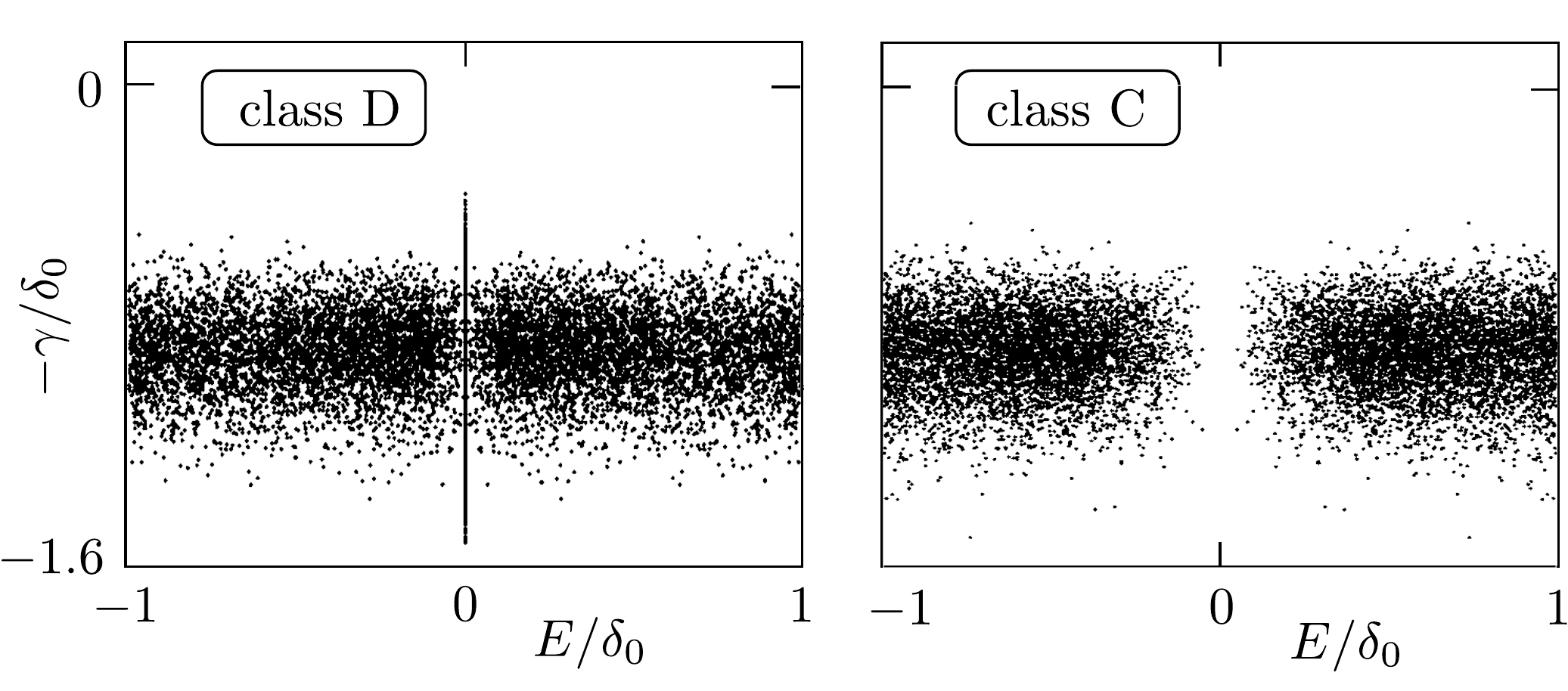}}
\caption{Reflection matrix poles $\varepsilon=E-i\gamma$ in the two Altland-Zirnbauer ensembles with broken time-reversal symmetry (parameters ${\cal N}=500$, $N=25$, $T=0.2$). Only a narrow energy range near $E=0$ is shown, to contrast the accumulation of the poles on the imaginary axis in class D and the repulsion in class C. Figure adapted from \textcite{Mi14}.
}
\label{fig_poles}
\end{figure}

Fig.\ \ref{fig_poles} is a scatter plot of the eigenvalues of ${\cal H}$ for the Gaussian distribution \eqref{PHGaussian} of the Hamiltonian $H$, with coupling matrix \cite{Guh98}
\begin{align}
&W_{nm}=w_n \delta_{nm},\;\;1\leq n\leq {\cal N},\;\;1\leq m\leq 2N,\label{Wdef}\\
&|w_n|^2 = \frac{{\cal N}\delta_0}{\pi^2 T_n}\bigl( 2 - T_n - 2 \sqrt{1-T_n} \bigr),\label{wdef}
\end{align}
representing a tunnel barrier with transmission probability $T_n$. (The plot is for a mode-independent $T_n\equiv T=0.2$.) The two Altland-Zirnbauer ensembles with broken time-reversal symmetry are contrasted, with and without spin-rotation symmetry (class C and class D). For $|E|\gtrsim\delta_0$ the poles have a uniform density in a strip parallel to the real axis, familiar from the Wigner-Dyson ensembles \cite{Fyo97}. For smaller $|E|$ the poles are repelled from the imaginary axis in class C, while in class D they accumulate on that axis \cite{Mi14}.

As pointed out in  \textcite{Pik11}, a nondegenerate pole $\varepsilon=-i\gamma$ on the imaginary axis has a certain stability, it cannot acquire a nonzero real part $E$ without breaking the $\varepsilon\leftrightarrow-\varepsilon^{\ast}$ symmetry imposed by particle-hole conjugation. To see why this stability is not operative in class C, we note that on the imaginary axis $\gamma$ is a real eigenvalue of a matrix $i{\cal H}$ that {\em commutes} with the charge-conjugation operator: ${\cal C}i{\cal H}=-i{\cal C}{\cal H}=i{\cal H}{\cal C}$. In class C the anti-unitary operator ${\cal C}$ squares to $-1$, see Table \ref{table_AZ}, so Kramers theorem\footnote{The usual Kramers degeneracy refers to the eigenvalues of a Hermitian matrix that commutes with an anti-unitary operator squaring to $-1$. Here the matrix is not Hermitian, but the degeneracy still applies to real eigenvalues.} 
forbids nondegenerate poles on the imaginary axis. In class D, in contrast, the operator ${\cal C}$ squares to $+1$, Kramers degeneracy is inoperative and a number $N_{\rm Y}$ of nondegenerate poles is allowed on the imaginary axis.

The reflection matrix $r(0)$ at the Fermi level is a real orthogonal matrix in class D, with determinant $\pm 1$. Because
\begin{equation}
(-1)^{N_{\rm Y}}=\lim_{E\rightarrow 0}\,{\rm Det}\,r(E)\equiv Q
\end{equation}
is the same class-D topological quantum number as in Eq.\ \eqref{QclassD}, the nanowire is topologically trivial or nontrivial depending on whether $N_{\rm Y}$ is even or odd. One can now distinguish two types of transitions \cite{Pik11,Pik13}: At a topological phase transition $N_{\rm Y}$ changes by $\pm 1$, which requires closing of the excitation gap in the nanowire and breaking of the unitarity of the reflection matrix $r$. At a ``pole transition'' $N_{\rm Y}$ changes by $\pm 2$, the excitation gap remains closed and $r$ remains unitary. Both types of transitions produce the same Y-shaped conductance profile of two peaks that merge and stick together for a range of parameter values --- distinct from the X-shaped profile that happens without a change in $N_{\rm Y}$. Other similarities of the two types of transitions are discussed by \textcite{San14}
 
In experiments one can use a variety of methods to distinguish the pole transition from the topological phase transition: As calculated in  \textcite{Mi14} the average number of poles on the imaginary axis is $\langle N_{\rm Y}\rangle \simeq T^{3/2}\sqrt{N}$ for $T\ll 1$, so one way to suppress the pole transitions is to couple the metal to the superconductor via a small number of modes $N$ with a small transmission probability $T$. The pole transitions are a sample-specific effect, while the topological phase transition is expected to be less sensitive to microscopic details of the disorder. One would therefore not expect the pole transitions to reproduce in the same sample upon thermal cycling. Most convincingly, if one can measure from both ends of a nanowire, one could search for correlations: The $\pm 2$ changes in $N_{\rm Y}$ at the two ends are uncorrelated, while $\pm 1$ changes should happen jointly at both ends --- provided that the wire is not broken into disjunct segments.

\subsection{Shot noise of Majorana edge modes}
\label{shotMajorana}

\begin{figure}[tb]
  \begin{center}
	 \includegraphics[width=0.9\columnwidth]{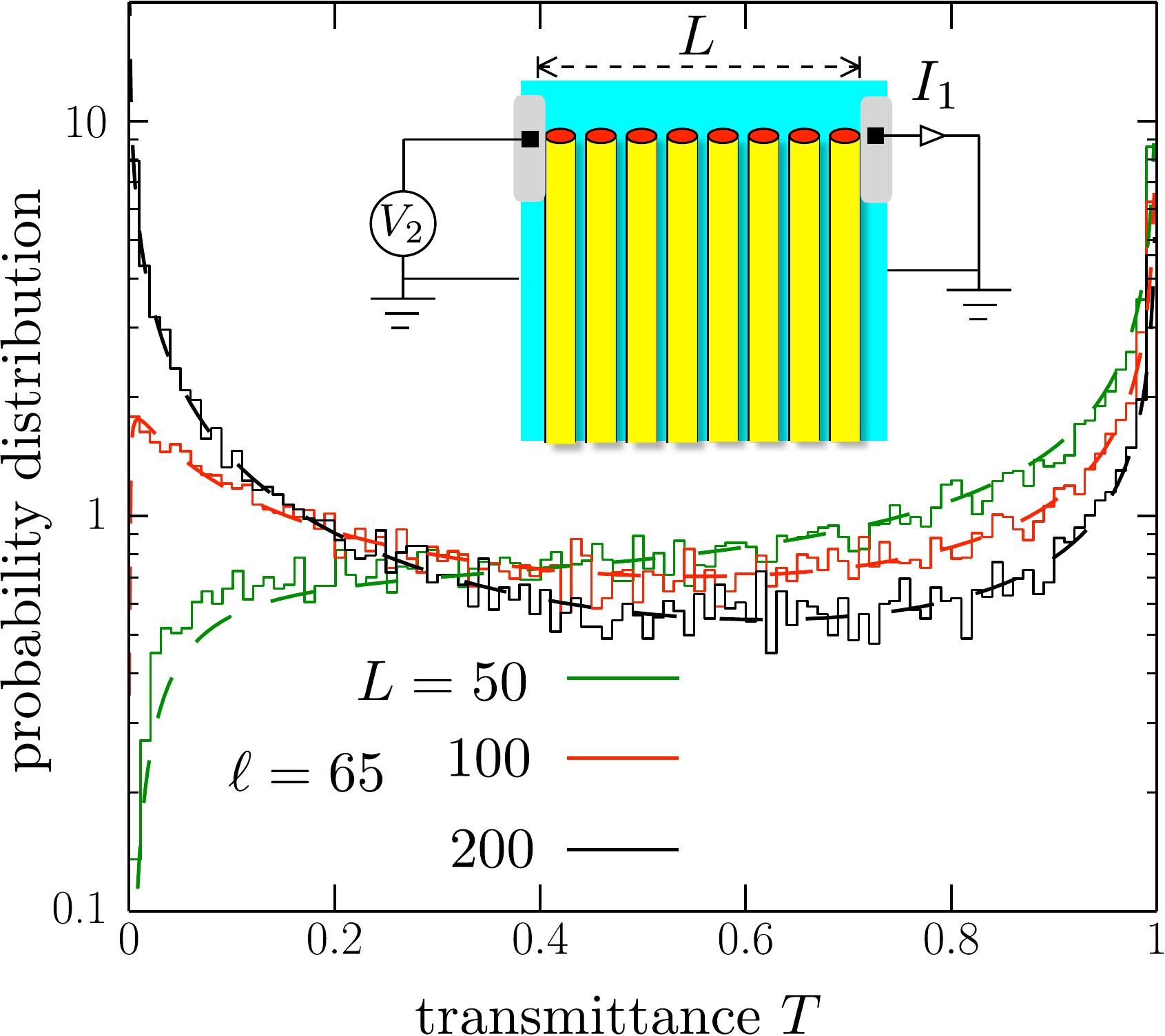}
  \end{center}
  \caption{Distribution of the transmittance of a Majorana edge mode in a chain of parallel nanowires on a superconducting substrate. The histograms are calculated numerically for a model Hamiltonian of an anisotropic \textit{p}-wave superconductor, for different lengths $L$ of the edge. Dashed lines show the analytical result \eqref{PTbimodal}, with the mean free path $\ell$ as the single fit parameter. With increasing $L$, a bimodal distribution evolves, which produces a slow $1/\sqrt{L}$ decay of the shot noise power $P_{\rm shot}=\tfrac{1}{2}T(e^3 V_2/h)$. Figure adapted from \textcite{Die14}.
}
  \label{fig_array}
\end{figure}

When Majorana zero-modes are in close proximity and overlap, a Majorana mode is formed with a linear dispersion. This happens at the edge of an array of parallel nanowires, as illustrated in Fig.\ \ref{fig_array}. Such a Majorana edge mode carries heat (see Sec.\ \ref{Majorana_heat}) but it carries no charge, as a consequence of particle-hole symmetry. Electrical detection remains possible via the time-dependent fluctuations $\delta I_1(t)$ in the electrical current transmitted along the edge in response to a bias voltage $V_2$.

An unpaired Majorana mode with transmittance $T$ has shot noise power \cite{Akh11}
\begin{equation}
P_{\rm shot}=\int_{-\infty}^\infty dt\,\langle\delta I_1(0)\delta I_1(t)\rangle=\frac{e^3 V_2}{2h}\,T,\label{Pshotnoise}
\end{equation}
see App.\ \ref{majoranashotapp}. What is remarkable about this formula is the coexistence of shot noise and unit transmittance. For electrons, the shot noise power is proportional to $T(1-T)$, so the shot noise vanishes for $T=1$. The difference for Majorana fermions is that they are not in an eigenstate of charge: The average charge is zero but the variance is $e^2$, so the current can fluctuate even if $T=1$. A fully transmitted Majorana mode has a \textit{quantized} shot noise power of $\tfrac{1}{2}\times e^2/h$ per electron volt, the factor $1/2$ expressing the fact that an incident electron has overlap  $1/2$ with the Majorana mode.

The transmittance can be calculated from the edge mode Hamiltonian
\begin{equation}
H_{\rm edge}=\sum_{n}i\kappa_n\gamma_{n}\gamma_{n+1},\label{HKdef}
\end{equation}
describing the random coupling $\kappa_n$ of adjacent Majorana operators $\gamma_n$ and $\gamma_{n+1}$. The transmittance $T=1/\cosh^2 \alpha$ is determined by the Lyapunov exponent $\alpha$, which has a Gaussian distribution \cite{Bro00,Gru05}. The variance ${\rm Var}\,\alpha=L/\ell$ equals the ratio of the length $L$ of the edge and the mean free path $\ell$. For statistically independent $\kappa_n$'s the average $\langle\alpha\rangle$ vanishes, resulting in a {\em bimodal} distribution of the transmission probability \cite{Die14},
\begin{align}
P(T)&=\int_{-\infty}^{\infty}d\alpha\,\delta(T-1/\cosh^2\alpha)(2\pi L/\ell)^{-1/2}e^{-\alpha^2\ell/2L}\nonumber\\
&=(\ell/2\pi L)^{1/2}\,T^{-1}(1-T)^{-1/2}\nonumber\\
&\qquad\times\exp\bigl[-(\ell/2L)\,{\rm arcosh}^{2}(T^{-1/2})\bigr],\label{PTbimodal}
\end{align}
peaked near $T=0$ and $T=1$. It follows that the average shot noise power decays algebraically as
\begin{equation}
\langle P_{\rm shot}\rangle=\frac{e^3 V_2}{h}\sqrt{\frac{\ell}{2\pi L}}.\label{Paveragedecay}
\end{equation}

The absence of localization of the Majorana edge mode is a consequence of the statistical equivalence of the coupling between any pair of neigboring Majorana operators \cite{Ful14}. This is the crucial distinction with the Kitaev chain formed out of magnetic nanoparticles, discussed in Sec.\ \ref{Kitaevintro}. There Majorana operators on the same nanoparticle have a different coupling strength than those on adjacent nanoparticles. The Lyapunov exponent $\alpha$ then still has a Gaussian distribution, but with a nonzero mean $\langle\alpha\rangle=L/\xi$ corresponding to a finite localization length $\xi$. It follows that the transmittance of the Kitaev chain has a log-normal distribution peaked at $T=e^{-2L/\xi}$, with an exponentially decaying average transmission \cite{Mot01,Gru05}.

\section{Thermal conduction}
\label{thG}

\subsection{Topological phase transitions}
\label{topphasetrans}

\begin{figure}[tb]
\centerline{\includegraphics[width=0.9\linewidth]{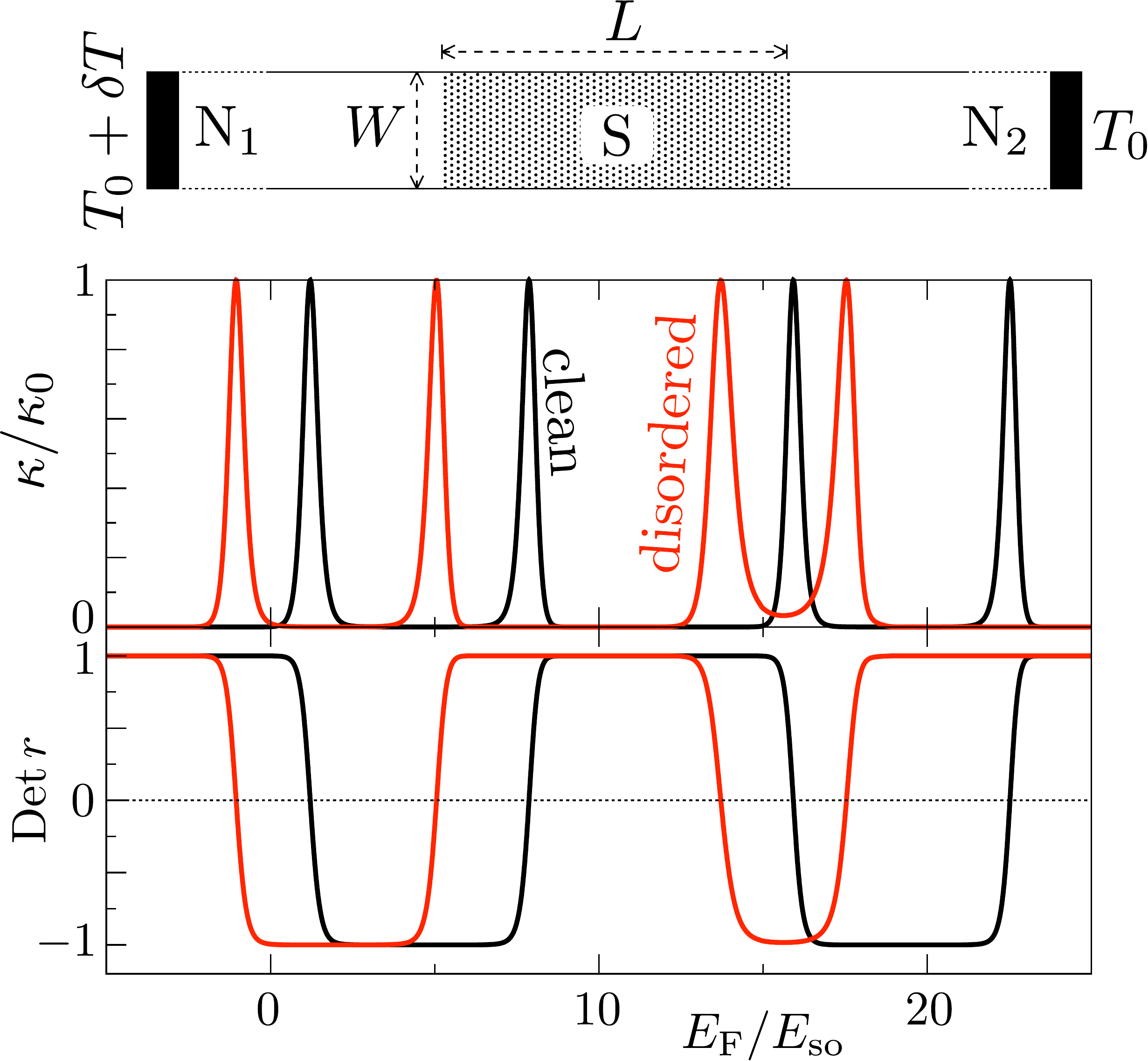}}
\caption{Thermal conductance and determinant of reflection matrix of a clean and disordered superconducting nanowire. The curves are calculated for the Rashba-Zeeman Hamiltonian \eqref{H0RashbaZeeman}, with parameters $W =L/10=l_{\rm so}$, $\Delta_0 = 10\,E_{\rm so}$, and $E_{\rm Z} = 10.5\,E_{\rm so}$. The number of propagating modes $N$ varies from 0 to 4 in the range of Fermi energies shown. The thermal conductance $\kappa=I_{\rm heat}/\delta T$ gives the heat current between the two contacts at temperature $T_0$ and $T_0+\delta T$. Figure adapted from \textcite{Akh11}.
}
\label{fig_Gthermal}
\end{figure}

While the electrical conductance $G$ of a superconducting nanowire gives information on the topological quantum number $Q$, the thermal conductance $\kappa$ signals the topological phase transitions, where $Q$ changes from one value to another. This is illustrated in Fig.\ \ref{fig_Gthermal}, for the class-D Majorana nanowire of Section \ref{RZH}. The peak in $\kappa$ at the topological phase transition has the quantized value
\begin{equation}
\kappa_0=\frac{\pi^2 k_{\rm B}^{2}T_{0}}{6h},\label{kappa0def}
\end{equation}
without any finite-size or disorder corrections \cite{Akh11}.

The quantization follows directly from the relation
\begin{equation}
\kappa/\kappa_{0}=2N-{\rm Tr}\,r^{\vphantom{\dagger}}r^{\dagger}=\sum_{n=1}^{2N}(1-R_{n})\label{kappaRnrelation}
\end{equation}
between the thermal conductance and the eigenvalues $R_{n}$ of the reflection matrix product $rr^{\dagger}$. Far from the topological phase transition the reflection matrix $r$ is unitary, hence $R_{n}=1$ for all $n$ and $\kappa$ vanishes. In a finite system the reflection matrix is a continuous function of external parameters, such as the Fermi energy $E_{\rm F}$, so if the class-D topological quantum number $Q={\rm Det}\,r$ changes sign, it must go through zero: ${\rm Det}\,r=0\Rightarrow\prod_{n}R_{n}=0$ at the transition point.  Generically, one single $R_{n}$ will vanish, producing a peak in $\kappa$ of amplitude $\kappa_{0}$. 

In class BDI the argument is similar \cite{Ful11}: A change in the topological quantum number $Q=\tfrac{1}{2}\,{\rm Tr}\,\tau_{x}r$ by one unit happens when one eigenvalue of $\tau_{x}r$ switches between $\pm 1$, so it must go through zero, hence ${\rm Det}\,\tau_{x}r=0\Rightarrow{\rm Det}\,r=0\Rightarrow\prod_{n}R_{n}=0$ at the transition point, again resulting in a quantized thermal conductance peak.

\subsection{Super-universality}
\label{superuniv}

The two symmetry classes D and BDI in an $N$-mode superconducting wire are distinguished by the number $\nu$ of stable zero-modes at each end point: $\nu\in\{0,1\}$ in class D versus $\nu\in\{0,1,2,\ldots N\}$ in class BDI. One speaks of a $\mathbb{Z}_2$ versus a $\mathbb{Z}$ topological quantum number. At the topological phase transition itself, where $\nu$ changes by one unit as some control parameter $\alpha$ (such as the Fermi energy) passes through $\alpha_{\rm c}$, the thermal conductance peak has the universal line shape \cite{Ful11}
\begin{equation}
\kappa=\frac{\kappa_0}{\cosh^2\delta}\;\;\delta=(\alpha-\alpha_{\rm c})/\Gamma,\label{Guniversal}
\end{equation}
in both class D and class BDI. (The width $\Gamma$ of the peak is not universal.) One cannot, therefore, distinguish the $\mathbb{Z}_2$ and $\mathbb{Z}$ topological phases by studying a single phase transition. 

This super-universality \cite{Gru05} extends to the average density of states near a topological phase transition \cite{Rie14},
\begin{align}
\rho(E)&=\frac{L}{2l}\frac{d}{dE}|K_{l\delta/L}(2iE v_{\rm F}/\hbar l)|^{-2}\nonumber\\
&\propto|E|^{2l|\delta|/L-1}\;\;{\rm for}\;\;E\rightarrow 0,\label{rhoEresultPB}
\end{align}
in a disordered wire of length $L$, mean free path $l$, and Fermi velocity $v_{\rm F}$. (The function $K_{a}(x)$ is a Bessel function.) At the critical point $\delta=0$ the power-law singularity has an additional logarithmic factor \cite{Dys53}: $\rho(E)\propto 1/(E\ln^{3} E)$ for $E\rightarrow 0$.

\begin{figure}[tb]
\centerline{\includegraphics[width=0.6\linewidth]{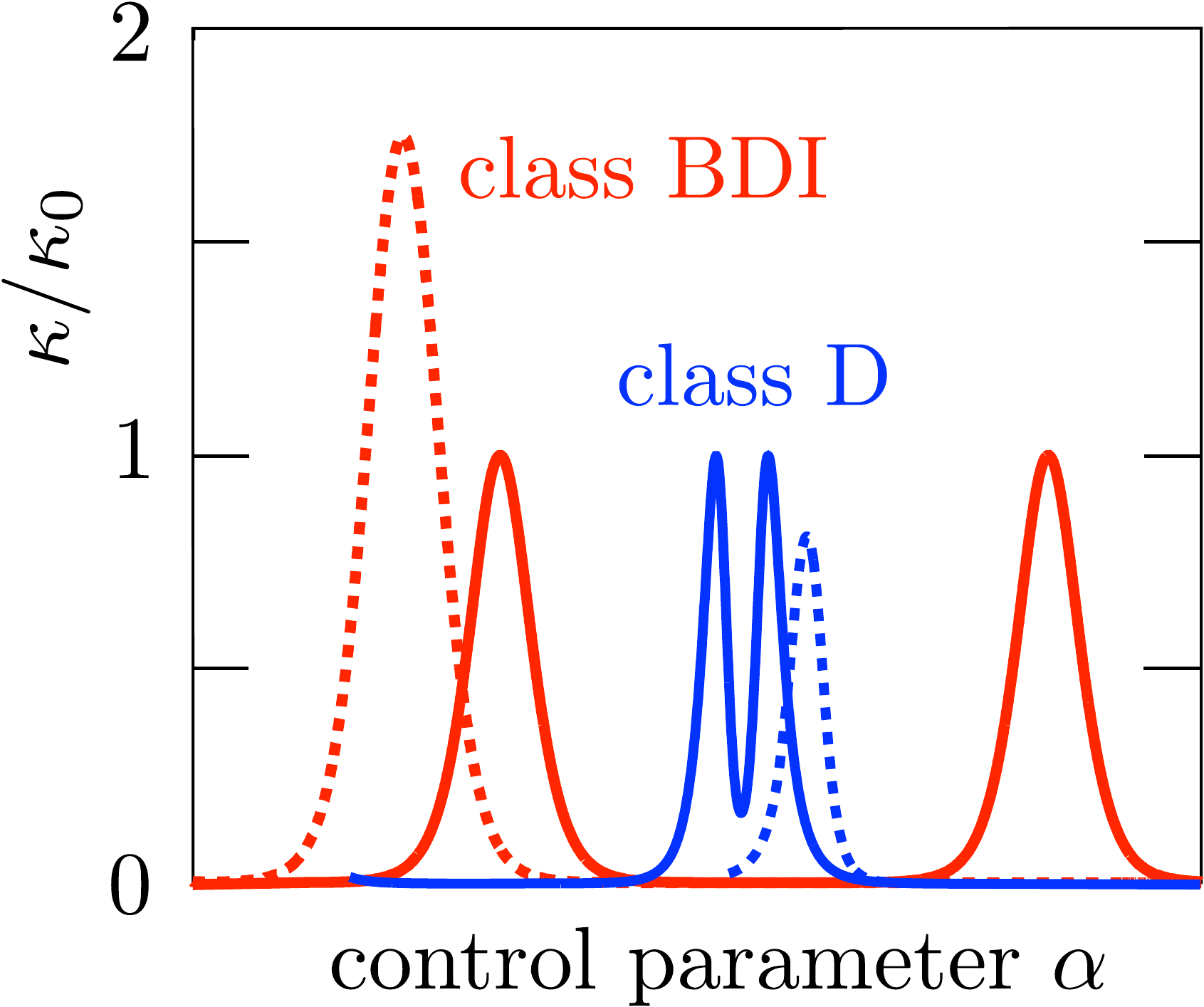}}
\caption{Thermal conductance of a $5$-mode superconducting nanowire, calculated for two disorder strengths (solid and dashed curves) without chiral symmetry (class D) and with chiral symmetry (class BDI). When two peaks merge, $\kappa$ drops below the quantized value $\kappa_0$ in class D but rises above it in class BDI. In either symmetry class the isolated conductance peaks have the same line shape \eqref{Guniversal}. Figure adapted from \textcite{Ful11}.
}
\label{fig_merger}
\end{figure}

All of this refers to well separated conductance peaks. The difference between the $\mathbb{Z}_2$ and $\mathbb{Z}$ topological phases becomes evident when conductance peaks merge: In class D the conductance peaks annihilate, while in class BDI a maximum of $N$ conductance peaks can reinforce each other, see Fig.\ \ref{fig_merger}.

\subsection{Heat transport by Majorana edge modes}
\label{Majorana_heat}

\begin{figure}[tb]
\centerline{\includegraphics[width=0.9\linewidth]{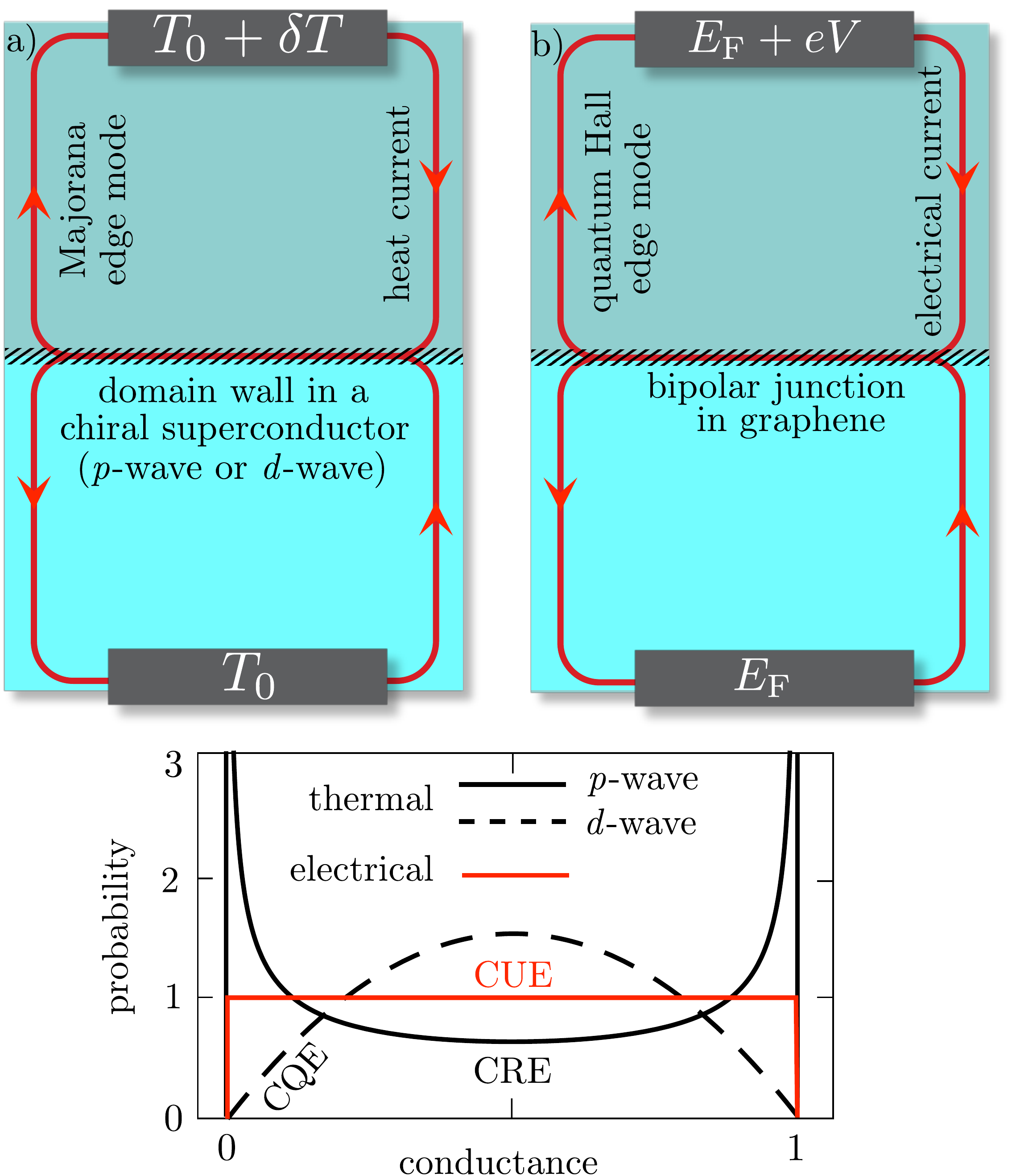}}
\caption{Thermal and electrical transport by chiral edge modes. Panel a) (from  \textcite{Ser10}) shows the chiral Majorana modes at the edge of a topological superconductor, with either spin-triplet \textit{p}-wave or spin-singlet \textit{d}-wave pairing. The shaded strip at the center represents a disordered boundary between two domains of opposite chirality of the order parameter. Panel b) shows the chiral edge modes of the quantum Hall effect in graphene, where modes of opposite chirality meet at a bipolar junction between electron-doped and hole-doped regions \cite{Aba07}. The probability distributions \eqref{Pgsingle} of the conductance in the corresponding circular ensembles are plotted at the bottom \cite{Dah10}.
}
\label{fig_chiral}
\end{figure}

Two-dimensional topological superconductors have gapless edge modes that allow for thermal transport. In the absence of time-reversal symmetry these are chiral (unidirectional) edge modes of Majorana fermions, analogous to the chiral electronic edge modes in the quantum Hall effect. One speaks of the \textit{thermal} quantum Hall effect \cite{Rea00,Vol89,Sen00,Vis01}. The quantization of the thermal conductance $\kappa$ in units of $\kappa_0=\pi^2 k_{\rm B}^{2}T_{0}/6h$ is the superconducting analogue of the quantization of the electrical conductance $G$ in units of $G_0=e^2/h$.

Chiral Majorana edge modes require particle-hole symmetry without time-reversal symmetry, so they exist in symmetry classes D and C (see Table \ref{table_AZ}). The difference between these two symmetry classes is that spin-rotation symmetry is broken in class D and preserved in class C. The corresponding symmetry of the superconducting pair potential is spin-triplet $p_x\pm ip_y$-wave pairing in class D (possibly realized in strontium ruthenate \cite{Mac03,Kal09}) and spin-singlet $d_{x^2-y^2}\pm id_{xy}$ pairing in class C (possibly realized in honeycomb lattice superconductors such as the pnictide SrPtAs \cite{Bis13,Fis14}). 

An alternative platform to realize this exotic pairing is offered by ultracold fermionic atoms \cite{Sat10}.  Quite generally, singlet and triplet pairing coexist and are mixed by Rashba spin-orbit coupling, but the edge modes remain topologically protected for sufficiently weak admixture \cite{Tan10,Sat11}.  

Edge modes of opposite chirality can meet at a domain wall \cite{Ser10}, as illustrated in Fig.\ \ref{fig_chiral}. Disorder at the boundary will mix the modes and remove the conductance quantization. Under the assumption of uniform mode mixing the probability distribution of the thermal conductance can be obtained from a circular ensemble \cite{Dah10}: the Circular Real Ensemble (CRE) of random orthogonal matrices in class D and the Circular Quaternion Ensemble (CQE) of random symplectic matrices in class C (see Table \ref{tab:chiral}).

\begin{table}
\centering
\begin{tabular}{ | l || c | c | c | c |}
\hline
Ensemble name & \textbf{CRE} & \textbf{CQE} & \textbf{CUE} \\ \hline
Symmetry class\ &  D &  C & A \\ \hline
$S$-matrix elements\ &  real &  quaternion & complex \\ \hline 
$S$-matrix space\ & orthogonal & symplectic & unitary \\ \hline
\qquad\qquad $d_T$  &  $1$ & $2$ & $1$ \\ \hline
\qquad\qquad $\alpha_T$&$-1$ &$2$&$0$\\ \hline
\qquad\qquad $\beta_T$ &$1$&$4$&$2$\\ \hline
\end{tabular}
\caption{The three ensembles that support chiral edge modes, Majorana modes in classes D and C, and electronic modes in class A.}
\label{tab:chiral}
\end{table}

For $M$ chiral Majorana modes at each edge the scattering matrix $S$ has dimension $2M\times 2M$, with $M\times M$ reflection and transmission subblocks,
\begin{equation}
S=\begin{pmatrix}
r&t\\
t'&r'
\end{pmatrix}.\label{Srtdef}
\end{equation}
Because of unitarity the transmission matrix products $tt^\dagger$ and $t't'^\dagger$ have the same set of eigenvalues $T_{1},T_{2},\ldots T_{M}$. These determine the thermal conductance
\begin{equation}
\kappa/\kappa_{0}=\sum_{n=1}^{M}T_{n}.\label{kappaTnrelation}
\end{equation}
We denote the degeneracy of the transmission eigenvalues by $d_T$, without counting spin degeneracy (since we may exclude that from the very beginning by considering only a single spin band for electron and hole).

Notice the distinction between the degeneracy factors $d_T$ for eigenvalues of $tt^\dagger$ and $d_E$ for eigenvalues of $H$, as listed in Table \ref{table_AZ}. In class D these are equal, $d_E=d_T=1$, but in class C we have $d_E=1$ while $d_T=2$ because of a twofold Kramers degeneracy of the electron-hole degree of freedom. Kramers degeneracy of the $T_n$'s appears when $tt^\dagger$ commutes with ${\cal C}$ squaring to $-1$. Because the Hamiltonian $H$ does not commute with ${\cal C}$ (it anticommutes), there is no Kramers degeneracy of the energy levels in class C. 

The probability distribution of the $M/d_T$ independent transmission eigenvalues in the CRE and CQE is given by \cite{Dah10}
\begin{equation}
P(\{T_{n}\}) \propto{} \prod_{1=i<j}^{M/d_T} \bigl|T_{i}-T_{j} \bigr|^{\beta_T}\prod_{k=1}^{M/d_T} T_{k}^{\beta_T/2-1}(1-T_{k})^{\alpha_T/2}, \label{pdfcircular}
\end{equation}
with exponents $\alpha_T,\beta_T$ listed in Table \ref{tab:chiral}. As one can see from comparison with Table \ref{table_AZ}, these exponents for transmission eigenvalue repulsion are different from the exponents $\alpha_E,\beta_E$ that govern the repulsion of energy eigenvalues in Eq.\ \eqref{PEnAZ}.\footnote{In the context of differential geometry, the parameters $\alpha_E+1\equiv m_{\ell}$ and $\beta_E\equiv m_{\rm o}$ are root multiplicities characterizing the symmetric space of Hamiltonians, while $\alpha_T+1$ and $\beta_T$ do the same for transfer matrices \cite{Bro05}.} 

Unlike the class-D distribution \eqref{PcircularD} of the Andreev reflection eigenvalues $A_n$, the distribution \eqref{pdfcircular} of the $T_n$'s does not depend on the number $\nu$ of Majorana zero-modes, it is therefore the same whether the superconductor is topologically trivial or nontrivial. Notice also that in class D the $T_n$'s are nondegenerate, while the $A_n$'s have a twofold B\'{e}ri degeneracy.

For a single ($d_T$-fold degenerate) Majorana edge mode the distribution of the dimensionless thermal conductance $g=\kappa/d_{T}\kappa_0\in[0,1]$ following from Eq.\ \eqref{pdfcircular} is
\begin{equation}
P(g)\propto\begin{cases}
g^{-1/2}(1-g)^{-1/2}&\text{in class D},\\
g(1-g)&\text{in class C},
\end{cases}\label{Pgsingle}
\end{equation}
as plotted in Fig.\ \ref{fig_chiral}. For comparison, we also show there the uniform distribution of the electrical conductance in the quantum Hall effect \cite{Bee97} (symmetry class A, Circular Unitary Ensemble).

\subsection{Thermopower and time-delay matrix}
\label{timedelaymatrix}

\begin{figure}[tb]
\centerline{\includegraphics[width=0.9\linewidth]{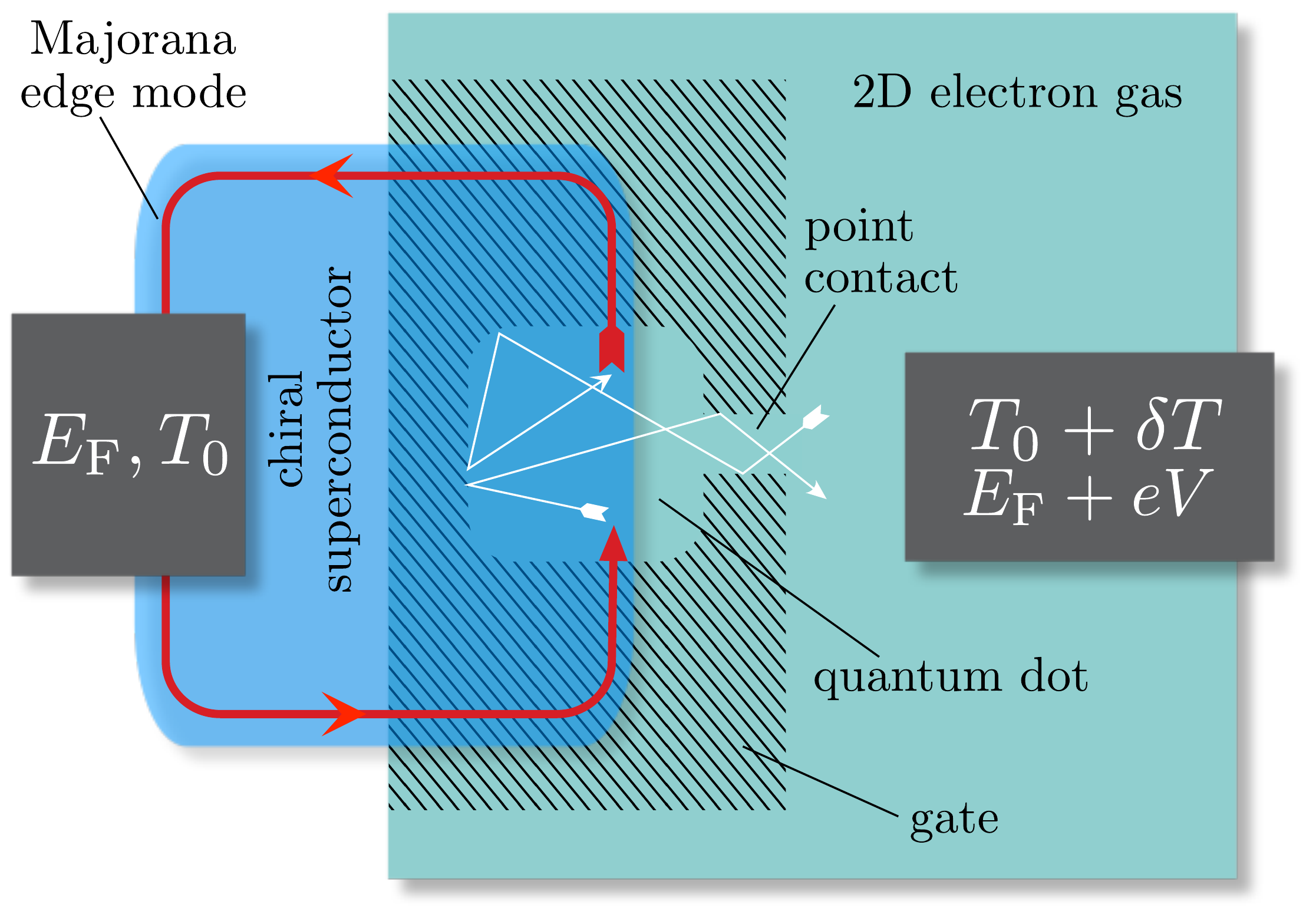}}
\caption{Geometry to measure the thermopower ${\cal S}$ of a semiconductor quantum dot (mean level spacing $\delta_0$) coupled to chiral Majorana modes at the edge of a topological superconductor. A temperature difference $\delta T$ induces a voltage difference $V=-{\cal S}\delta T$ under the condition that no electrical current flows between the contacts. For a random-matrix theory we assume that the Majorana modes are uniformly mixed with the modes in the point contact, via quasibound states in the quantum dot. Figure adapted from \textcite{Mar14}.
}
\label{fig_thermo}
\end{figure}

A temperature difference $\delta T$ may produce an electrical current in addition to a heat current. In an open circuit the electrical current is balanced by a voltage difference $V=-{\cal S}\delta T$, the Seebeck effect, with ${\cal S}$ the thermopower or Seebeck coefficient. Majorana edge modes allow for thermo-electricity --- even if they themselves carry only heat and no charge \cite{Hou13}. 

The thermopower geometry is shown in Fig.\ \ref{fig_thermo}. In a scattering formulation two matrices enter, the scattering matrix at the Fermi level $S_0\equiv S(E=0)$ and the Wigner-Smith time-delay matrix \cite{Wig55,Smi60,Fyo10}
\begin{equation}
D=-i\hbar \lim_{E\rightarrow 0}S^{\dagger}\frac{dS}{dE}.\label{Qdef} 
\end{equation}
We define transmission and reflection submatrices as in Eq.\ \eqref{Srtdef}, where the transmission matrices $t,t'$ couple the $N'$ Majorana edge modes to the $N$ electron-hole modes in the point contact, mediated by quasibound states in the quantum dot. 

In the circular ensembles the joint distribution of $S_0$ and $D$ follows from the invariance \cite{Bro97,Mar14}
\begin{equation}
P[S(E)]=P[U\cdot S(E)\cdot U']\label{PSEinvariance}
\end{equation}
of the distribution functional $P[S(E)]$ upon multiplication of the scattering matrix by a pair of energy-independent matrices $U,U'$, restricted by symmetry to a subset of the full unitary group (see Table \ref{tab:chiral}).

The time-delay matrix $D$ is a positive-definite Hermitian matrix. Its eigenvalues $D_n>0$ are the delay times, and $\gamma_n\equiv 1/D_n$ are the corresponding rates, each with the same degeneracy $d_T$ as the transmission eigenvalues. It follows from the invariance \eqref{PSEinvariance} that $D$ and $S_0$ are independent, so the distribution of the $D_n$'s can be considered separately from the distribution \eqref{pdfcircular} of the $T_n$'s. The distribution of the ${\cal N}=(N+N')/d_T$ distinct delay times is given by \cite{Mar14}
\begin{align}
P(\{\gamma_{n}\})\propto{}&\prod_{k=1}^{{\cal N}}\Theta(\gamma_k)\gamma_{k}^{\alpha_T+{\cal N}\beta_T/2}\exp\left(-\tfrac{1}{2}\beta_T t_{0}\gamma_{k}\right)\nonumber\\
&\times\prod_{1=i<j}^{{\cal N}}|\gamma_{i}-\gamma_{j}|^{\beta_T},\;\;t_0=\frac{d_{E}}{d_{T}}\frac{2\pi\hbar}{\delta_0},\label{Pgamman}
\end{align}
with coefficients from Table \ref{tab:chiral}. The unit step function $\Theta(\gamma)$ ensures that the probability vanishes if any $\gamma_{n}$ is negative.

The Cutler-Mott formula for the thermopower \cite{Cut69}
\begin{equation}
{\cal S}/{\cal S}_{0}=-\lim_{E\rightarrow 0}\frac{1}{G}\frac{dG}{dE},\;\;{\cal S}_{0}=\frac{\pi^{2}k_{\rm B}^{2}T_{0}}{3e},\label{CutlerMott}
\end{equation}
can be written in terms of the matrices $S_0$ and $D$,
\begin{equation}
{\cal S}/{\cal S}_{0}=i\hbar^{-1}\frac{{\rm Tr}\,{\cal P}\tau_{z}S_{0}(D{\cal P}-{\cal P}D)S_{0}^{\dagger}}{N-{\rm Tr}\,{\cal P}\tau_{z}S_0{\cal P}\tau_{z}S_{0}^{\dagger}}.\label{Presulteh}
\end{equation}
The Pauli matrix $\tau_z$ acts on the electron-hole degree of freedom, while ${\cal P}$ projects onto the $N$ modes at the point contact. The commutator of $D$ and ${\cal P}$ in the numerator ensures a vanishing thermopower in the absence of gapless modes in the superconductor, because then the projector ${\cal P}$ is just the identity.

\begin{figure}[tb]
\centerline{\includegraphics[width=0.8\linewidth]{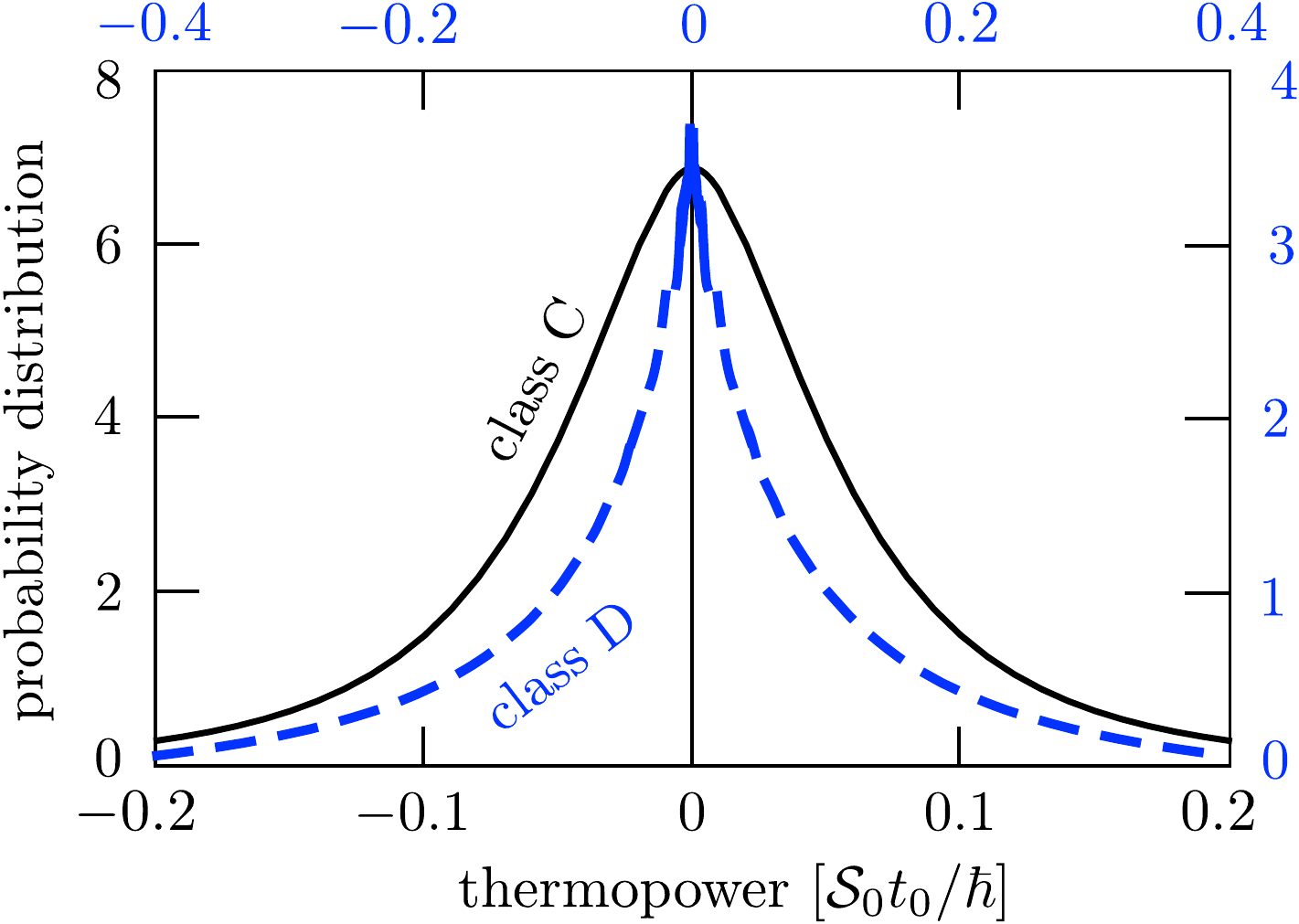}}
\caption{Probability distribution of the dimensionless thermopower in symmetry class C (solid curve, bottom and left axes), and in class D (dashed curve, top and right axes). These are results for the quantum dot of Fig.\ \ref{fig_thermo} connecting a single-channel point contact to the Majorana edge modes of a chiral superconductor. Figure adapted from \textcite{Mar14}.
}
\label{fig_classC}
\end{figure}

The resulting thermopower distributions, shown in Fig.\ \ref{fig_classC} for a single-channel point contact, are qualitatively different for Majorana edge modes in class C (\textit{d}-wave pairing) or class D (\textit{p}-wave pairing). Like the thermal conductance of Section \ref{Majorana_heat}, the thermopower does not feel the presence or absence of a Majorana zero-mode in the quantum dot --- to probe that one needs the electrical conductance (Section \ref{GprobeMzm}), or alternatively, introduce chiral symmetry.

\subsection{Andreev billiard with chiral symmetry}
\label{chiralsymm}

\begin{figure}[tb]
\centerline{\includegraphics[width=0.9\linewidth]{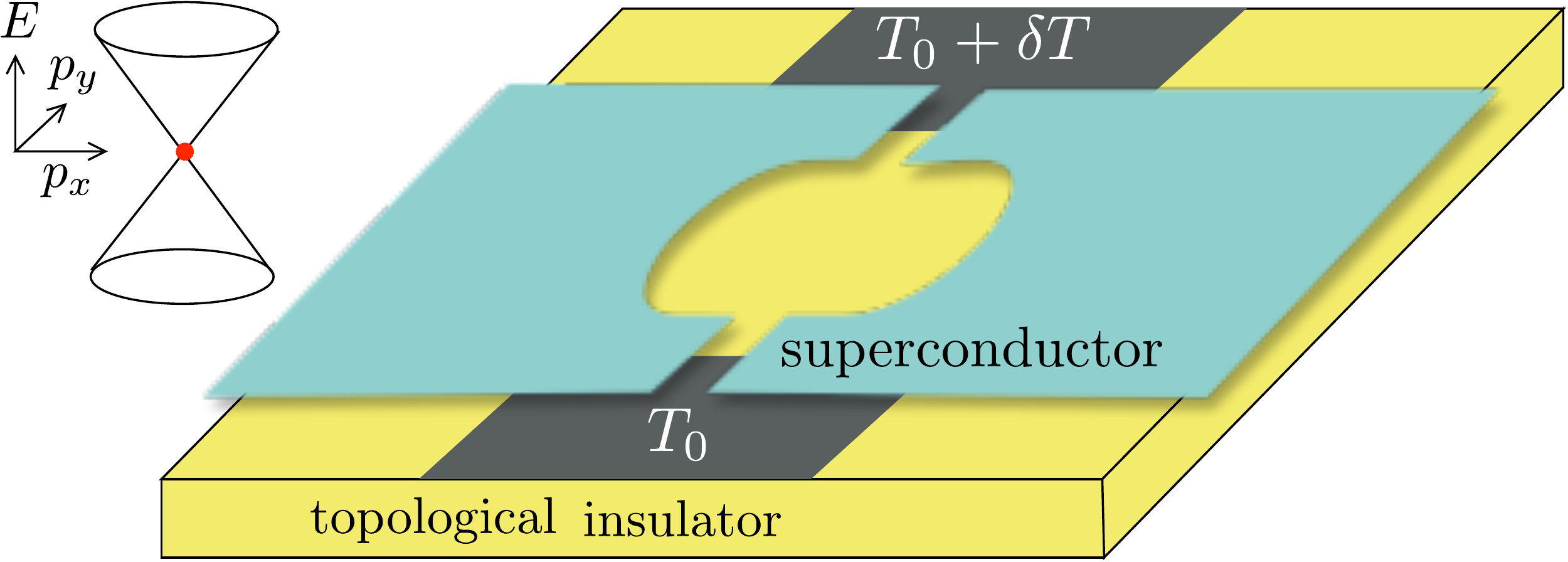}}
\caption{
Thermal conduction through an Andreev billiard on the surface of a topological insulator. Multiple Majorana zero-modes are stabilized by chiral symmetry, when the Fermi level lines up with the Dirac point joining the two cones of the band structure (marked by a dot in the left inset).}
\label{fig_billiard}
\end{figure}

Chiral symmetry (meaning the anticommutation $H{\cal CT}=-{\cal CT}H$), promotes the symmetry class from D to BDI, see Table \ref{table_AZ}, and allows for multiply degenerate Majorana zero-modes. A quantum dot (Andreev billiard) with chiral symmetry can be realized on the surface of a topological insulator, see Fig.\ \ref{fig_billiard}, because the Dirac Hamiltonian
\begin{equation}
H_0=v(p_x-eA_x)\sigma_x+v(p_y-eA_y)\sigma_y+V \label{DiracH0}
\end{equation}
anticommutes with $\sigma_z$ when the electrostatic potential $V\rightarrow 0$ is tuned to the Dirac point. We rely on random scattering by disorder to produce a finite density of states at $E=0$, but in order to preserve the chiral symmetry the disorder cannot be electrostatic ($V$ must remain zero). Scattering by a random vector potential is one possibility \cite{Lud94,Mot02}, or alternatively scattering by random surface deformations \cite{Lee09,Dah10b,Par11}.

The coupling to a superconductor at Fermi energy $E_{\rm F}\rightarrow 0$ introduces particle-hole symmetry without breaking the chiral symmetry\footnote{
Eq.\ \eqref{Hsigmazchiral} at $E_{\rm F}=0$ amounts to the chiral symmetry $H{\cal CT}=-{\cal CT}H$ with particle-hole symmetry ${\cal C}=\tau_x{\cal K}$ and fake time-reversal symmetry ${\cal T}=(\sigma_z\otimes\tau_x){\cal K}$. In the nanowire geometry of Sec.\ \ref{RZH} the chiral symmetry takes the different form \eqref{Htaux}, with ${\cal T}={\cal K}$. The symmetry class is BDI in both realizations, determined by ${\cal C}^2={\cal T}^2=+1$. In each case, the chiral symmetry of $H$ is inherited by $S(E)={\cal CT}S^\dagger(-E)({\cal CT})^{-1}$, provided that the coupling matrix $W$ in Eq.\ \eqref{Cayley} commutes with ${\cal CT}$.}
of the BdG Hamiltonian \eqref{HBdG},
\begin{equation}
H(E_{\rm F})\sigma_z=-\sigma_z H(-E_{\rm F}).\label{Hsigmazchiral}
\end{equation}
As a consequence, overlapping Majorana zero-modes in a superconductor/topological insulator heterostructure \cite{Fu08} will not split when the Fermi energy $E_{\rm F}\rightarrow 0$ lines up with the Dirac point \cite{Che10,Teo10}.

The zero-modes are broadened by coupling to the continuum through a scattering matrix $S$. At the Fermi level $E=E_{\rm F}$ this is a real orthogonal matrix in the Majorana basis. When $E_{\rm F}=0$ the chiral symmetry relation \eqref{chiralsymmetry}, which here takes the form
\begin{equation}
S(E)=\sigma_z S^\dagger(-E)\sigma_z,\label{sigmazchiralsym}
\end{equation}
implies that $S_0\equiv \sigma_z S(0)$ is orthogonal and symmetric. In the two-terminal geometry of Fig.\ \ref{fig_billiard}, with $2M$ scattering channels at each contact,\footnote{To preserve the chiral symmetry at each contact, the reflection and transmission matrices in Eq.\ \eqref{S0decomposition} should be even-dimensional.} the matrix $S_0$ has the decomposition \eqref{SOcalS}:
\begin{equation}
S_0=\begin{pmatrix}
\sigma_z r&\sigma_z t\\
t^{\rm T}\sigma_z & \sigma_z r'
\end{pmatrix}=O\,{\rm diag}(\pm 1,\pm 1,\ldots\pm 1)O^{\rm T}.
\label{S0decomposition}
\end{equation}
For chaotic scattering the statistics of $S_0$ is obtained by drawing $O$ uniformly from the orthogonal group (ensemble ${\rm T}_+{\rm CRE}$ of Table \ref{tab:table2}). The trace ${\rm Tr}\,S_0=2Q$ is fixed by the number $\nu=|Q|$ of zero-modes that are coupled to the continuum \cite{Ful11}.

The transmission eigenvalues at the Fermi level determine the thermal conductance
\begin{equation}
\kappa=\kappa_0\sum_{n=1}^{2M}T_n=\kappa_0\sum_{n=1}^{2M}(1-r_n^2),\;\; \kappa_0=\tfrac{1}{6}\pi^2 k_{\rm B}^{2}T_{0}/h.\label{kappa2M}
\end{equation}
The $T_n$'s are the eigenvalues of the transmission matrix product $tt^\dagger=1-rr^\dagger$, and since $rr^\dagger$ and $\sigma_z rr^\dagger\sigma_z$ have the same eigenvalues, we may equivalently write $T_n=1-r_n^2$, with $r_n\in[-1,1]$ a real eigenvalue of the Hermitian matrix $\sigma_z r$.

The probability distribution of the $r_n$'s in the ${\rm T}_+{\rm CRE}$ is given by \cite{Mac02}
\begin{align}
P(\{r_{n}\})\propto{}&\prod_{l=2M-\nu+1}^{2M}\delta(r_l-{\rm sign}\, Q)\prod_{1=i<j}^{2M-\nu} \bigl|r_{i}-r_{j} \bigr|^{\beta_T}\nonumber\\
&\times \prod_{k=1}^{2M-\nu}(1-r_k^2)^{\nu\beta_T/2+\beta_T/2-1},\label{pdfcirculardeltaN0}
\end{align}
with symmetry index $\beta_T=1$ in class BDI. The distribution \eqref{pdfcirculardeltaN0} holds also in the other two chiral ensembles from Table \ref{table_AZ} with a $\mathbb{Z}$ topological quantum number: class AIII ($\beta_T=2$) and class CII ($\beta_T=4$, each eigenvalue twofold degenerate).

\begin{figure}[tb]
\centerline{\includegraphics[width=0.8\linewidth]{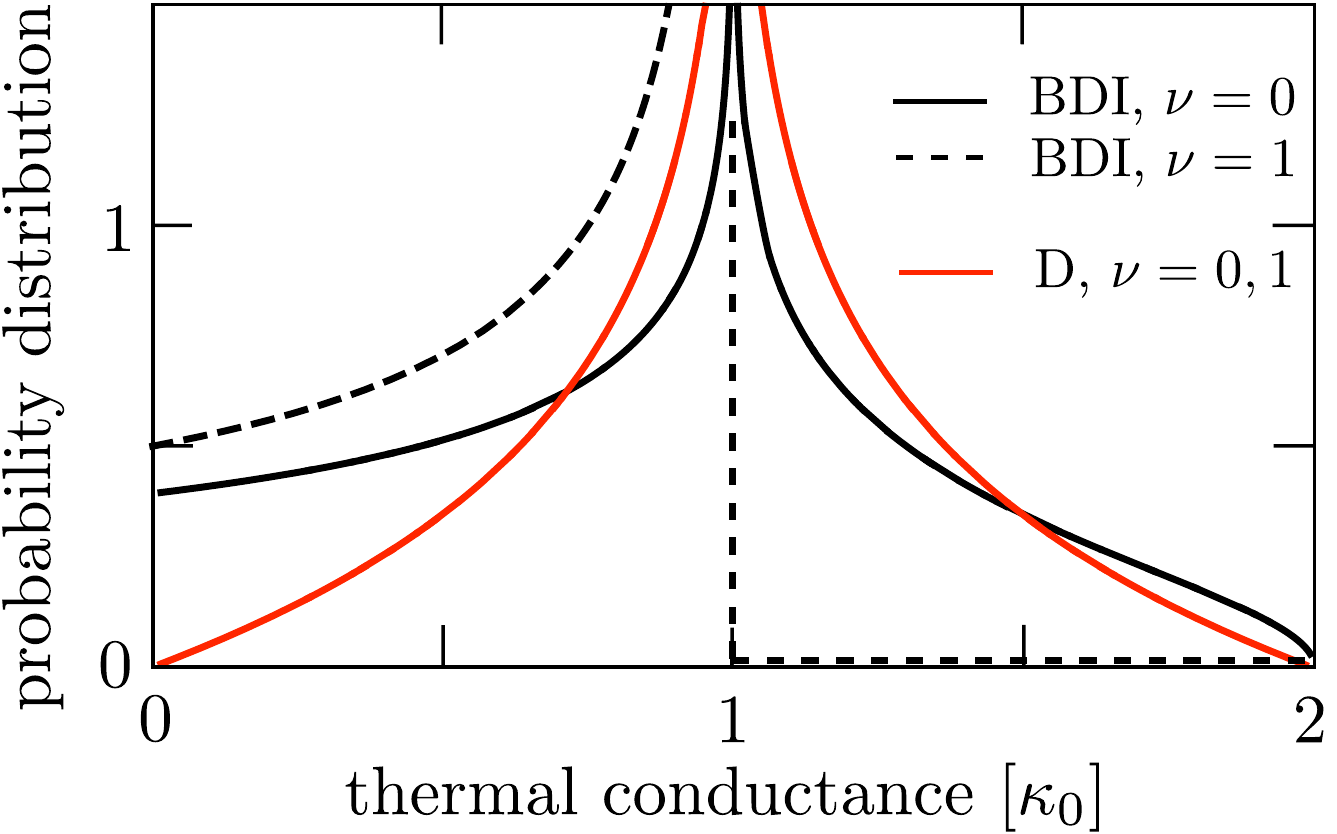}}
\caption{
Probability distribution of the dimensionless thermal conductance $\kappa/\kappa_0=T_1+T_2$ in an Andreev billiard, calculated from Eqs.\ \eqref{PchiralBDI} and \eqref{PnonchiralD} in symmetry class D (only particle-hole symmetry) and class BDI (particle-hole with chiral symmetry). In class D there is no dependence on the number $\nu$ of Majorana zero-modes, while in class BDI there is.
}
\label{fig_distributions}
\end{figure}

One pair of electron-hole channels in each contact produces two transmission probabilities $T_1$, $T_2$. We compare the $\nu$-dependent class-BDI distribution from Eq.\ \eqref{pdfcirculardeltaN0},
\begin{equation}
P_{\rm BDI}(T_1,T_2)\propto\begin{cases}
(T_1 T_2)^{-1/2}(1-T_1)^{-1/2}(1-T_2)^{-1/2}&\\
\qquad\times\sqrt{1-\min(T_1,T_2)},\;\;\text{for}\;\;\nu=0,&\\
(1-T_1)^{-1/2}\delta(T_2),\;\;\text{for}\;\;\nu=1,&\\
\delta(T_1)\delta(T_2),\;\;\text{for}\;\;\nu=2,&
\end{cases}\label{PchiralBDI}
\end{equation}
with the $\nu$-independent class-D distribution from Eq.\ \eqref{pdfcircular},
\begin{align}
P_{\rm D}(T_1,T_2) \propto{}&(T_1 T_2)^{-1/2}(1-T_1)^{-1/2}(1-T_2)^{-1/2}\nonumber\\
&\qquad\times\bigl|T_{1}-T_{2} \bigr|. \label{PnonchiralD}
\end{align}
The powers of $T_n$ and $1-T_n$ in class D are the same as in class BDI with $\nu=0$, but the eigenvalue repulsion in class D does not carry over to class BDI --- where $P(T_1,T_2)$ remains finite when $T_1\rightarrow T_2$. The corresponding distributions of the thermal conductance are plotted in Fig.\ \ref{fig_distributions}.

It is noteworthy that the sensitivity to the presence or absence of a Majorana zero-mode in class BDI appears already for $\nu=1$, so for the single nondegenerate zero-mode that is also stable in class D. A qualitatively similar sensitivity applies to the thermopower and time-delay matrix of Sec.\ \ref{timedelaymatrix}, see  \textcite{Sch14}. 

\section{Josephson junctions}
\label{Jeffect}

The electrical and thermal conductance discussed in the previous two sections probe the system out of equilibrium, in response to a voltage or temperature difference. A superconductor can also support a persistent electrical current $I(\phi)$ in equilibrium, in response to a phase difference $\phi$ of the pair potential (Josephson effect).

A nanowire is a Josephson junction if it connects two superconducting electrodes, as shown in Fig.\ \ref{fig_quantumdot}. The segment of the wire between the two superconductors forms a quantum dot, a confined region with quasiparticle excitation spectrum $0<E_0<E_1<E_2<\cdots$. The spectrum depends on the phase difference $\phi$ of the pair potential across the junction, which can be controlled by the magnetic flux $\Phi$. Because a $2\pi$ increment of $\phi$ corresponds to a variation of $\Phi$ by $h/2e$, the excitation spectrum has the flux periodicity $E_n(\Phi)=E_n(\Phi+h/2e)$.

The presence of Majorana zero-modes in the quantum dot induces a period-doubling of the flux dependence, with a free energy $F(\Phi)$ that has $h/e$ rather than $h/2e$ flux-periodicity, or equivalently, a $4\pi$ rather than $2\pi$ phase-periodicity \cite{Kit01}. The mechanism behind the period doubling is a switch in the fermion parity of the superconducting condensate, as we now discuss.

\subsection{Fermion parity switches}
\label{parityswitch}

\begin{figure}[tb]
\centerline{\includegraphics[width=1\linewidth]{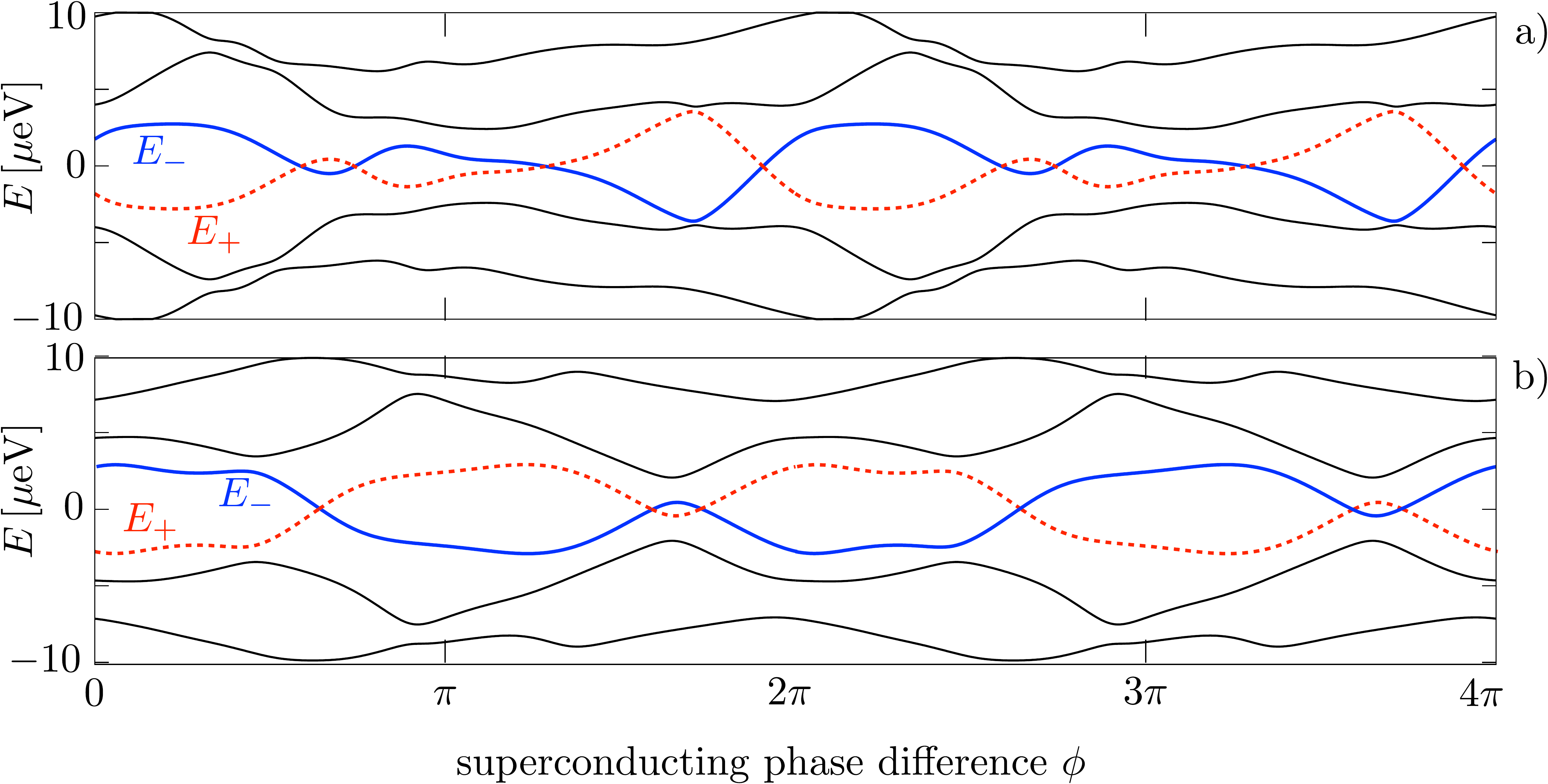}}
\caption{Excitation spectrum of an InSb Josephson junction, calculated using the Rashba-Zeeman Hamiltonian \eqref{H0RashbaZeeman} in a weak parallel magnetic field (panel a, Zeeman energy $V_{\rm Z}=0.35\,{\rm meV}$) and in a stronger field (panel b, $V_{\rm Z}=0.52\,{\rm meV}$). The other parameters are the same in both panels ($\Delta_0=0.4\,{\rm meV}$, $E_{\rm F}=2.4\,{\rm meV}$, $N=19$). The dashed and solid levels $E_{\pm}$ that cross at $E=0$ correspond to different parity ${\cal P}=\pm 1$ of the number of electrons in the junction. The phase periodicity of the spectrum at fixed parity is $2\pi$ in panel a (topologically trivial junction) and $4\pi$ in panel b (topologically nontrivial). Data provided by M. Wimmer.
}
\label{fig_singlecross}
\end{figure}

We mentioned in Section \ref{platforms} that a quantum dot in symmetry class D (no time-reversal or spin-rotation symmetry) supports level crossings at the Fermi energy when we vary the phase difference $\phi$ of the superconducting electrodes. As indicated in Fig.\ \ref{fig_singlecross}, we denote by $E_\pm(\phi)$ the smooth (adiabatic) $\phi$-dependence of the pairs of crossing levels, arranged such that $E_+<0<E_-=E_0$ at $\phi=0$.

Level crossings are a signal of a conserved quantity, preventing transitions between the levels that would convert the crossing into an avoided crossing. In this case the conserved quantity is the parity ${\cal P}=\pm 1$ of the number of electrons in the quantum dot. At low temperatures and for small charging energy\footnote{Existing experiments \cite{Cha12,Lee14} have mainly explored the opposite regime $e^{2}/C>\Delta_0$ of large charging energy.} ($k_{\rm B}T$ and $e^{2}/C\ll \Delta_0$), the quantum dot can only exchange {\em pairs\/} of electrons with the superconducting electrodes, so ${\cal P}$ is conserved. A transition between $E_+$ and $E_-$ would add or remove an unpaired quasiparticle, which is forbidden and hence the crossing is protected.

The free energy $F_{\cal P}$ of the quantum dot depends on whether the number of electrons is even (${\cal P}=+1$) or odd (${\cal P}=-1$) \cite{Ave92,Tuo92}. In the zero-temperature limit, this dependence can be written as \cite{Law11}
\begin{equation}
F_{\cal P}=\tfrac{1}{2}\bigl(E_{\cal P}-E_1-E_2-\cdots\bigr).\label{FPdef}
\end{equation}
(The factor $1/2$ ensures that the switch from $E_+$ to $E_-$ properly introduces an excitation energy $E_0$ rather than $2E_0$.) At $\phi=0$ the level $E_+$ lies below $E_-$, so the superconducting condensate favors fully paired electrons and is said to be of even fermion parity. At the first crossing $E_-$ drops below $E_+$, so now a single unpaired electron is favored (odd fermion parity). The fermion-parity switch signaled by a level crossing is a topological phase transition of the superconducting condensate \cite{Kit01}.

\begin{figure}[tb]
\centerline{\includegraphics[width=0.9\linewidth]{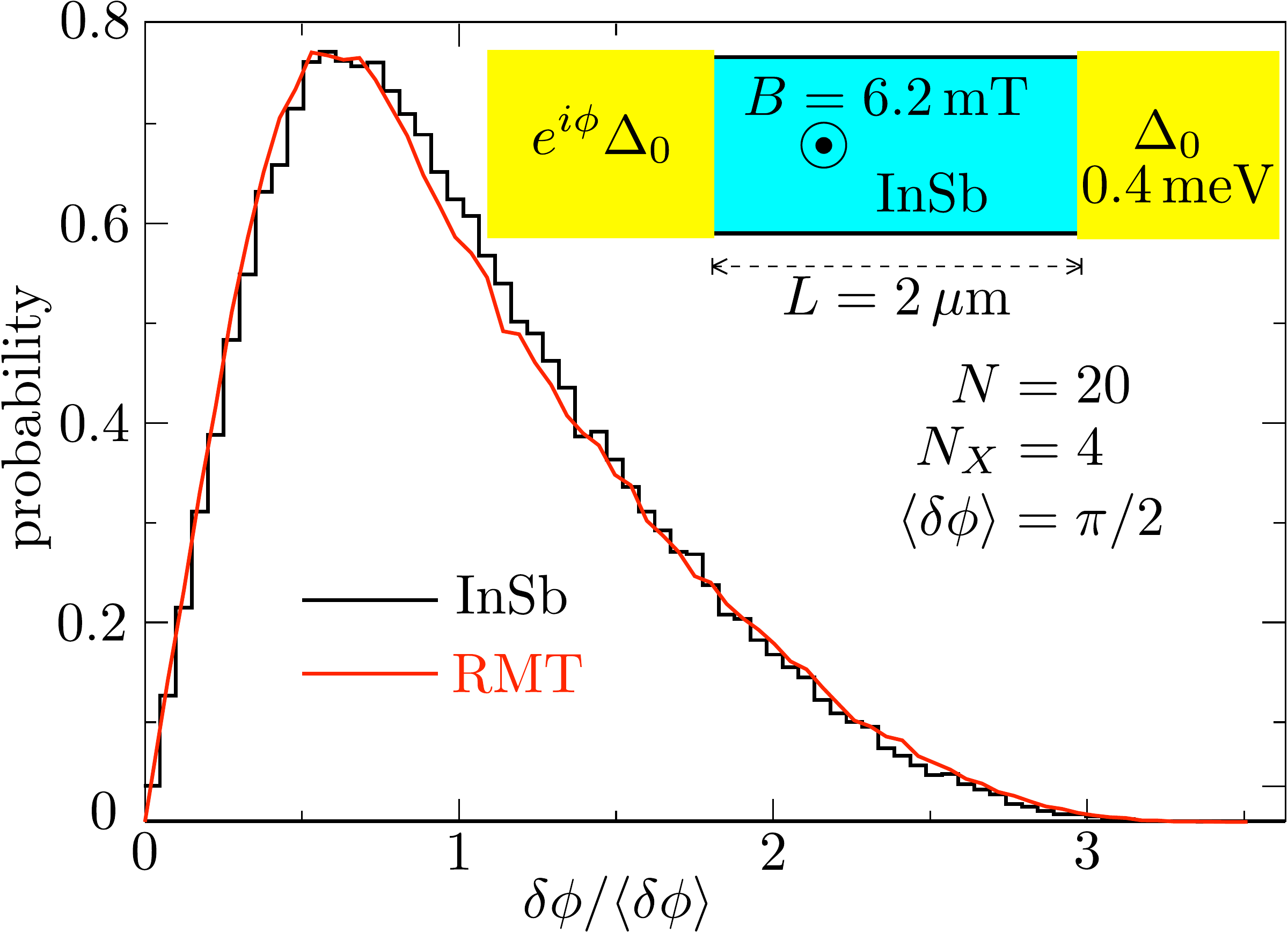}}
\caption{Spacing distribution of level crossings in a Josephson junction. The smooth curve shows the prediction from RMT. The histogram results from a microscopic calculation using the Rashba-Zeeman Hamiltonian \eqref{H0RashbaZeeman}, for a Josephson junction formed out of a disordered InSb wire in a weak perpendicular magnetic field, sampled over different impurity configurations. The excitation spectrum for one impurity configuration is shown in Fig.\ \ref{fig_repulsion}. This system is topologically trivial ($N_X=4$ crossings in a $2\pi$ phase interval). Figure adapted from \textcite{Bee13b}.
}
\label{fig_cross}
\end{figure}

Sequences of fermion-parity switches are not independent. As illustrated in Fig.\ \ref{fig_cross} for an InSb Josephson junction, the level crossings show an anti-bunching effect, with a spacing distribution that vanishes linearly at small spacings \cite{Bee13b}. So while in symmetry class D there is no energy level repulsion at $E=0$, there is a ``level crossing repulsion''.

This effect can be connected to a classic problem in non-Hermitian random-matrix theory \cite{Ede97,Kho09}: {\em How many eigenvalues of a real random matrix are real?} The connection is made by identifying the phase $\phi_n$ of a level crossing with the real eigenvalue $\varepsilon_n=\tan(\phi_n/2)$ of the matrix
\begin{equation}
{\cal M}=(1-O)(1+O)^{-1}J,\label{MOJ}
\end{equation}
with $J$ defined in Eq.\ \eqref{SOJ} and $O\in {\rm SO}(2N)$. The $2N\times 2N$ orthogonal matrix $O=r_{\rm L}r_{\rm R}$ is the product of the reflection matrices from the left and right ends of the Josephson junction, each end supporting $N$ electron modes and $N$ hole modes. The matrix $O$ is real because both $r_{\rm L}$ and $r_{\rm R}$ are taken in the Majorana basis \eqref{RMajoranabasis}, and ${\rm Det}\,O=+1$ because ${\rm Det}\,r_{\rm L}={\rm Det}\,r_{\rm R}=\pm 1$. 

Fig.\ \ref{fig_cross} shows good agreement between the spacing distribution from a random-matrix ensemble for $O$ and from a computer simulation of the InSb Josephson junction. Once the spacing is normalized by the average spacing, there are no adjustable parameters in this comparison. The linearly vanishing spacing distribution is reminiscent of the Wigner distribution in the Gaussian orthogonal ensemble \cite{Wig67}, but for larger spacings the distribution has approximately the Poisson form of uncorrelated eigenvalues  \cite{Bee13b,For13}. This difference can be understood as a ``screening'' by intervening complex eigenvalues of ${\cal M}$ of the eigenvalue repulsion on the real axis.

\subsection{$\bm{4\pi}$-periodic Josephson effect}
\label{4pi}

One can distinguish topologically trivial from nontrivial Josephson junctions by counting the number $N_{X}$ of level crossings when $\phi$ is incremented by $2\pi$. For the spectrum shown in Fig.\ \ref{fig_singlecross}a the number $N_{X}=4$ is even, this is the topologically trivial case. Alternatively, if $N_X$ odd the Josephson junction is topologically nontrivial, see Fig.\ \ref{fig_singlecross}b.

The free energy \eqref{FPdef} is $2\pi$-periodic when $N_X$ is even and $4\pi$-periodic when $N_X$ is odd. This period doubling can be observed via the supercurrent flowing through the ring,
\begin{equation}
I_{\pm}(\phi)=\frac{2e}{h}\frac{dF_{\pm}}{d\phi}=\frac{e}{h}\frac{dE_{\pm}}{d\phi}+\mbox{$2\pi$-periodic terms}.\label{Ipmdef}
\end{equation}
The dependence on the enclosed flux $\Phi=\phi\hbar/e$ has $h/2e$ periodicity when $N_X$ is even, but a doubled $h/e$ periodicity when $N_X$ is odd. This is the $4\pi$-periodic Josephson effect \cite{Kit01,Kwo04}.

Because $I_{+}(\phi)=I_{-}(\phi+2\pi)$, when $N_X$ is odd, the $2\pi$ periodicity is restored when a quasiparticle can enter or leave the junction during the measurement time, which severely complicates the observation of the effect \cite{Rok12}.

\subsection{Discrete vortices}
\label{discrete}

The class-D subgap states in a superconducting vortex core discussed in Section \ref{spectralpeak} can be realized in a quantum dot, if Josephson junctions are configured to produce a $2\pi$-winding of the superconducting order parameter. This is a \textit{discrete} vortex, with stepwise increments of the phase at each Josephson junction. The class-D, $\nu=1$ vortex needs a topological insulator to allow for an unpaired Majorana zero-mode \cite{Fu08}, while for the class-D, $\nu=0$ vortex a conventional semiconductor quantum dot suffices. 

\begin{figure}[tb]
\centerline{\includegraphics[width=0.8\linewidth]{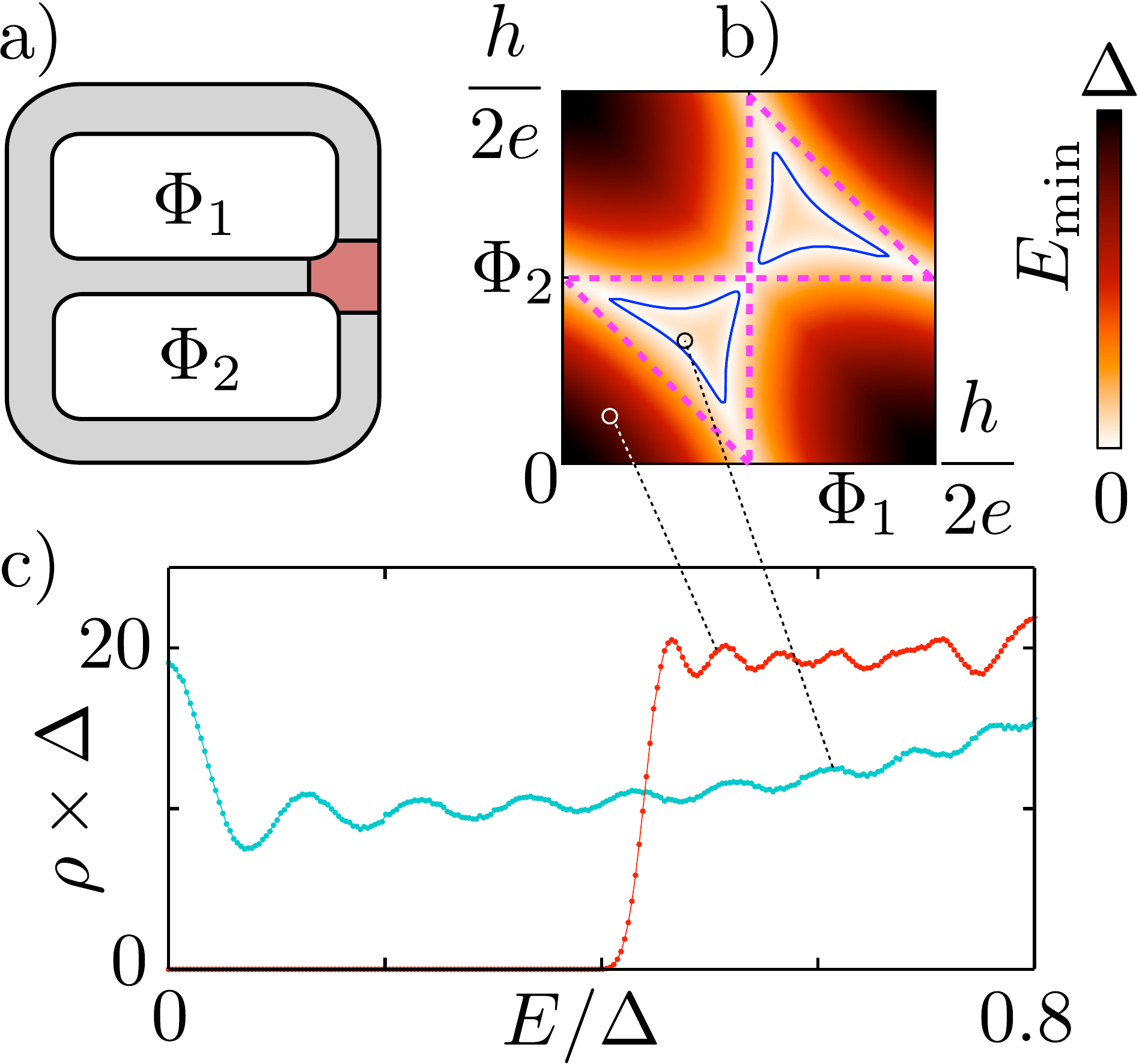}}
\caption{Discrete vortex in a three-terminal quantum-dot Josephson junction. Panel a) shows the geometry. The fluxes $\Phi_1$ and $\Phi_2$ produce a $2\pi$-winding of the order parameter within the dashed triangles of panel b). The data in panels b) and c) is obtained numerically from the symplectic ensemble of chaotic quantum dots ($N=10$ modes per lead, level spacing $\gg\Delta_0/N$). Panel b) shows the energy $E_{\rm min}$ of the lowest subgap state for a single sample, with level crossings appearing on the blue contour. The ensemble averaged density of states is plotted in panel c) for two arrangements of fluxes. The gap closes with a midgap peak when there is a $2\pi$-winding. Figure adapted from \textcite{Hec14}.
}
\label{fig_discrete}
\end{figure}

In Fig.\ \ref{fig_discrete} we illustrate the latter system, studied in  \textcite{Hec14}. A minimum of two independent fluxes $\Phi_1,\Phi_2$ is needed to produce a discrete vortex, so the minimal configuration consists of a quantum dot connected to three superconducting leads (panel a). Level crossings at $E=0$ appear along closed contours in the $(\Phi_1,\Phi_2)$ plane, encircling domains of \textit{odd} fermion parity (panel b). The density of states of the discrete vortex, averaged over the symplectic ensemble of chaotic quantum dots, has the characteristic midgap spectral peak of the class-D, $\nu=0$ vortex core (compare Figs.\ \ref{fig_spectralpeak}c and \ref{fig_discrete}c). The peak requires the $2\pi$-winding of the order parameter, without it the density of states is gapped.

\section{Conclusion}
\label{conclude}

C. N. Yang famously spoke of the three ``thematic melodies of twentieth century theoretical physics'', \textit{quantization, symmetry, and phase factor}, each of which ``evolved from primordial concepts in the cognitive history of mankind'' \cite{Yan03}. For the twenty-first century one might add \textit{topology} as a fourth melody, and one could say the same about its primordial origin. The prediction of states of matter that are distinguished by topology rather than by symmetry is a triumph of theoretical physics, and a reversal of the usual course of events where theory follows experiment.

The theory of random matrices is very much a twentieth century production, but turns out to be well suited as a framework for the classification of topologically distinct single-particle Hamiltonians. We have reviewed how the tools of random-matrix theory can be adapted to account for topology in addition to symmetry, focusing on superconductors --- where the topological quantum number counts the number of Majorana modes.

The ten-fold way classification exhausts the possibilities for topologically distinct states of non-interacting fermions, providing for two symmetry classes with a $\mathbb{Z}_2$ and three with a $\mathbb{Z}$ topological quantum number. Much recent work aims at a similarly complete classification for interacting electrons \cite{Che11,Tur11,Fid11,Lu12,Man12,Che13,Mei13,Neu14}. It remains to be seen whether the methods of RMT can be of use in that context.

The nontopological Wigner-Dyson ensembles were tested extensively in GaAs quantum dots. The topological Altland-Zirnbauer ensembles still lack an experimental platform of similar versatility, but there is a great variety of candidate systems. We are particularly excited by the recent progress in quantum spin-Hall insulators (HgTe or InAs quantum wells) with superconducting electrodes \cite{Kne12,Har14,Pri14}. The quantum dot geometry of Fig.\ \ref{fig_DIIIbilliard} seems within reach, and would provide an ideal testing ground for RMT.

\acknowledgments

My own research on this topic was carried out in collaboration with A. R. Akhmerov, B. B\'{e}ri, P. W. Brouwer, E. Cobanera, J. P. Dahlhaus, M. Diez, J. M. Edge, F. Hassler, M. Marciani, S. Mi, D. I. Pikulin, I. Serban, H. Schomerus, and M. Wimmer. It was supported by the Foundation for Fundamental Research on Matter (FOM), the Netherlands Organization for Scientific Research (NWO/OCW), and an ERC Synergy Grant.

\appendix

\section{B\'{e}ri degeneracy}
\label{AppBeri}

The twofold degeneracy of the Andreev reflection eigenvalues $A_{n}\neq 0,1$ in class D and BDI is a consequence of an anti-unitary symmetry ${\cal C}$ that squares to $+1$. This distinguishes it from the more familiar Kramers degeneracy, resulting from an anti-unitary symmetry ${\cal T}$ that squares to $-1$. We give a self-contained derivation of this unexpected degeneracy discovered by \textcite{Ber09b}.\footnote{
The proof of B\'{e}ri degeneracy published in \textcite{Wim11} is incomplete (the case of a noninvertible $r_{he}$ is not properly accounted for). The complete proof given here follows arXiv:1101.5795.
}

We recall from Sec.\ \ref{countingMzm} that particle-hole symmetry in class D and BDI implies that the reflection matrix $r$ at the Fermi level has the block decomposition
\begin{equation}
r=\begin{pmatrix}
r_{ee}&r_{eh}\\
r_{he}&r_{hh}
\end{pmatrix},\;\;r_{hh}^{\vphantom{\ast}}=r_{ee}^{\ast},\;\;r_{eh}^{\vphantom{\ast}}=r_{he}^{\ast}.\label{ehsymmetry}
\end{equation}
The Andreev reflection eigenvalues $A_n\in[0,1]$ are the eigenvalues of the matrix product $r_{he}^{\vphantom{dagger}}r_{he}^\dagger$.  (Equivalently, $\sqrt{A_n}$ is the singular value of $r_{he}$.) 

Let us first assume that all $A_{n}$'s are nonzero, so that the matrix $r_{he}$ is invertible. Unitarity $r^{\dagger}r=\openone$ requires that $r_{eh}^{\dagger}r_{ee}^{\vphantom{\dagger}}+r_{hh}^{\dagger}r_{he}^{\vphantom{\dagger}}=0$, hence
\begin{equation}
{\cal A}\equiv r_{ee}^{\vphantom{-1}}r_{he}^{-1} =-{\cal A}^{\rm T}\label{AT}
\end{equation}
is an antisymmetric matrix. We focus on the product
\begin{equation}
{\cal A}^{\dagger}{\cal A}=(r_{he}^{\vphantom{\dagger}}r_{he}^{\dagger})^{-1}-1,\label{AA}
\end{equation}
with eigenvalues $a_{n}=1/A_{n}-1\geq 0$. Let $\Psi$ be an eigenvector of ${\cal A}^{\dagger}{\cal A}$, so ${\cal A}^{\dagger}{\cal A}\Psi=a\Psi$. Then $\Psi'=({\cal A}\Psi)^{\ast}$ satisfies
\[
{\cal A}^{\dagger}{\cal A}\Psi'=-{\cal A}^{\ast}{\cal A}{\cal A}^{\ast}\Psi^{\ast}={\cal A}^{\ast}({\cal A}^{\dagger}{\cal A}\Psi)^{\ast}=(a{\cal A}\Psi)^{\ast}=a\Psi'.
\]
The eigenvalue $a$ is therefore twofold degenerate if $\Psi'$ and $\Psi$ are linearly independent.

Suppose they are not independent, meaning that $\Psi'=\lambda\Psi$ for some $\lambda$, then
\[
a\Psi={\cal A}^{\dagger}{\cal A}\Psi=-{\cal A}^{\ast}(\lambda\Psi)^{\ast}=-|\lambda|^{2}\Psi,
\] 
hence $a=0$. So any eigenvalue $1/A_{n}-1\neq 0$ of ${\cal A}^{\dagger}{\cal A}$ is twofold degenerate. Since we assumed from the beginning that $A_n\neq 0$, this implies that the Andreev reflection eigenvalues $A_{n}\neq 0,1$ are twofold degenerate.

To extend the proof to the case that $r_{he}$ is not invertible, we regularize the inverse and consider the matrix
\begin{equation}
{\cal A}_{\epsilon}=X^{\rm T}r_{he}^{\rm T}r_{ee}^{\vphantom{\rm T}}X,\;\;X=r_{he}^{\dagger}(r_{he}^{\vphantom{\dagger}}r_{he}^{\dagger}+\epsilon)^{-1},\label{Aregularized}
\end{equation}
with $\epsilon$ a positive infinitesimal. Since ${\cal A}_{\epsilon}$ remains antisymmetric, we can follow the same steps to conclude that the nonzero eigenvalues of ${\cal A}_{\epsilon}^{\dagger}{\cal A}_{\epsilon}^{\vphantom{\dagger}}$ are twofold degenerate. Evaluation of this matrix product using the identity
\begin{equation}
r_{ee}^{\dagger}(r_{he}^{\vphantom{\dagger}}r_{he}^{\dagger})^{\rm T}=(r_{he}^{\dagger}r_{he}^{\vphantom{\dagger}})r_{ee}^{\dagger}\label{reerhe}
\end{equation}
gives the expression
\begin{equation}
{\cal A}_{\epsilon}^{\dagger}{\cal A}_{\epsilon}^{\vphantom{\dagger}}=(1-r_{he}^{\vphantom{\dagger}}r_{he}^{\dagger})(r_{he}^{\vphantom{\dagger}}r_{he}^{\dagger})^{3}(r_{he}^{\vphantom{\dagger}}r_{he}^{\dagger}+\epsilon)^{-4},\end{equation}
which has eigenvalues 
\begin{equation}
a_{\epsilon,n}=(1-A_{n})A_{n}^{3}(A_{n}+\epsilon)^{-4}.\label{aepsilondef}'
\end{equation} 
These are either zero or twofold degenerate, hence we conclude that $A_{n}\neq 0,1$ is twofold degenerate.

In the absence of time-reversal and spin-rotation symmetry, only the B\'{e}ri degeneracy of the Andreev reflection eigenvalues applies. If one or both of these symmetries are present, then all $A_{n}$'s are twofold degenerate --- including those equal to $0$ or $1$. Kramers degeneracy then comes in the place of B\'{e}ri degeneracy, it is not an additional degeneracy \cite{Bee11}. This is probably the reason that B\'{e}ri degeneracy was not noticed during half a century of studies of Andreev reflection --- it was hidden by Kramers degeneracy.

Notice that B\'{e}ri degeneracy immediately explains the stability of an unpaired unit Andreev reflection eigenvalue: It cannot be displaced from unity by disorder without a partner. This is in essence why a Majorana resonance in the conductance remains stable no matter how strongly the zero-mode is coupled to the continuum \cite{Wim11}.

\section{Shot noise of a Majorana mode}
\label{majoranashotapp}

A fully transmitted Majorana edge mode combines zero electrical conductance with a quantized shot noise power of $e^2/2h$ per electron volt. This result was derived by \textcite{Akh11} starting from general scattering formulas in the literature \cite{Ana96}. Because of its significance for the characterization and detection of Majorana modes, we give here a self-contained derivation.

The zero-frequency shot noise power is defined by the correlator
\begin{equation}
P_{\rm shot}=\int_{-\infty}^\infty dt\,\langle \delta I_1(0)\delta I_1(t)\rangle,\label{Pshotdefapp}
\end{equation}
of the time-dependent fluctuations $\delta I_1(t)$ of the electrical current into the grounded terminal 1. Terminal 2 is biased at voltage $V$ and connected to terminal 1 via the Majorana edge mode, see Fig.\ \ref{fig_array}.

We introduce vectors of fermion operators
\[
a_{\rm in}=(a_{1e},a_{1h},a_{2e},a_{2h})_{\rm in},\;\; a_{\rm out}=(a_{1e},a_{1h},a_{2e},a_{2h})_{\rm out}
\]
for incoming and outgoing quasiparticles, labeled by the terminal number 1,2 and electron-hole degree of freedom $e,h$. Incoming and outgoing operators are related by the scattering matrix,
\begin{align}
&a_{\rm out}(t)=\frac{1}{\sqrt{4\pi}}\int_{-\infty}^\infty dE\,e^{-iEt} S(E)a_{\rm in}(E),\\
&S=\begin{pmatrix}
r&t\\
t'&r'
\end{pmatrix},\;\;SS^\dagger=1,\;\;S(E)=\tau_x S^\ast(-E)\tau_x,\label{Sdefapp}
\end{align}
composed out of reflection and transmission blocks $r,r',t,t'$. The Pauli matrix $\tau_x$ acts on the electron-hole degree of freedom. (The spin degree of freedom plays no role here.) Since we will be restricting ourselves to the linear response limit $V\rightarrow 0$, we can evaluate the scattering matrix at the Fermi level $E=0$ (and will omit the energy argument of $S$ in what follows).

The current operator at terminal $1$ is given by
\begin{equation}
I_1(t)=ea_{\rm out}^\dagger(t){\cal P}_1 a_{\rm out}(t),\;\;{\cal P}_1=\begin{pmatrix}
\tau_z&0\\
0&0
\end{pmatrix}.
\end{equation}
The projector ${\cal P}_1$ projects onto terminal 1 and the Pauli matrix $\tau_z$ accounts for the opposite charge of electrons and holes. (For notational simplicity we take the electron charge $e>0$ and we have also set $\hbar=1$.) Substitution into Eq.\ \eqref{Pshotdefapp} gives the shot noise power
\begin{align}
&P_{\rm shot}=\frac{e^2}{h}\int_{0}^\infty dE\int_{0}^\infty dE'\int_{0}^\infty dE''\nonumber\\
&\quad\times\left[\langle a_{\rm in}^\dagger(E'){\cal M}a_{\rm in}(E'')a_{\rm in}^\dagger(E){\cal M}a_{\rm in}(E)\rangle\right.\nonumber\\
&\quad\left.-\langle a_{\rm in}^\dagger(E'){\cal M}a_{\rm in}(E'')\rangle\langle a_{\rm in}^\dagger(E){\cal M}a_{\rm in}(E)\rangle\right],\\
&{\cal M}=S^\dagger{\cal P}_1 S=\begin{pmatrix}
r^\dagger\tau_z r&r^\dagger\tau_z t\\
t^\dagger\tau_z r&t^\dagger\tau_z t
\end{pmatrix}.
\end{align}

The fermion operators $a_{\rm in}$ originate from a reservoir in termal equilibrium, so the expectation values are given by the Fermi distribution. In the zero-temperature limit we need to consider only energies in the interval $0<E<eV$, where
\begin{align}
&\langle a_{{\rm in},n}^\dagger(E)a_{{\rm in},m}^{\vphantom{\dagger}}(E')\rangle={\cal F}_{nm}(E)\delta(E-E'),\nonumber\\
&{\cal F}=\begin{pmatrix}
0&0\\
0&(1+\tau_z)/2
\end{pmatrix}.
\end{align}
The term $(1+\tau_z)/2$ in the definition of ${\cal F}$ ensures that terminal 2 injects electrons and not holes (assuming $V>0$). 

The equilibrium expectation value that we need is given by
\begin{align}
&\langle a_{\rm in}^\dagger(E') {\cal M} a_{\rm in}(E'')a_{\rm in}^\dagger(E) {\cal M} a_{\rm in}(E)\rangle\nonumber\\
&\quad-\langle a_{\rm in}^\dagger(E') {\cal M} a_{\rm in}(E'')\rangle\langle a_{\rm in}^\dagger(E) {\cal M} a_{\rm in}(E)\rangle\nonumber\\
&\quad=\delta(E-E')\delta(E-E'') \,{\rm Tr}\,{\cal F}{\cal M}(1-{\cal F}){\cal M},
\end{align}
so we arrive at the shot noise power
\begin{equation}
P_{\rm shot}=\frac{e^2}{h}eV\,{\rm Tr}\,{\cal F}S^\dagger{\cal P}_1 S(1-{\cal F})S^\dagger {\cal P}_1S.
\end{equation}
Using that $(S^\dagger {\cal P}_2 S)^2=S^\dagger {\cal P}_2^2 S$, this evaluates to
\begin{align}
&P_{\rm shot}/P_0=\tfrac{1}{2}\,{\rm Tr}\,(1+\tau_z)t^\dagger t-\tfrac{1}{4}\,{\rm Tr}\,[(1+\tau_z)t^\dagger\tau_z t]^2\nonumber\\
&\quad={\rm Tr}\,(t_{ee}^\dagger t_{ee}^{\vphantom{\dagger}}+t_{he}^\dagger t_{he}^{\vphantom{\dagger}})-{\rm Tr}\,(t_{ee}^\dagger t_{ee}^{\vphantom{\dagger}}-t_{he}^\dagger t_{he}^{\vphantom{\dagger}})^2.\label{Pshotresultapp}
\end{align}
We have defined $P_0=e^3 V/h$, the shot noise quantum of $e^2/h$ per electron volt.

So far we have not yet used that terminals 1 and 2 are connected by an unpaired Majorana mode. To account for that, it is convenient to switch from the electron-hole representation to the Majorana representation, by means of the unitary transformation
\begin{equation}
S_{\rm M}= \Omega S\Omega^{\dagger},\;\;\Omega=\sqrt{\tfrac{1}{2}}\begin{pmatrix}
1&1\\
i&-i
\end{pmatrix}.
\end{equation}
The particle-hole symmetry relation \eqref{Sdefapp} implies that $S_{\rm M}$ is a real orthogonal matrix at the Fermi level. The Pauli matrix $\tau_z$ transforms into $\tau_y$, so that the shot noise expression \eqref{Pshotresultapp} becomes
\begin{align}
P_{\rm shot}/P_0&=\tfrac{1}{2}\,{\rm Tr}\,(1+\tau_y)t_{\rm M}^{\rm T} t_{\rm M}-\tfrac{1}{4}\,{\rm Tr}\,[(1+\tau_y)t_{\rm M}^{\rm T}\tau_y t_{\rm M}]^2\nonumber\\
&=\tfrac{1}{2}\,{\rm Tr}\,t_{\rm M}^{\rm T} t_{\rm M}-\tfrac{1}{4}\,{\rm Tr}\,[(1+\tau_y)t_{\rm M}^{\rm T}\tau_y t_{\rm M}]^2.\label{PMajoranaapp}
\end{align}
(The trace of the antisymmetric matrix $t_{\rm M}\tau_y\tau_{\rm M}^{\rm T}$ vanishes.)

The transmission matrix for an unpaired Majorana mode has rank 1, meaning that $t_{\rm M}$ has only 1 nonzero singular value. We make the polar decomposition
\begin{equation}
t_{\rm M}=U\,{\rm diag}\,(\sqrt{T},0,0,\ldots 0)V
\end{equation} 
in terms of a pair of orthogonal matrices $U,V$ and the transmittance $T$ of the Majorana edge mode. The matrix
\begin{equation}
(t_{\rm M}^{\rm T}\tau_y t_{\rm M})_{nm}=TV_{1n}(U^{\rm T}\tau_y U)_{11}V_{1m}=0
\end{equation}
vanishes identically, because it is proportional to the diagonal element of an antisymmetric matrix. The shot noise power \eqref{PMajoranaapp} thus reduces to
\begin{equation}
P_{\rm shot}/P_0=\tfrac{1}{2}\,{\rm Tr}\,t_{\rm M}^{\rm T} t_{\rm M}=\tfrac{1}{2}T.
\end{equation}
In the absence of backscattering, $T=1$, the shot noise power of the Majorana edge mode equals $P_{\rm shot}=e^3 V/2h$.

\end{document}